\documentclass[11pt]{article}
\usepackage{eurosym}
\usepackage{graphicx}
\usepackage[utf8]{inputenc}
\usepackage[T1]{fontenc}
\usepackage{indentfirst}
\usepackage{float}
\usepackage[margin=0.6in,nomarginpar]{geometry}
\usepackage[final]{hyperref}
\usepackage{amsmath}
\usepackage{hyperref}
\usepackage{cite}
\usepackage{physics}
\usepackage{subcaption}
\usepackage{caption}
\usepackage{amssymb}
\usepackage{multirow}
\usepackage[table]{xcolor}
\usepackage{orcidlink}
\usepackage{multirow,stackengine}
\usepackage{float}
\usepackage{booktabs}

\hypersetup{colorlinks=true, linkcolor=blue, citecolor=blue, urlcolor=blue}
\begin{document}

\title{Thermodynamic and Dynamical Properties of Phantom Charged Black Holes in 4D Einstein–Gauss–Bonnet Gravity}
\author{B. Hamil \orcidlink{0000-0002-7043-6104} \thanks{%
hamilbilel@gmail.com; bilel.hamil@umc.edu.dz } \\
Laboratoire de Physique Math\'{e}matique et Subatomique,\\
Facult\'{e} des Sciences Exactes, Universit\'{e} Constantine 1, Constantine,
Algeria. }
\date{\today }
\maketitle

\begin{abstract}
We present charged black hole solutions of regularized four dimensional Einstein Gauss Bonnet gravity coupled to a phantom electromagnetic field. We investigate the combined effects of higher curvature corrections and phantom charge on the horizon structure, thermodynamics, geodesic motion, and scalar perturbations. The phantom sector exhibits properties that differ qualitatively from those of the ordinary Maxwell case, including the absence of critical thermodynamic behavior and persistent thermal instability. Circular geodesics and accretion efficiency are also significantly modified. Quasinormal modes are computed using the sixth-order WKB approximation and verified through time-domain evolution. The results reveal characteristic signatures of both the Gauss Bonnet coupling and phantom electrodynamics, while confirming linear stability against scalar perturbations.
\end{abstract}

\section{Introduction}

\label{sec1}

The predictive success of General Relativity (GR) in describing gravitational phenomena has been extensively confirmed by a wide range of astrophysical and cosmological observations. Nevertheless, several theoretical and observational challenges suggest that GR may not represent the final theory of gravity. On the theoretical side, the occurrence of spacetime singularities and the absence of a consistent quantum formulation of gravity motivate the search for extensions of Einstein's theory. Observationally, phenomena commonly attributed to dark matter and dark energy continue to stimulate the development of alternative gravitational frameworks \cite{Austin,Leor}. The growing catalog of gravitational-wave detections has strengthened this effort by providing new tests of gravity in strong-field regimes, where deviations from GR may become observable \cite{Abbott,BAbbott,BPAbbott,RAbbott}. Consequently, compact objects and their observational signatures have become pivotal tools for investigating the viability of modified gravity theories.

Higher curvature modifications of gravity can be introduced without sacrificing the second-order nature of the gravitational equations. The most general realization of this idea is provided by Lovelock gravity \cite{Lovelock}, whose action extends the Einstein-Hilbert form through a series of geometric invariants. The simplest nontrivial contribution in the Lovelock hierarchy is the Gauss Bonnet (GB) term. A notable feature of Lovelock gravity is that the higher curvature corrections do not introduce higher-order derivatives in the equations of motion, thereby avoiding many of the pathologies encountered in generic higher curvature theories. The stability of black hole solutions in GB gravity has been widely investigated under scalar, vector, and tensor perturbations \cite{Gustavo,Dott,Gleiser,Beroiz,Zhidenko}, revealing a stability behavior that depends on the spacetime dimension.

Higher curvature terms arise naturally in the low energy effective framework of string theory \cite{PCandelas,Barton,Bruno}. In general, such corrections give rise to ghost-like degrees of freedom, which can affect the consistency of the theory at the quantum level. However, the GB combination represents a notable exception, as it preserves the second-order structure of the gravitational field equations \cite{Gross}. Owing to this distinctive property, GB gravity has received considerable attention as a well-motivated extension of GR for exploring higher curvature effects. Static, spherically symmetric black hole solutions are well established within GR. In four spacetime dimensions, however, the GB term reduces to a topological invariant and does not contribute to the gravitational dynamics. Consequently, the GB contribution does not lead to additional four-dimensional black hole solutions, and the theory remains equivalent to Einstein gravity at the level of classical black hole geometries. This equivalence severely restricts the possibility of testing GB effects via astrophysical observations

To address this limitation, Glavan and Lin \cite{Lin} introduced a four-dimensional Einstein-Gauss-Bonnet (4D-EGB) framework by rescaling the coupling constant as $\alpha \rightarrow \alpha/(D-4)$ and subsequently taking the limit $D \rightarrow 4$. This prescription effectively renders the GB term dynamical in four dimensions and results in novel black hole solutions with nontrivial curvature corrections. A related regularization approach was earlier suggested by Tomozawa \cite{Tomozawa} in the context of quantum corrections to Einstein gravity, where static, spherically symmetric solutions were also derived. This construction was subsequently reformulated by Cognola et al. \cite{Tomozawa} within a classical Lagrangian framework, yielding effective contributions associated with the GB invariant. The resulting four-dimensional black hole geometries have been shown to be related to solutions arising in semiclassical gravity with conformal anomaly \cite{Ohta}, in regularized Lovelock models \cite{Lovelock}, and in specific subclasses of Horndeski scalar-tensor theories \cite{Pang}. Since its proposal, 4D-EGB gravity has attracted substantial attention, and a wide range of studies has explored its black hole structure, thermodynamic behavior, gravitational lensing, quasinormal spectra, and cosmological implications \cite%
{Aoki,CLiu,Guo,Wei,Zubair,Malafarina,Mansoori,XHGe,Rayimbaev,Chakraborty,Odintsov,KYang, BAhmedov,RKumar,SGGhosh,Larranaga,Hamil,Ayoub}.
 
In recent years, black hole solutions sourced by exotic forms of matter have attracted considerable interest. Among these, phantom fields are particularly noteworthy because they violate the null and weak energy conditions and are characterized by an equation-of-state parameter $\omega_{\rm phantom}<-1$ \cite{Caldwell}. Although their physical interpretation remains uncertain, phantom fields arise in various cosmological and modified-gravity models and remain compatible with current observational constraints \cite{Nojiri,Elizalde,Melchiorri}. The presence of phantom matter significantly modifies the geometry and thermodynamics of black hole spacetimes. In particular, it alters the horizon structure and causal properties of the solution, leading to deviations from standard vacuum and electrovacuum geometries \cite{Clement,ronnikov}. For charged black holes, especially those of Reissner--Nordström type, phantom fields modify the effective electromagnetic contribution and consequently affect the global spacetime structure \cite{Jamil,Jardim}. Moreover, the accretion of phantom energy can reduce the black hole mass, with potential implications for cosmic censorship and black hole thermodynamics \cite{Izquierdo,Sadjadi}. Phantom fields also influence observable phenomena such as photon trajectories and gravitational lensing in the strong-field regime \cite{Ding,Maria}. Their negative-energy contribution modifies the Hawking temperature and evaporation process and may even change the horizon configuration of charged black holes \cite{MJamil,Jardim}. In extended black hole thermodynamics, phantom matter can suppress the usual Van der Waals-like critical behavior and give rise to unconventional phase structures and stability properties \cite{Quevedo,Zotos,MAkbar,Eslam,BEslam}. These features make phantom black holes an interesting framework for exploring the interplay between exotic matter, spacetime geometry, and thermodynamics.

The interplay between higher curvature gravity and exotic matter fields provides a useful framework for probing deviations from General Relativity in the strong-field regime. In particular, EGB gravity introduces ultraviolet curvature corrections that become relevant near black hole horizons, while phantom fields violate standard energy conditions and induce effective repulsive gravitational effects. Their coexistence leads to a nontrivial competition between higher curvature terms and negative-energy matter sources. This combination can significantly modify black hole geometry and dynamics. The Gauss–Bonnet term affects the short-distance structure of spacetime, whereas the phantom electromagnetic sector alters the effective charge in the metric function. Consequently, the horizon structure, thermodynamic behavior, geodesic motion, and stability properties may deviate markedly from standard Einstein-Maxwell solutions. In some cases, phantom contributions may also soften or modify singular structures arising in higher curvature backgrounds, potentially leading to less singular configurations. An additional motivation arises from black hole perturbations. Quasinormal modes are highly sensitive to both the background geometry and the matter content. Therefore, the combined effects of EGB corrections and phantom electrodynamics may induce characteristic shifts in oscillation spectra. These signatures are particularly relevant for gravitational-wave observations, where quasinormal modes probe near-horizon physics and possible deviations from General Relativity.

In the present work, we adopt the regularized 4D EGB construction of Glavan and Lin \cite{Lin} as an effective description of higher curvature gravitational effects in four dimensions. Regardless of the ongoing discussions concerning the fundamental consistency of the regularization scheme, the corresponding black hole solutions have been extensively used as phenomenological models for probing deviations from General Relativity. Motivated by this approach, the present work explores the geometric, 
thermodynamic, and dynamical features of such black hole configurations embedded 
in a phantom electromagnetic field background. The manuscript is outlined as 
follows. In Sect.\ref{sec2}, the exact black hole solution is established. The corresponding thermodynamic behavior is analyzed in Sect.\ref{sec3}. Sect.\ref{sec4} is concerned with the investigation of geodesic motion, whereas Sect.\ref{sec5} addresses the calculation of quasinormal modes (QNMs). Phenomenological constraints from gravitational waves are discussed in Sect.\ref{sec6}. Finally, a summary of our primary findings and concluding remarks are presented in Sect.\ref{sec7}.

\section{Derivation and Geometry of 4D EGB Phantom Black Holes}

\label{sec2}
This section focuses on deriving phantom black hole solutions within the framework of 4D EGB gravity. We consider a spacetime geometry characterized by the following gravitational action:

\begin{equation}
S=\frac{1}{16\pi }\int d^{D}x\sqrt{-g}\left[ R+\alpha \mathcal{L}^{\text{GB}%
}-4\eta \mathcal{L}^{\text{EM}}\right] ,  \label{eq1}
\end{equation}
where $g$ denotes the metric tensor determinant, $R$ represents the Ricci scalar, and $\alpha$ stands for the GB coupling parameter. The parameter $\eta $
distinguishes the ordinary Maxwell sector ($\eta =1$) from the phantom
electromagnetic sector ($\eta =-1$). 

Following \cite{Lanczos}, the explicit form of the GB Lagrangian density reads:

\begin{equation}
\mathcal{L}^{\text{GB}}=R^{2}-4R_{ab}R^{ab}+R_{abcd}R^{abcd}.
\end{equation}

In action (\ref{eq1}), the term $\mathcal{L}^{\text{EM}}$ denotes the electromagnetic contribution, defined as:
\begin{equation}
\mathcal{L}^{\text{EM}}=\frac{F_{\mu \nu }F^{\mu \nu }}{4}.  \label{eq3}
\end{equation}%
Here, the field strength tensor is anti-symmetric and related to the gauge potential $A_{\mu}$ via:

\begin{equation}
F_{\mu \nu }=\partial _{\mu }A_{\nu }-\partial _{\nu }A_{\mu }.
\end{equation}
Varying the action functional with respect to the spacetime metric $g_{\mu\nu}$ and the gauge field $A_{\mu}$ leads to the coupled gravitational and electromagnetic field equations:

\begin{equation}
R_{ab}-\frac{1}{2}g_{ab}R+\alpha H_{ab}^{\left( \text{GB}\right) }=8\pi \eta
T_{ab}^{^{\left( \text{EM}\right) }},  \label{Einstein}
\end{equation}%
\begin{equation}
\frac{1}{\sqrt{-g}}\partial _{\mu }\left( \sqrt{-g}F^{\mu \nu }\right) =0.
\label{conserved}
\end{equation}

The contribution of the Lanczos tensor $H_{ab}^{\left( \text{GB}\right) }$ to these field equations can be expressed as \cite{Lanczos}:

\begin{equation}
H_{ab}^{\left( \text{GB}\right) }=2\left( RR_{ab}-R_{acd\lambda
}R_{b}^{d\lambda c}-2R_{adbc}R^{dc}-2R_{ac}R_{b}^{c}\right) -\mathcal{L}^{%
\text{GB}}\text{ }g_{ab},
\end{equation}
while the energy-momentum tensor corresponding to the phantom electromagnetic configuration takes the form:

\begin{equation}
T_{\mu \nu }^{\left( \text{EM}\right) }=\frac{1}{4\pi }\left( g^{\rho \sigma
}F_{\mu \rho }F_{\nu \sigma }-\frac{1}{4}g_{\mu \nu }F_{\rho \sigma }F^{\rho
\sigma }\right) .
\end{equation}%
To investigate static, spherically symmetric black holes, we employ a $D$-dimensional line element of the form:
\begin{equation}
ds^{2}=-\mathcal{F}\left( r\right) dt^{2}+\frac{1}{\mathcal{F}\left(
r\right) }dr^{2}+r^{2}d\Omega _{D-2}^{2}.  \label{metric}
\end{equation}
For the gauge field sector, we assume a purely electrostatic configuration described by:
\begin{equation}
A_{\mu }=\varepsilon \left( r\right) \delta _{\mu }^{t}.  \label{gauge}
\end{equation}%
Inserting this ansatz into the Maxwell equation (\ref{conserved}) leads to the differential relation:
\begin{equation}
\frac{d}{dr}\left( r^{D-2}\frac{d}{dr}\varepsilon \left( r\right) \right) =0,
\end{equation}%
which integrates directly to yield:
\begin{equation}
\varepsilon \left( r\right) =-\frac{\left( D-3\right) q}{r^{D-3}}.
\end{equation}%
Here, the integration constant $q$ represents the total electric charge.

It is known that the GB term acts as a topological invariant in four dimensions, meaning it fails to contribute dynamically to the field equations. To circumvent this, we utilize the regularization scheme proposed in \cite{Lin}, where a non-trivial 4D EGB theory is recovered by rescaling the coupling parameter:

\begin{equation}
\alpha \rightarrow \frac{\alpha }{D-4},  \label{constant}
\end{equation}%
followed by the limit $D\rightarrow 4$. Evaluating the field equations under this limiting procedure yields:

\begin{equation}
\frac{\mathcal{G}\left( r\right) ^{\prime }}{r}+\frac{\mathcal{G}\left(
r\right) }{r^{2}}-\alpha \left[ \frac{2\mathcal{G}\left( r\right) \mathcal{G}%
\left( r\right) ^{\prime }}{r^{3}}-\frac{\mathcal{G}\left( r\right) ^{2}}{%
r^{4}}\right] +\eta \frac{q^{2}}{r^{4}}=0.  \label{12}
\end{equation}%
where the metric profile is parameterized as:
\begin{equation}
\mathcal{F}\left( r\right) =1+\mathcal{G}\left( r\right) .
\end{equation}%
Solving Eq. (\ref{12}) leads to the general expression:
\begin{equation}
\mathcal{G}\left( r\right) =\frac{r^{2}}{2\alpha }\left( 1\pm \sqrt{1+\frac{%
4\alpha }{r^{2}}\left( \frac{2M}{r}-\eta \frac{q^{2}}{r^{2}}\right) }\right).
\end{equation}%
Consequently, the full metric function for the 4D phantom EGB black hole is given by:
\begin{equation}
\mathcal{F}\left( r\right) =1+\frac{r^{2}}{2\alpha }\left( 1\pm \sqrt{1+%
\frac{4\alpha }{r^{2}}\left( \frac{2M}{r}-\eta \frac{q^{2}}{r^{2}}\right) }%
\right) .  \label{20}
\end{equation}%
To understand the physical structure of the solution,  we examine the general relativity limit ($\alpha \rightarrow 0$). For the negative branch, this yields:

\begin{equation}
\mathcal{F}\left( r\right) =1-\frac{2M}{r}+\eta \frac{q^{2}}{r^{2}},
\end{equation}%
whereas the positive branch reduces to:
\begin{equation}
\mathcal{F}\left( r\right) =1+\frac{r^{2}}{\alpha }+\frac{2M}{r}-\eta \frac{%
q^{2}}{r^{2}}.
\end{equation}

The negative branch successfully recovers the standard Reissner-Nordstr\"{o}m geometry for both  Maxwell and phantom electromagnetic fields. The positive branch is
generally regarded as unphysical due to the appearance of graviton
instabilities associated with the positive mass term \cite{Wheeler}.
Therefore, only the negative branch will be considered in the present
analysis.

To guarantee a real-valued spacetime metric, the term inside the square root of  (\ref{20}), defined as:
\begin{equation}
\sum \left( r\right) =1+\frac{4\alpha }{r^{2}}\left( \frac{2M}{r}-\eta \frac{%
q^{2}}{r^{2}}\right) ,
\end{equation}%
must be strictly non-negative. A branch singularity emerges from these higher curvature corrections at the radius $r_{b}$ where $\sum \left( r_{b}\right) =0$. Setting this argument to zero produces the quartic polynomial

\begin{equation}
r^{4}+8\alpha Mr-4\eta \alpha q^{2}=0.  \label{meq}
\end{equation}

\begin{enumerate}
\item \textbf{Standard Maxwell case ($\eta =+1$):} Eq. (\ref{meq}) yields a single positive real root, $r_{b}>0$. This restricts the physically meaningful domain of the spacetime to $r>r_{b}$. For radii smaller than $r_{b}$, the metric function turns complex, indicating a breakdown of the geometric description. Note that in neutral limit $q=0$, the branch singularity reduces to $r_{b}=0$, recovering the Schwarzschild-like central singularity.

\item \textbf{Phantom electromagnetic case ($\eta =-1$):} Here, the square root argument takes the form:
\begin{equation}
\sum \left( r\right) =1+\frac{4\alpha }{r^{2}}\left( \frac{2M}{r}+\frac{q^{2}%
}{r^{2}}\right) .
\end{equation}%
For positive values of the parameters $\alpha $, $M$, and $q$, all terms in $\sum \left( r\right) $ stay positive throughout the region $r>0$. Because $\sum \left( r\right) >0$ everywhere, the corresponding quartic relation
\begin{equation}
r^{4}+8\alpha Mr+4\alpha q^{2}=0,
\end{equation}%
has no real positive roots. As a result, the phantom configuration remains entirely free of branch singularities.
\end{enumerate}

\begin{figure}[H]
\begin{minipage}[t]{0.5\textwidth}
        \centering
        \includegraphics[width=\textwidth]{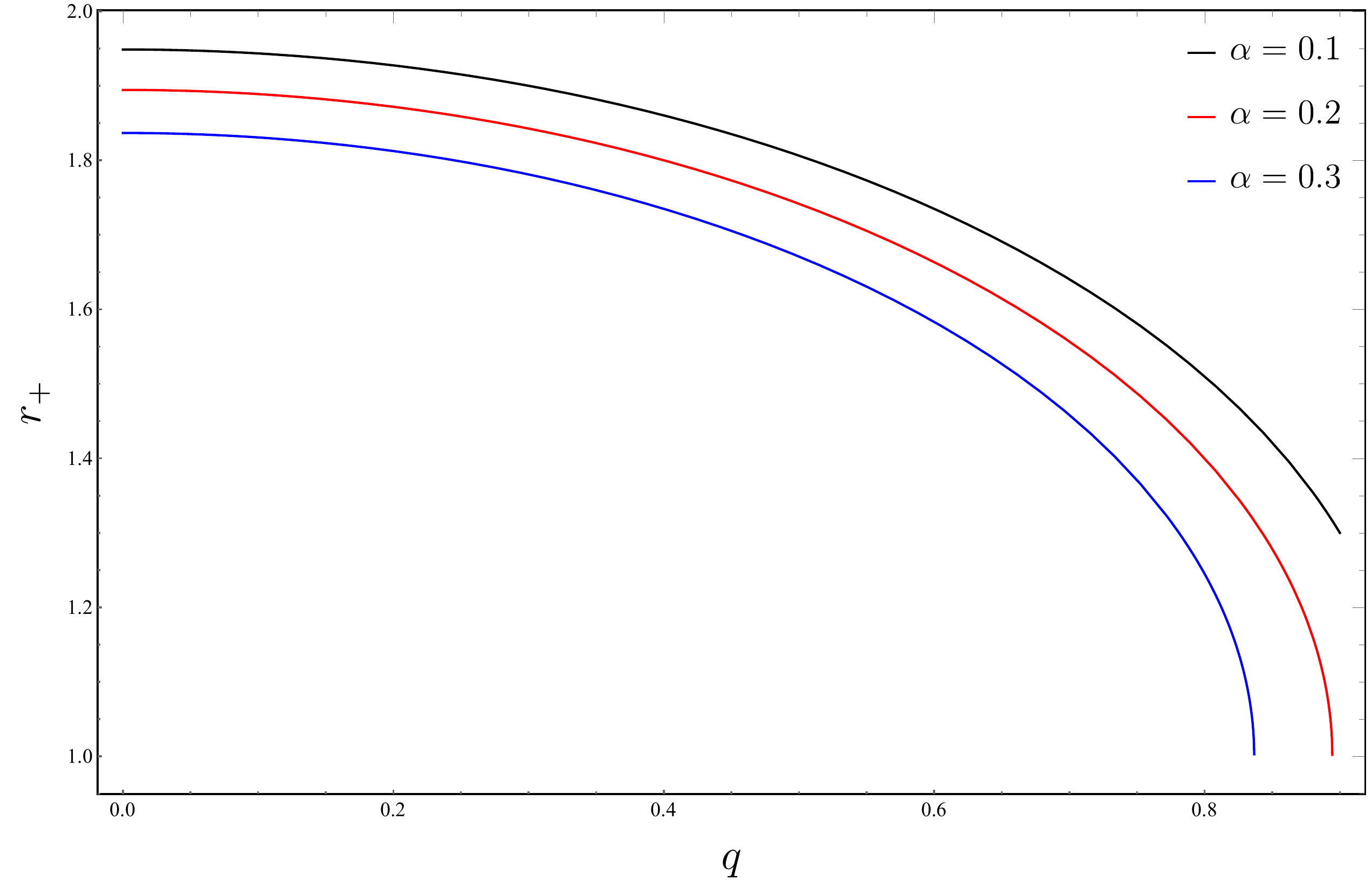}
                \subcaption{$\eta=1$}
        \label{fig:me1}
\end{minipage}
\begin{minipage}[t]{0.5\textwidth}
        \centering
        \includegraphics[width=\textwidth]{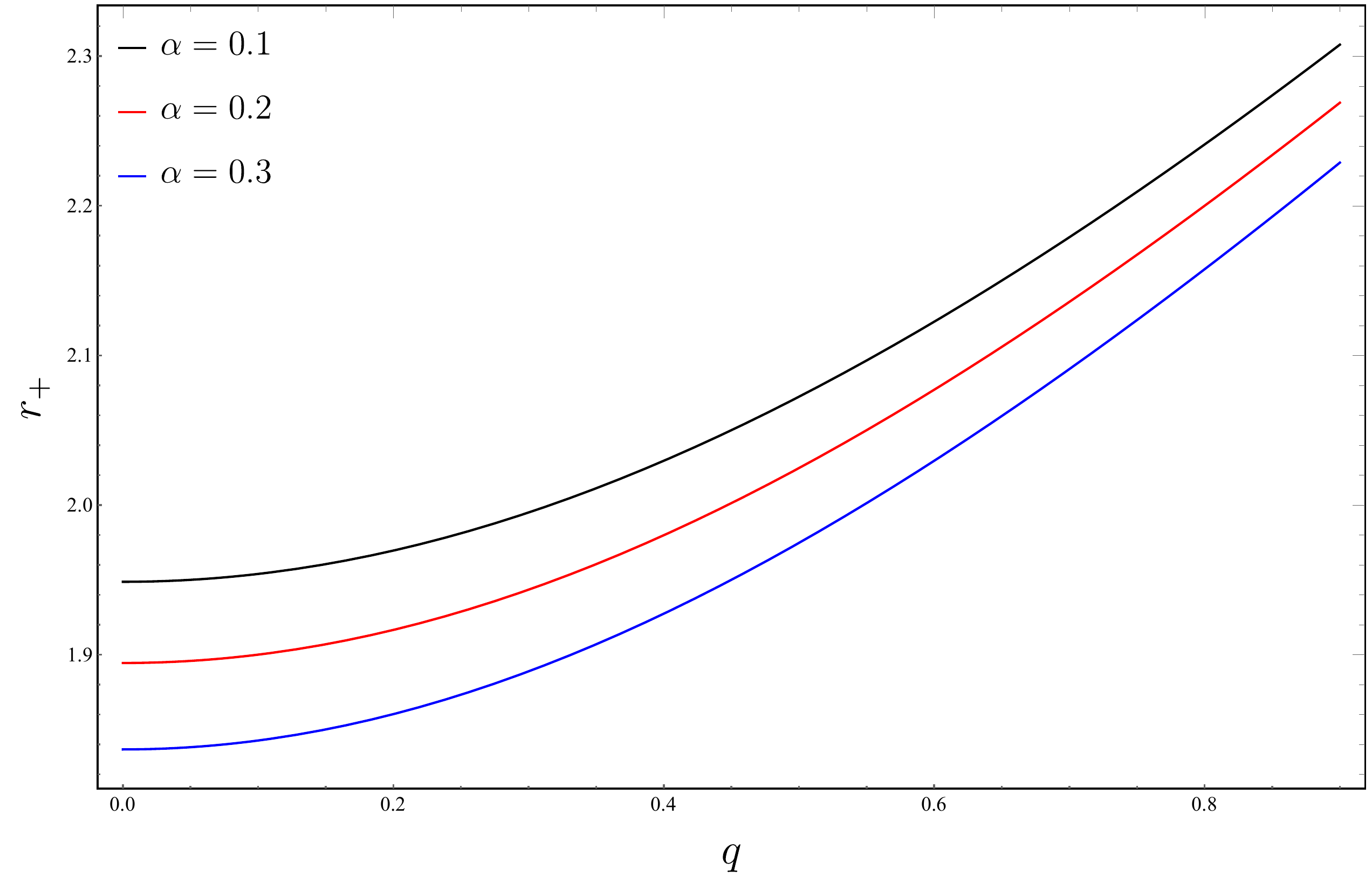}
                \subcaption{$\eta=-1$}
        \label{fig:me2}
\end{minipage}
\caption{The event horizon radius $r_{+}$ as a function of the electric
charge $q$ for different values of the GB coupling parameter.}
\label{fig:metric1}
\end{figure}

\begin{figure}[H]
\begin{minipage}[t]{0.5\textwidth}
        \centering
        \includegraphics[width=\textwidth]{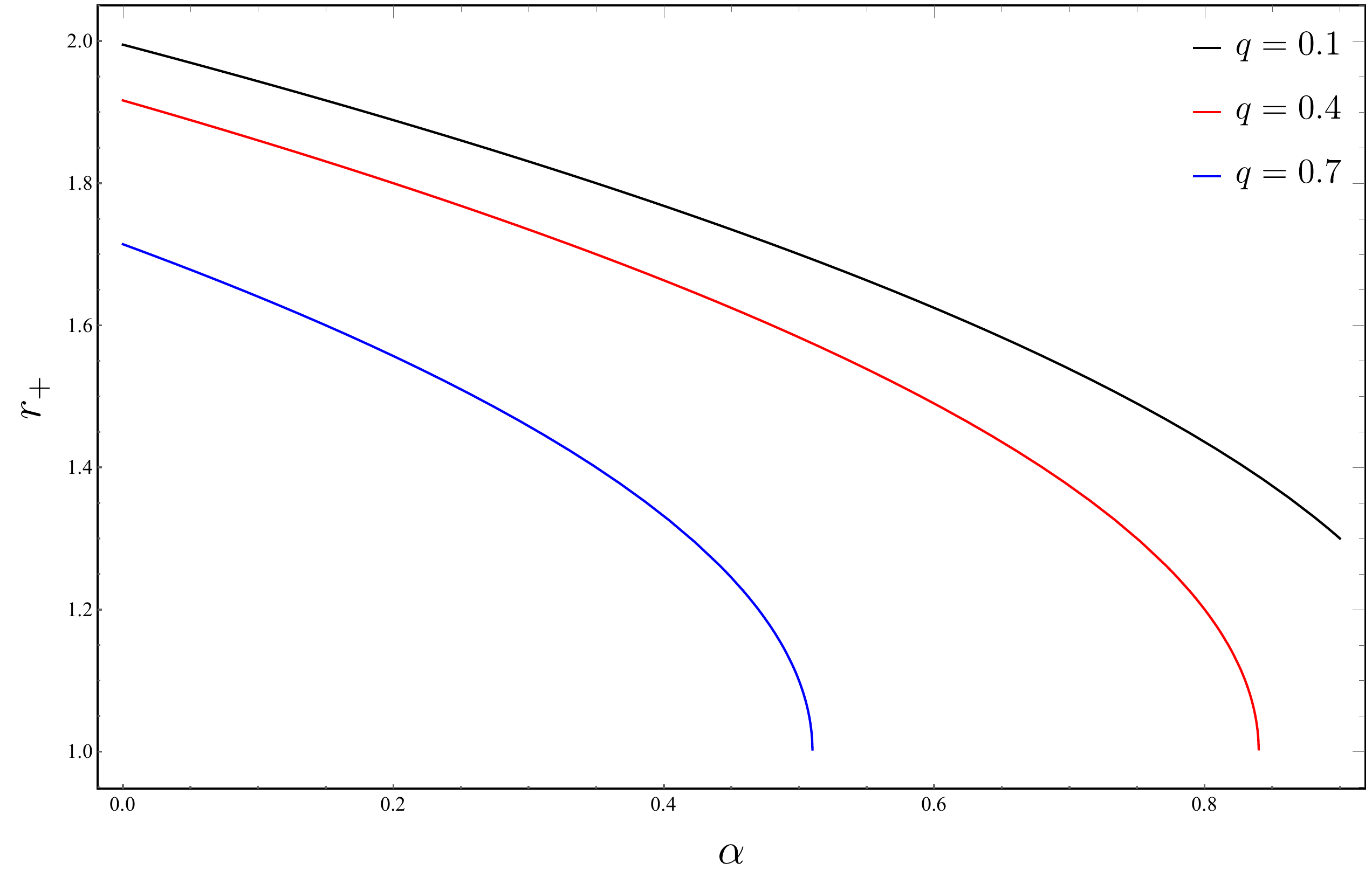}
                \subcaption{$\eta=1$}
        \label{fig:me11}
\end{minipage}
\begin{minipage}[t]{0.5\textwidth}
        \centering
        \includegraphics[width=\textwidth]{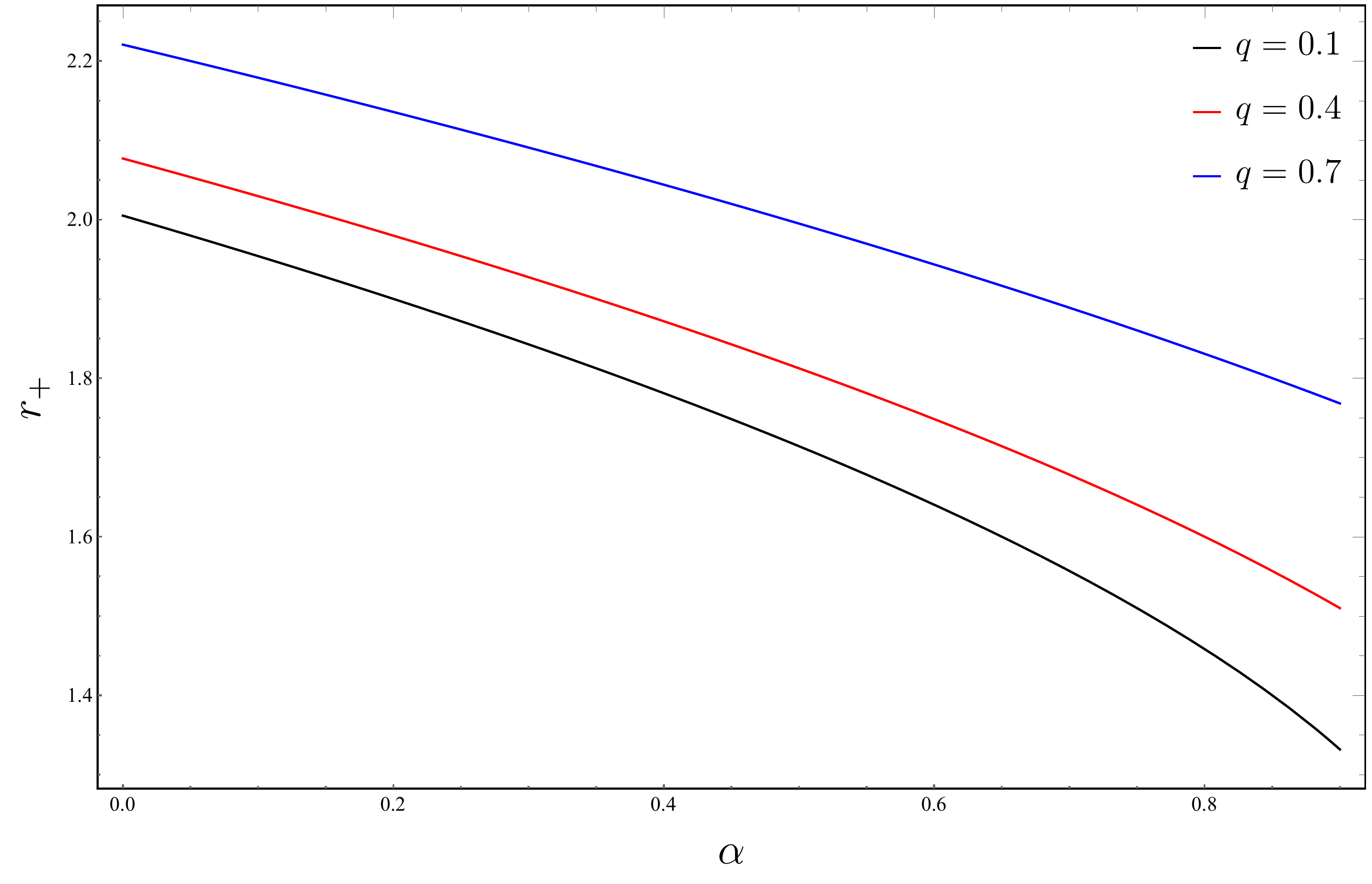}
                \subcaption{$\eta=-1$}
        \label{fig:me22}
\end{minipage}
\caption{The event horizon radius $r_{+}$ as a function of the GB coupling
parameter $\protect\alpha $ for different values of the electric charge $q$. 
}
\label{fig:metric2}
\end{figure}

We locate the event horizons by finding the roots of the condition $\mathcal{F}(r)=0$. Solving this explicitly gives the horizon radii:
\begin{equation}
r_{+}=M+\sqrt{M-\left( \alpha +\eta q^{2}\right) }\text{; \ \ \ }r_{-}=M-%
\sqrt{M-\left( \alpha +\eta q^{2}\right) }.
\end{equation}
where $r_+$ and $r_-$ identify the outer event horizon and the inner Cauchy horizon, respectively.
The obtained geometry therefore resembles a Reissner-Nordström-type
black hole modified by both higher curvature GB corrections and phantom field configurations.
The general behavior of the outer horizon radius is illustrated in Figs. (\ref{fig:metric1}) and (\ref{fig:metric2}).

Specifically, Fig. (\ref{fig:metric1}) illustrates the functional dependence of $r_+$ on the electric charge $q$ for varying choices of the GB coupling parameter $\alpha$. Within the standard Maxwell framework, the outer horizon radius exhibits a monotonic decrease as the net charge increases. On the contrary, in the phantom sector, the  horizon radius increases with $q$. In both cases, larger values of the GB parameter
lead to smaller horizon radii for fixed charge. This behavior can be
understood from the effective contribution of the charge term $\eta q^{2}$
in the metric function.  When $\eta = +1$, the classical electromagnetic sector induces a repulsive effect that diminishes the black hole boundary, whereas the phantom sector ($\eta = -1$) acts to amplify the local gravitational attraction, effectively increasing the horizon radius.

Furthermore, Fig. (\ref{fig:metric2}) illustrates the variations of the event horizon radius as a function of the GB parameter $\alpha $ for different values of the electric
charge. In both sectors, increasing $\alpha $ decreases the horizon radius,
indicating that higher curvature corrections strengthen the effective
gravitational interaction at short distances. However, the effect of the
charge remains qualitatively different in the two sectors: increasing $q$
decreases the horizon radius in the ordinary Maxwell case, while it enlarges
the horizon in the phantom configuration.


\section{Thermodynamic Properties and Stability}

\label{sec3}
The thermodynamic analysis of these black hole solutions requires a systematic expression for the mass $M$ in terms of the independent state parameters $r_{+}$, $q$, $\alpha$ and $\eta$. By imposing the horizon condition $\mathcal{F}(r_+) = 0$ subject to the physical lower bound $r_{+} > r_{\text{b}}$, the mass can be written as:

\begin{equation}
M=\frac{r_{+}}{2}\left( 1+\frac{\alpha }{r_{+}^{2}}+\eta \frac{q^{2}}{%
r_{+}^{2}}\right) .  \label{25}
\end{equation}%
We examine the behavior of this mass function in limiting regimes. In the small-horizon limit $r_{+}\rightarrow 0$, one finds%

\begin{equation}
\lim_{r_{+}\rightarrow 0}M=\frac{\alpha +\eta q^{2}}{2r_{+}},
\end{equation}%
showing  that the mass is an explicit function of the electric charge, the GB coupling, and the specific nature (Maxwell or phantom) of the electromagnetic sector. Furthermore, in the asymptotic regime $r_{+}\rightarrow \infty $, the mass behaves as%

\begin{equation}
\lim_{r_{+}\rightarrow \infty }M=\frac{r_{+}}{2},
\end{equation}%
indicating that the effects of $\alpha $, $q$ and $\eta $ become negligible
for large black holes.

The mass function possesses a minimum value,

\begin{equation}
M_{\text{min}}=\sqrt{\alpha +\eta q^{2}},
\end{equation}

which characterizes the extremal configuration where the two horizons merge, 
$r_{+}=r_{-}$. Consequently, the existence of a black-hole horizon requires

\begin{equation}
M^{2}\geq \alpha+\eta q^{2}.
\end{equation}

For the ordinary Maxwell sector ($\eta =1$), this condition reduces to $%
M^{2}\geq \alpha +q^{2}$ and always admits an extremal black hole with $M_{%
\text{min}}=\sqrt{\alpha +q^{2}}$. In contrast, for the phantom sector ($%
\eta =-1$), the horizon condition becomes

\begin{equation}
M^{2}\geq \alpha-q^{2}.
\end{equation}

Hence, the allowed parameter space depends on the relative magnitude of $%
\alpha$ and $q^{2}$. For $\alpha>q^{2}$, an extremal configuration exists
with

\begin{equation}
M_{\text{min}}=\sqrt{\alpha -q^{2}},
\end{equation}

whereas for $q^{2}>\alpha $ the right-hand side of the horizon condition is
negative and the inequality is automatically satisfied for any positive
mass. In this latter regime no extremal solution exists. The case $\alpha
=q^{2}$ defines the transition between these two behaviors and corresponds
to $M_{\text{min}}=0$.

\begin{figure}[H]
\begin{minipage}[t]{0.5\textwidth}
        \centering
        \includegraphics[width=\textwidth]{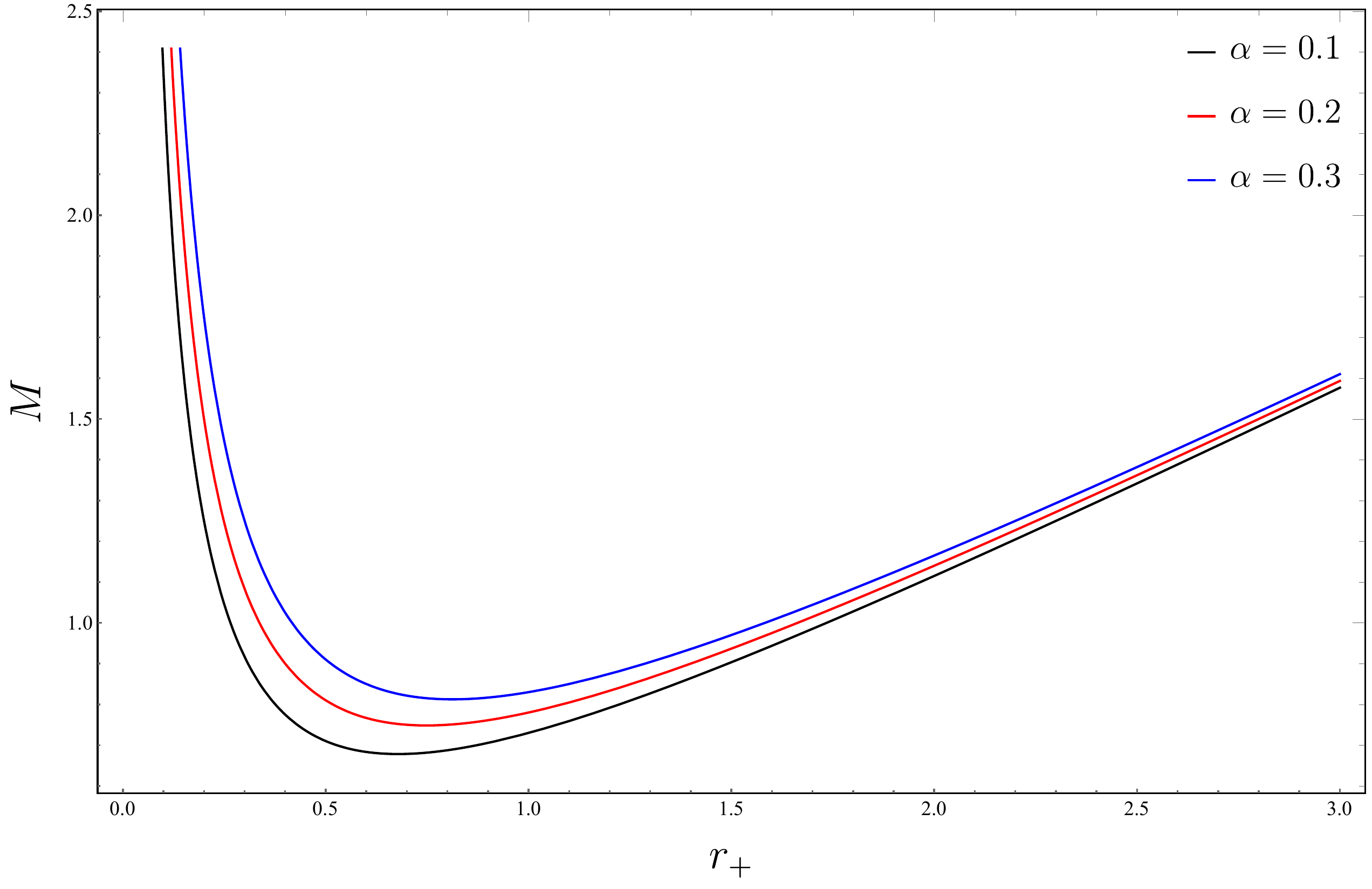}
                \subcaption{$\eta=1$}
        \label{fig:m1}
\end{minipage}
\begin{minipage}[t]{0.5\textwidth}
        \centering
        \includegraphics[width=\textwidth]{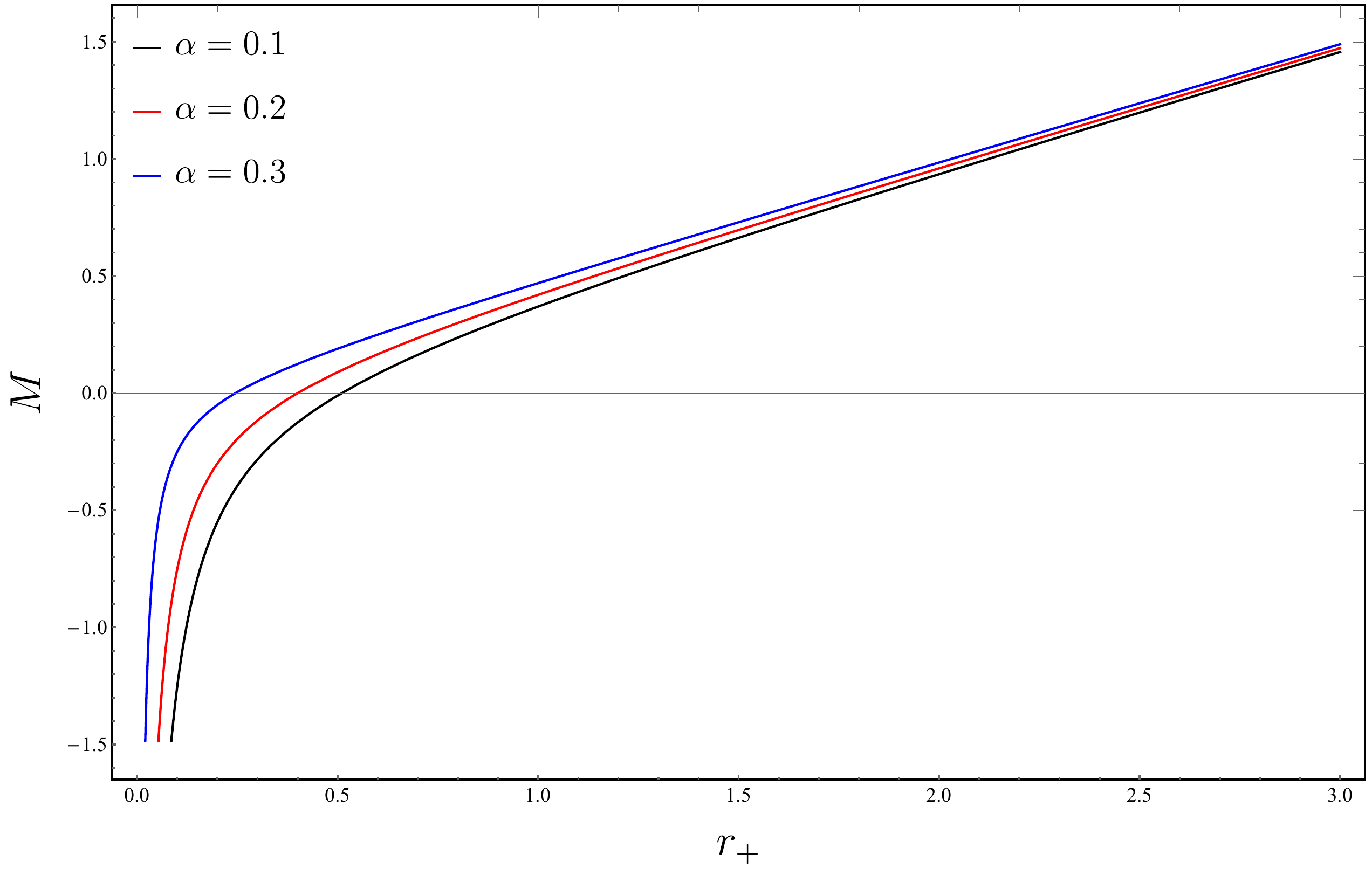}
                \subcaption{$\eta=-1$}
        \label{fig:m11}
\end{minipage}
\caption{The black hole mass $M$ versus $r_{+}$ for a fixed charge $q=0.5$ and various choices of the GB parameter $\alpha$.}
\label{fig:mass1}
\end{figure}

Figure \ref{fig:mass1} illustrates how the black hole mass varies with the outer horizon radius. In the standard Maxwell regime ($\eta = +1$), the mass curve displays a local minimum $M_{\min }$ at a critical radius defined by $r_{0}=\sqrt{\alpha +q^{2}}$. Increasing the GB parameter $\alpha$ systematically shifts this minimum toward larger horizon scales. Additionally, the mass diverges as the horizon radius approaches zero ($r_{+} \rightarrow 0$).

In contrast, in the phantom regime, the mass increases monotonically with
the horizon radius. However, for $M<M_{\min }$, the mass becomes negative,
indicating the absence of physically acceptable black hole configurations in
this regime.

The Hawking temperature is evaluated using the surface gravity relation, $T = \kappa / (2\pi)$, where $\kappa = \frac{1}{2} \left. \frac{\partial \mathcal{F}(r)}{\partial r} \right|_{r=r_{+}}$. Applying this definition to the metric function in Eq. (\ref{20}) yields:

\begin{equation}
T=\frac{1}{4\pi r_{+}}\frac{1-\frac{\alpha +\eta q^{2}}{r_{+}^{2}}}{1+\frac{%
2\alpha }{r_{+}^{2}}}.  \label{28}
\end{equation}

\begin{figure}[H]
\begin{minipage}[t]{0.5\textwidth}
        \centering
        \includegraphics[width=\textwidth]{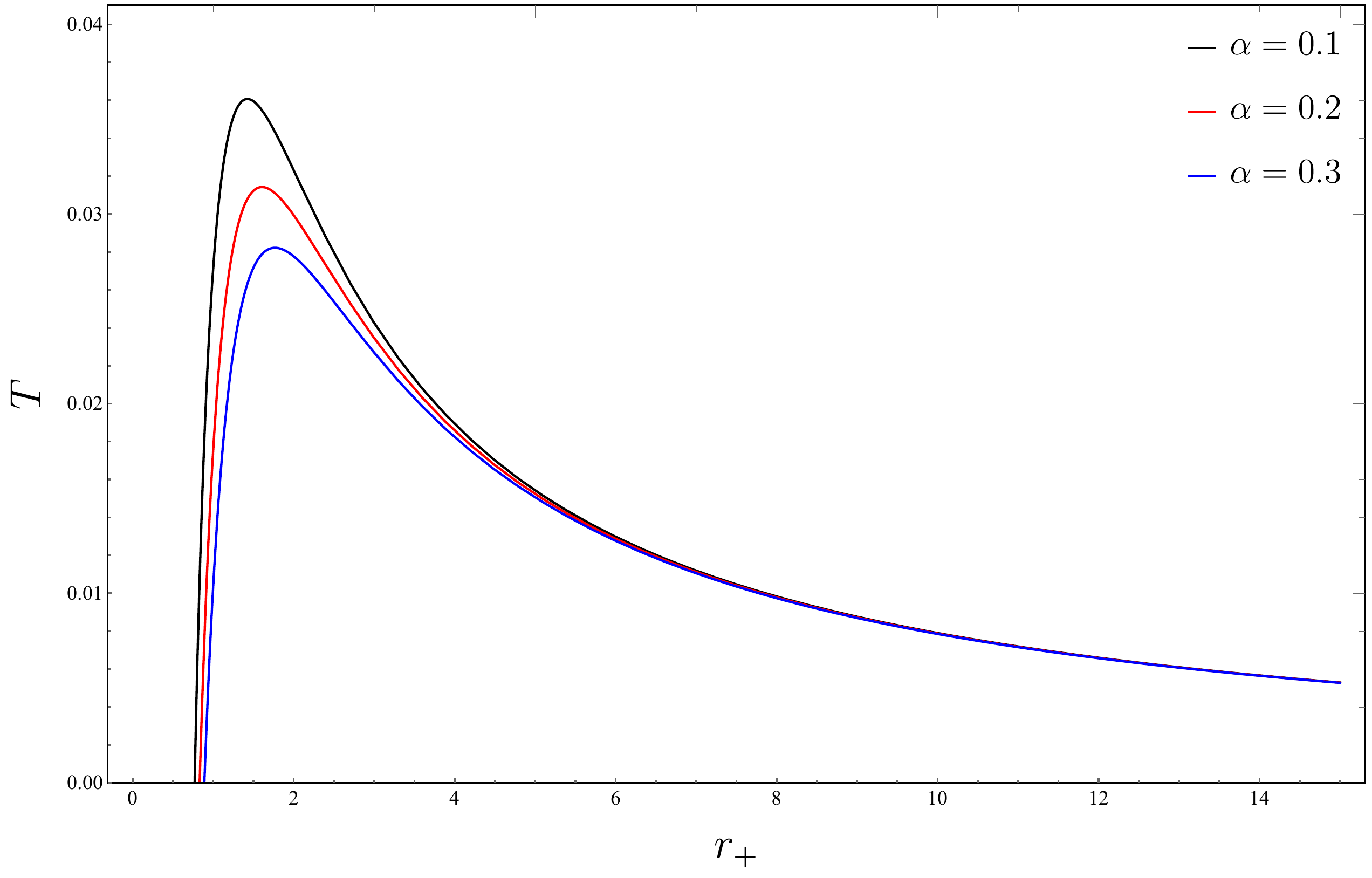}
                \subcaption{$\eta=1$}
        \label{fig:t1}
\end{minipage}
\begin{minipage}[t]{0.5\textwidth}
        \centering
        \includegraphics[width=\textwidth]{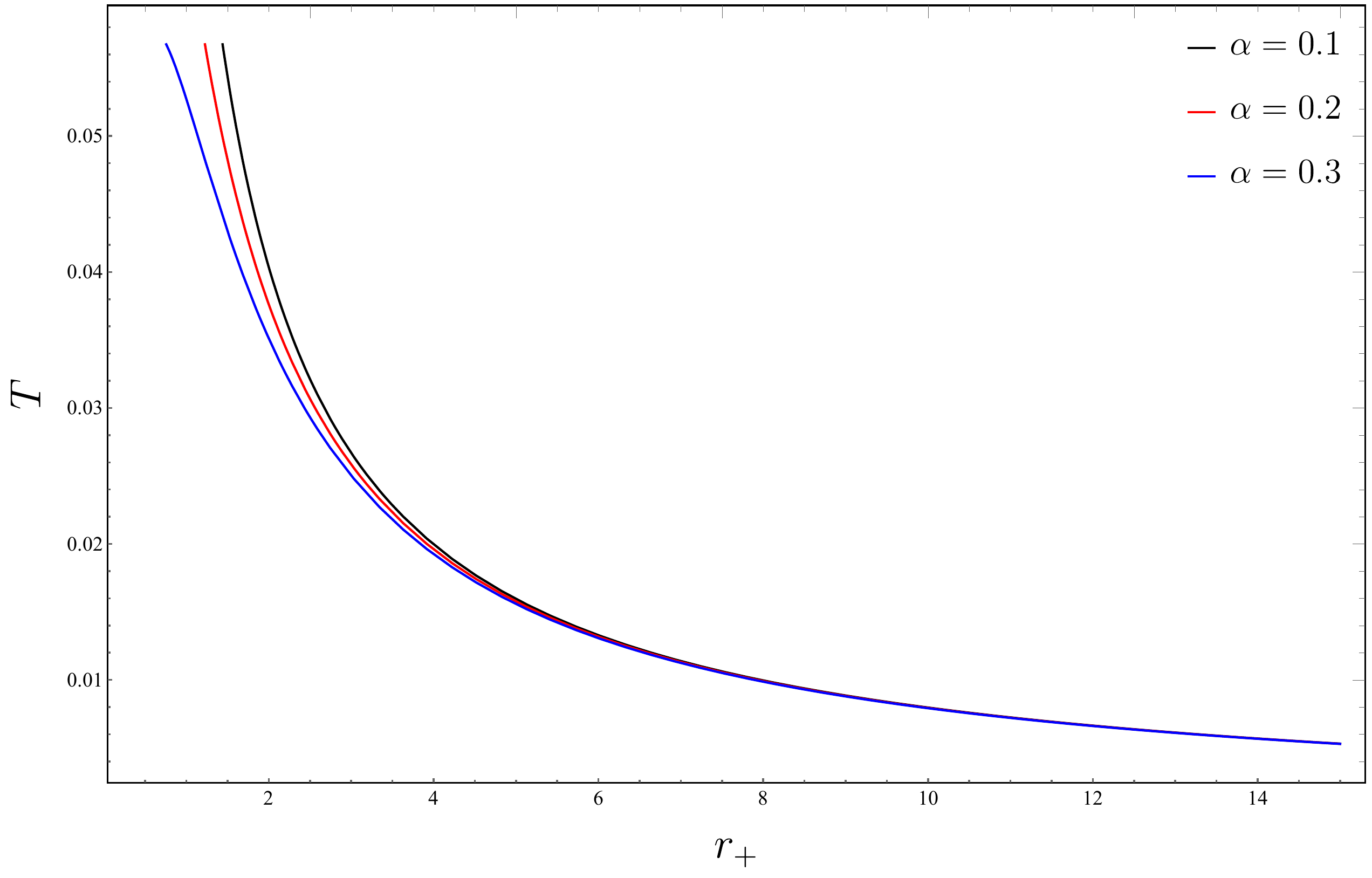}
                \subcaption{$\eta=-1$}
        \label{fig:t11}
\end{minipage}
\caption{Hawking temperature $T$ as a function of the event horizon radius $r_{+}$ for a fixed charge $q=0.5$ and different GB coupling parameters}
\label{fig:temp1}
\end{figure}

Notably, small black holes with $r_{+}<\sqrt{\alpha +\eta q^{2}}$, are not physical, as the temperature becomes negative in this region.  The functional dependence of the temperature on the horizon scale across different values of $\alpha$ is plotted in Fig. \ref{fig:temp1} for both electromagnetic sectors.

Panel (\ref{fig:t1}) illustrates the thermal behavior of the Maxwell case.
As the black hole evaporates, the temperature initially increases until it
reaches a maximum value $T^{\max }$ at a critical horizon radius $r_{\text{%
crt}}$ . Beyond this point, the temperature decreases rapidly. Furthermore,
increasing the GB parameter shifts the critical radius toward larger values
of $r_{+}$. This maximum temperature point corresponds to a second-order phase transition where the heat capacity diverges.

The corresponding behavior for the phantom sector is shown in Panel (\ref%
{fig:t11}). In this case, the phantom contribution significantly alters the
thermal evolution of the black hole. The Hawking temperature diverges in the
limit $r_{+}\rightarrow 0$, and decreases monotonically as the horizon
radius increases. Unlike the ordinary Maxwell case, the phantom sector exhibits neither a maximum temperature nor a second-order phase transition.

Next, we evaluate the black hole entropy. The first law of black hole thermodynamics takes the form:
\begin{equation}
dM=TdS+\eta \Phi dq,
\end{equation}%
where $\Phi =\left( \frac{\partial M}{\partial q}\right) _{S}$ denotes the
electric potential. In this study, we restrict ourselves to fixed
electric charge, namely $dq=0$, such that the first law reduces to

\begin{equation}
dM=TdS.
\end{equation}%
The entropy can therefore be obtained from 
\begin{equation}
S=\int \frac{dM}{T}=\int \frac{\partial M}{\partial r_{+}}\frac{dr_{+}}{T}.
\end{equation}%
Substituting Eqs. (\ref{25}), and (\ref{28}) yields

\begin{equation}
S=\pi r_{+}^{2}+4\pi \alpha \ln r_{+}.  \label{23}
\end{equation}

To determine the local thermodynamic stability of these solutions, we evaluate the specific heat capacity. Positive values ($C_{q}>0$) imply local thermal stability, while negative values ($C_{q}<0$) indicate that the configuration is unstable. The specific heat for constant charge is given by:
\begin{equation}
C_{q}=T\left( \frac{\partial S}{\partial T}\right) _{q}.
\end{equation}%
Employing Eqs. (\ref{28}) and (\ref{23}), the specific heat is obtained as%
\begin{equation}
C_{q}=-2\pi r_{+}^{2}\frac{\left( 1-\frac{\alpha +q^{2}}{%
r_{+}^{2}}\right) \left( 1+\frac{2\alpha }{r_{+}^{2}}\right) ^{2}}{\left( 1-%
\frac{2\alpha }{r_{+}^{2}}\right) \left( 1-\frac{\alpha +q^{2}}{%
r_{+}^{2}}\right) -2\frac{q^{2}+\alpha }{r_{+}^{2}}\left( 1+%
\frac{2\alpha }{r_{+}^{2}}\right) }.
\end{equation}%
The roots of the heat capacity $C_{q}=0$ correspond to the physical
limitation points separating physical $T>0$ and non-physical $T<0$ black
holes. Solving Eq. (\ref{28}), gives the positive root

\begin{equation}
r_{\mathrm{root}}=\sqrt{\alpha +\eta q^{2}}.
\end{equation}

The behavior of the heat capacity as a function of the event horizon radius
for different values of the GB coupling parameter is displayed in Fig.(\ref%
{fig:heat}) for both the Maxwell and phantom sectors.

For the Maxwell case, the heat capacity reveals the existence of a locally
stable thermodynamic phase. In particular, one finds $C_{q}>0$ for $r_{+}<r_{%
\text{crt}}$, indicating thermal stability, whereas $C_{q}<0$ for $r_{+}>r_{%
\text{crt}}$ corresponding to an unstable phase. At the critical radius $%
r_{+}=r_{\mathrm{crt}}$ the heat capacity diverges, signaling a second-order
phase transition.

By contrast, in the phantom sector, the heat capacity remains negative for
all values of $r_{+}$, implying that the black hole is thermodynamically
unstable throughout its entire evolution.

\begin{figure}[H]
\begin{minipage}[t]{0.5\textwidth}
        \centering
        \includegraphics[width=\textwidth]{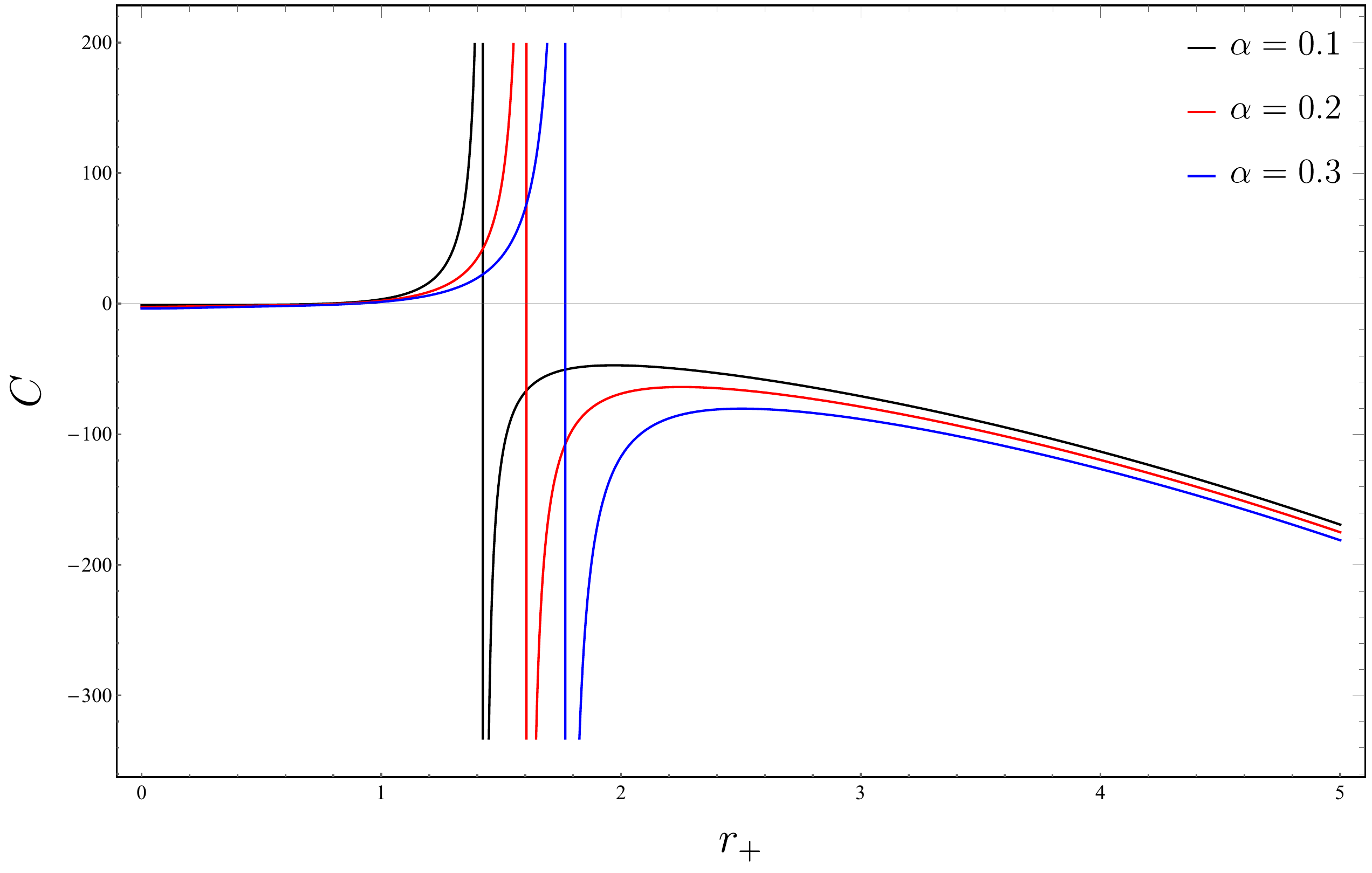}
                \subcaption{$\eta=1$}
        \label{fig:h1}
\end{minipage}
\begin{minipage}[t]{0.5\textwidth}
        \centering
        \includegraphics[width=\textwidth]{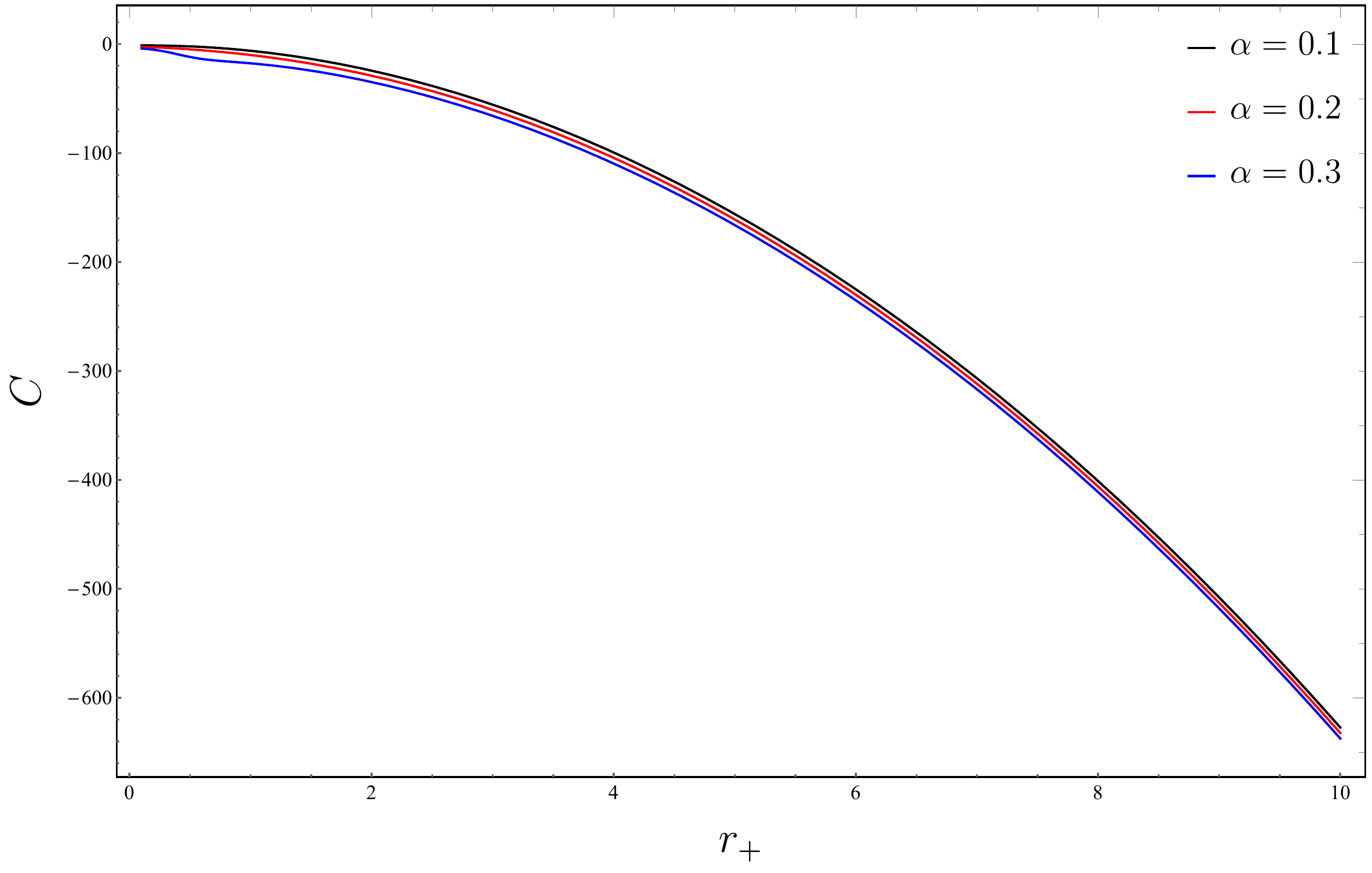}
                \subcaption{$\eta=-1$}
        \label{fig:h2}
\end{minipage}
\caption{The specific heat capacity plotted versus $r_{+}$ for a fixed charge $q=0.5$.}
\label{fig:heat}
\end{figure}

\section{Geodesic Dynamics and Particle Orbits}

\label{sec4}

The motion of neutral test particles is described by the geodesic Lagrangian

\begin{equation}
\mathcal{L}=\frac{1}{2}g_{\mu \nu}\dot{x}^{\mu}\dot{x}^{\nu},
\end{equation}

which, for the metric (\ref{20}), becomes

\begin{equation}
\mathcal{L}=\frac{1}{2}\left[ -\mathcal{F}(r)\dot{t}^{,2}+\frac{\dot{r}^{,2}%
}{\mathcal{F}(r)}+r^{2}\dot{\theta}^{2}+r^{2}\sin ^{2}\theta \dot{\varphi}%
^{2}\right] .
\end{equation}

Because the spacetime admits the Killing vectors $\partial_t$ and $%
\partial_\varphi$, the corresponding first integrals are

\begin{equation}
E=\mathcal{F}(r)\dot{t},\qquad L=r^{2}\sin ^{2}\theta \dot{\varphi},
\end{equation}

where $E$ and $L$ denote the conserved energy and angular momentum per unit
mass. The four-velocity additionally satisfies

\begin{equation}
g_{\mu \nu}\dot{x}^{\mu}\dot{x}^{\nu}=-\epsilon,
\end{equation}

with $\epsilon=1$ and $\epsilon=0$ corresponding to timelike and null
trajectories, respectively.

Restricting the motion to the equatorial plane $(\theta=\pi/2)$ and using the conserved quantities, the radial equation reduces to

\begin{equation}
\dot{r}^{2}+V_{\text{eff}}\left( r\right) =E^{2},
\end{equation}%
where the effective potential is
\begin{equation}
V_{\text{eff}}\left( r\right) =\left( \epsilon +\frac{L^{2}}{r^{2}}\right) 
\mathcal{F}\left( r\right) .
\end{equation}
The structure of the effective potential determines the allowed particle trajectories as well as the existence and stability of circular orbits.

\subsection{Orbital Dynamics and Timelike Geodesics}

For massive particles ($\epsilon=1$), the effective potential is

\begin{equation}
V_{\text{eff}}(r)=\left( 1+\frac{L^{2}}{r^{2}}\right) \mathcal{F}(r).
\label{41}
\end{equation}

\begin{figure}[H]
\begin{minipage}[t]{0.5\textwidth}
        \centering
        \includegraphics[width=\textwidth]{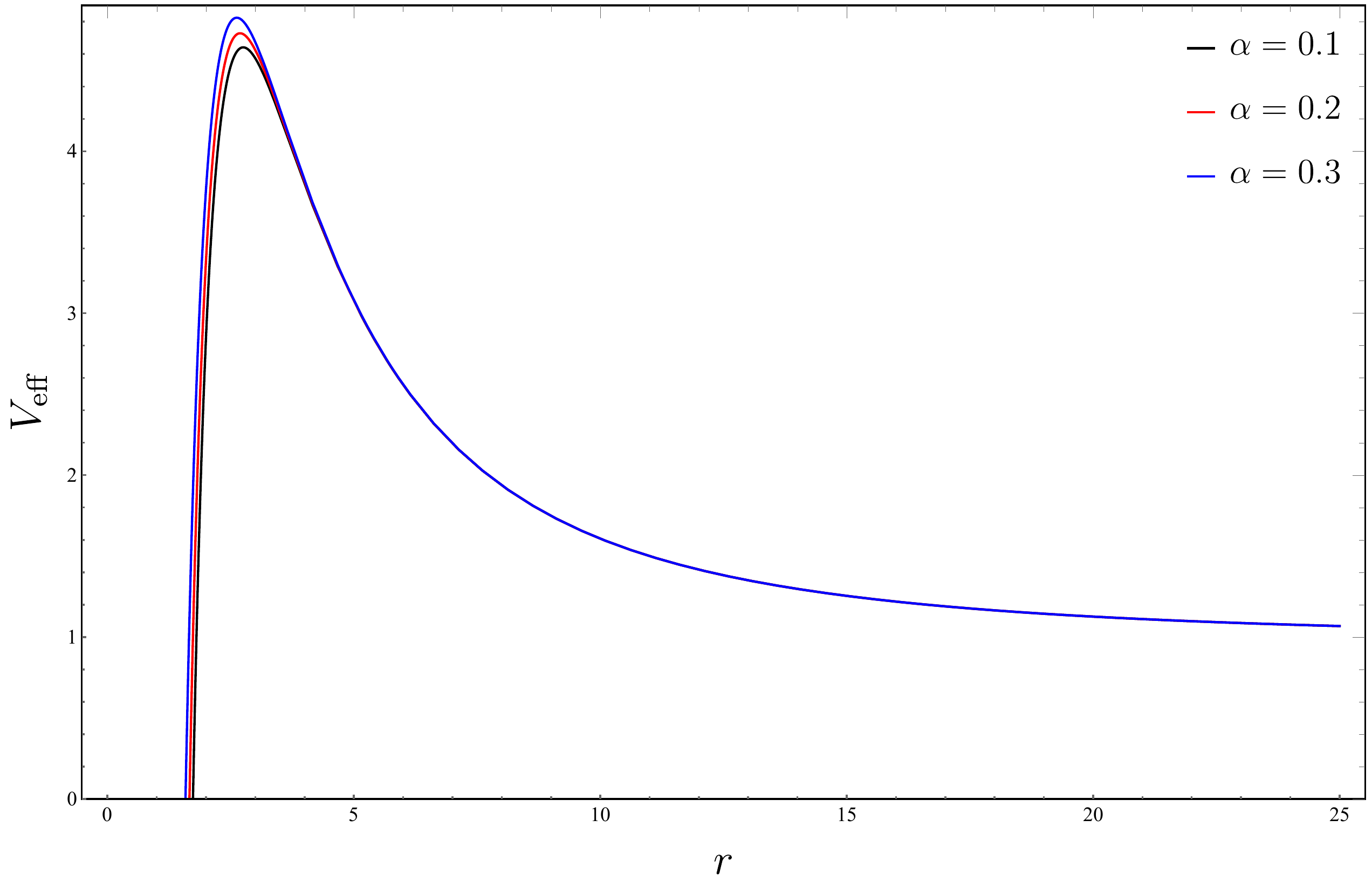}
                \subcaption{$\eta=+1$}
        \label{fig:V1}
\end{minipage}%
\begin{minipage}[t]{0.5\textwidth}
        \centering
        \includegraphics[width=\textwidth]{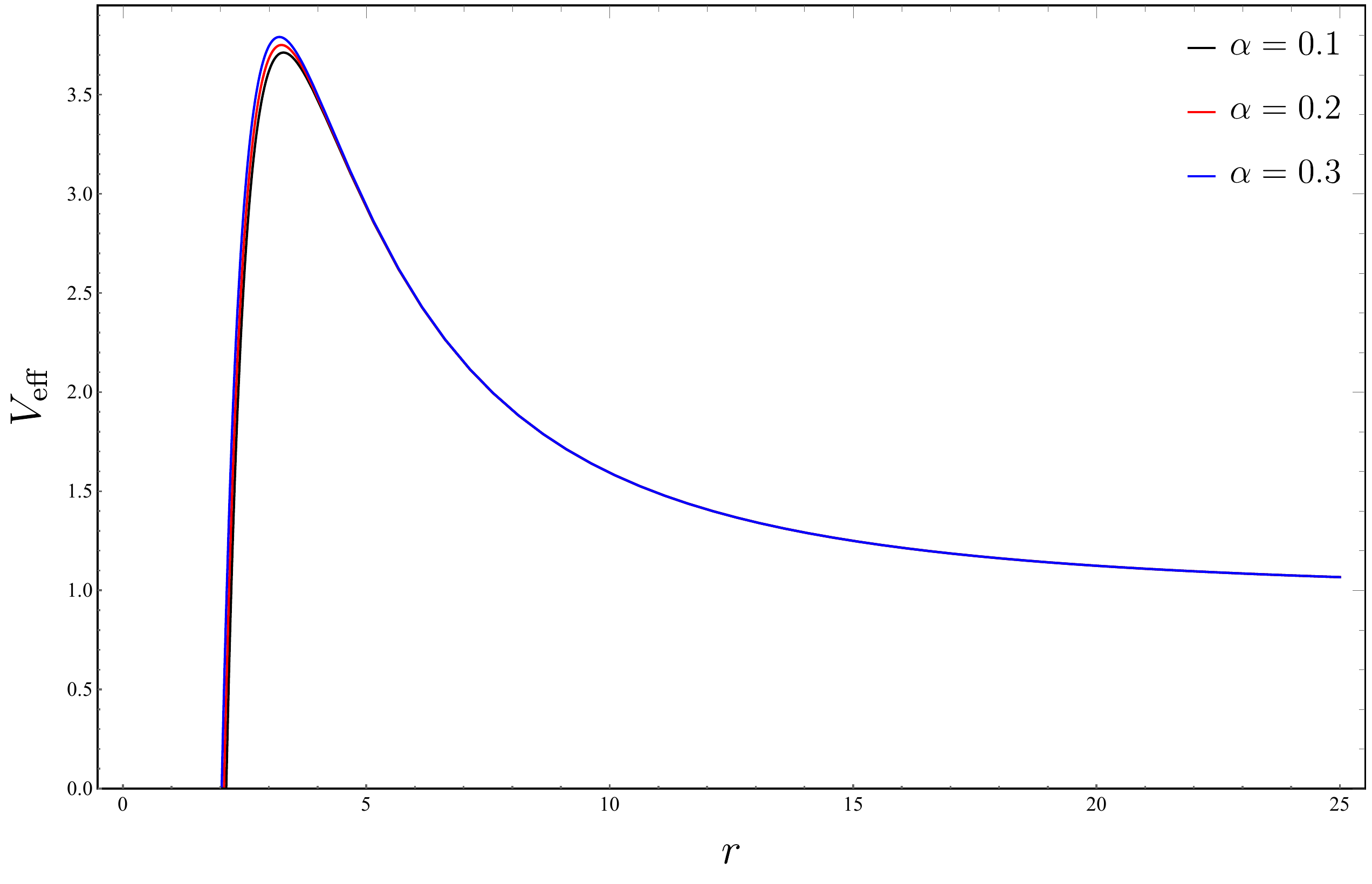}
                \subcaption{$\eta=-1$}
        \label{fig:V2}
\end{minipage}
\caption{The effective potential as a function of radial distance for $M=1$, $q=0.6$ and $\ell =10$ }
\label{fig:veff}
\end{figure}

Since $\mathcal{F}(r_{+})=0$, the potential vanishes at the event horizon
and its profile is entirely determined by the spacetime geometry. The
corresponding radial dependence is displayed in Fig.(\ref{fig:veff}). In
both electromagnetic sectors, the effective potential develops a local
maximum associated with an unstable circular orbit. Increasing the GB
coupling raises the height of the barrier and shifts its location toward
larger radii. For fixed values of the parameters, the Maxwell branch
produces a higher barrier than the phantom field branch, indicating stronger
confinement of massive particles.

\begin{figure}[H]
\begin{minipage}[t]{0.3\textwidth}
        \centering
        \includegraphics[width=\textwidth]{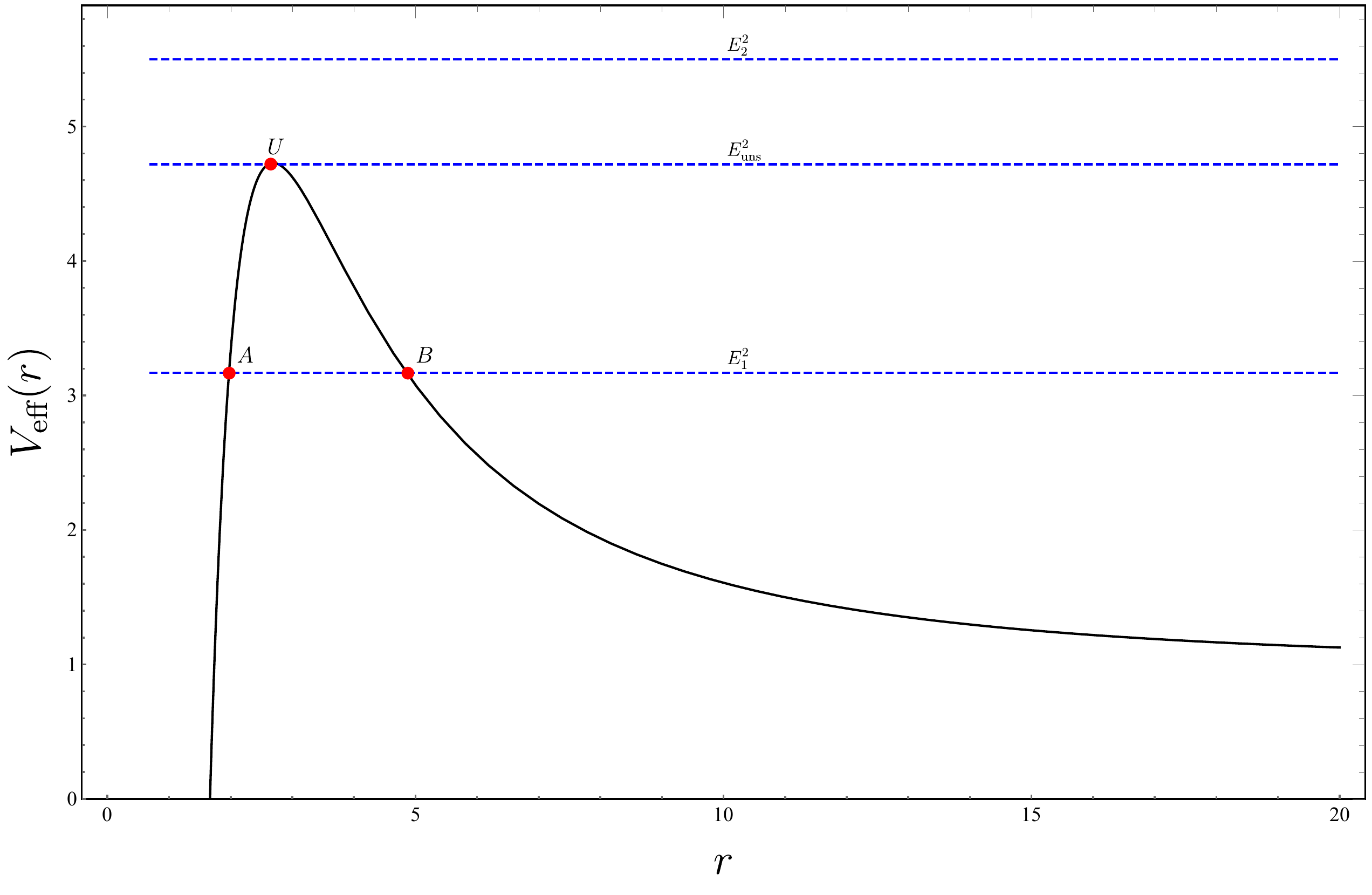}
                \subcaption{The effective potential}
        \label{fig:VE2}
\end{minipage}
\begin{minipage}[t]{0.3\textwidth}
        \centering
        \includegraphics[width=\textwidth]{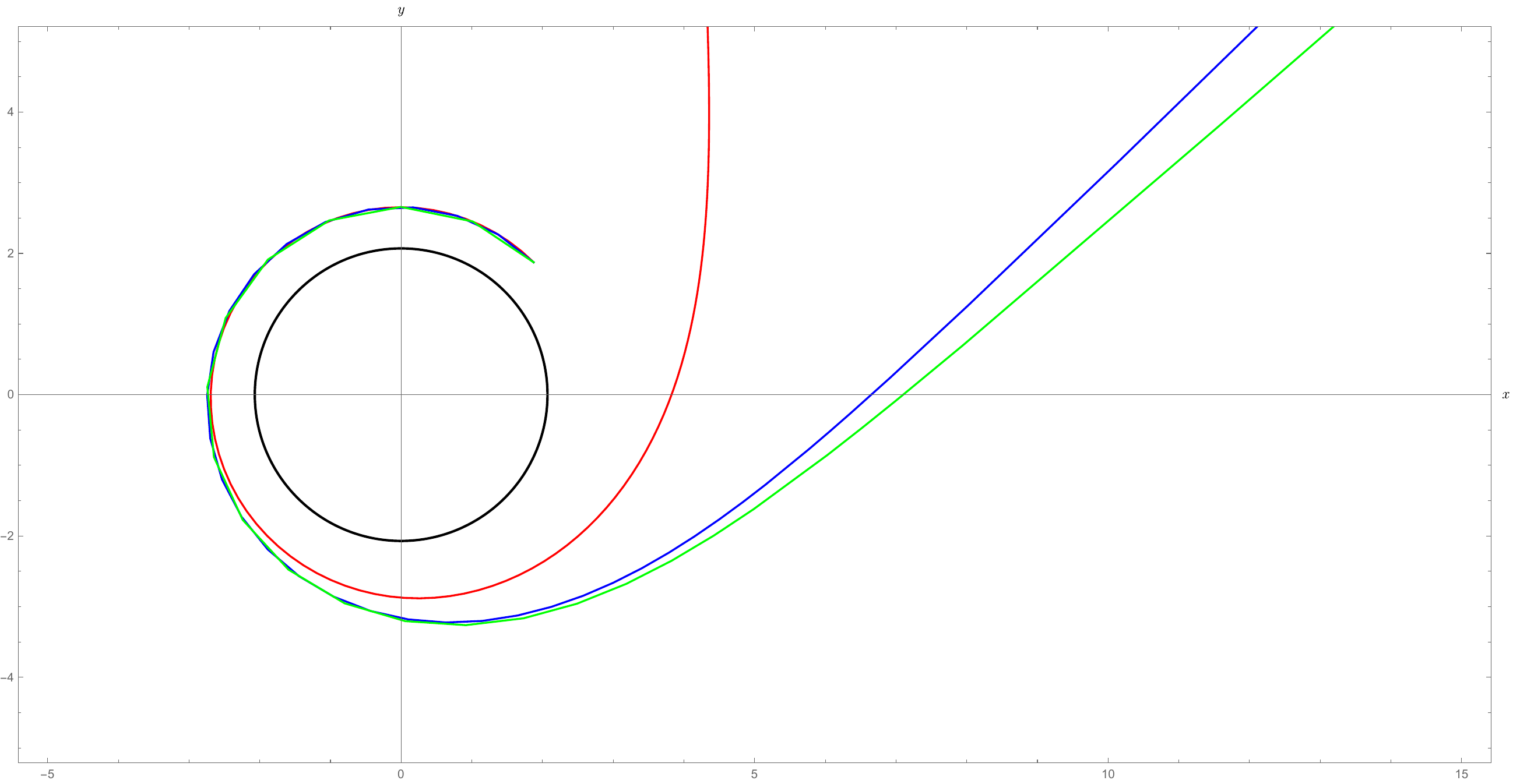}
                \subcaption{$\eta=+1$}
        \label{fig:tr1}
\end{minipage}
\begin{minipage}[t]{0.3\textwidth}
        \centering
        \includegraphics[width=\textwidth]{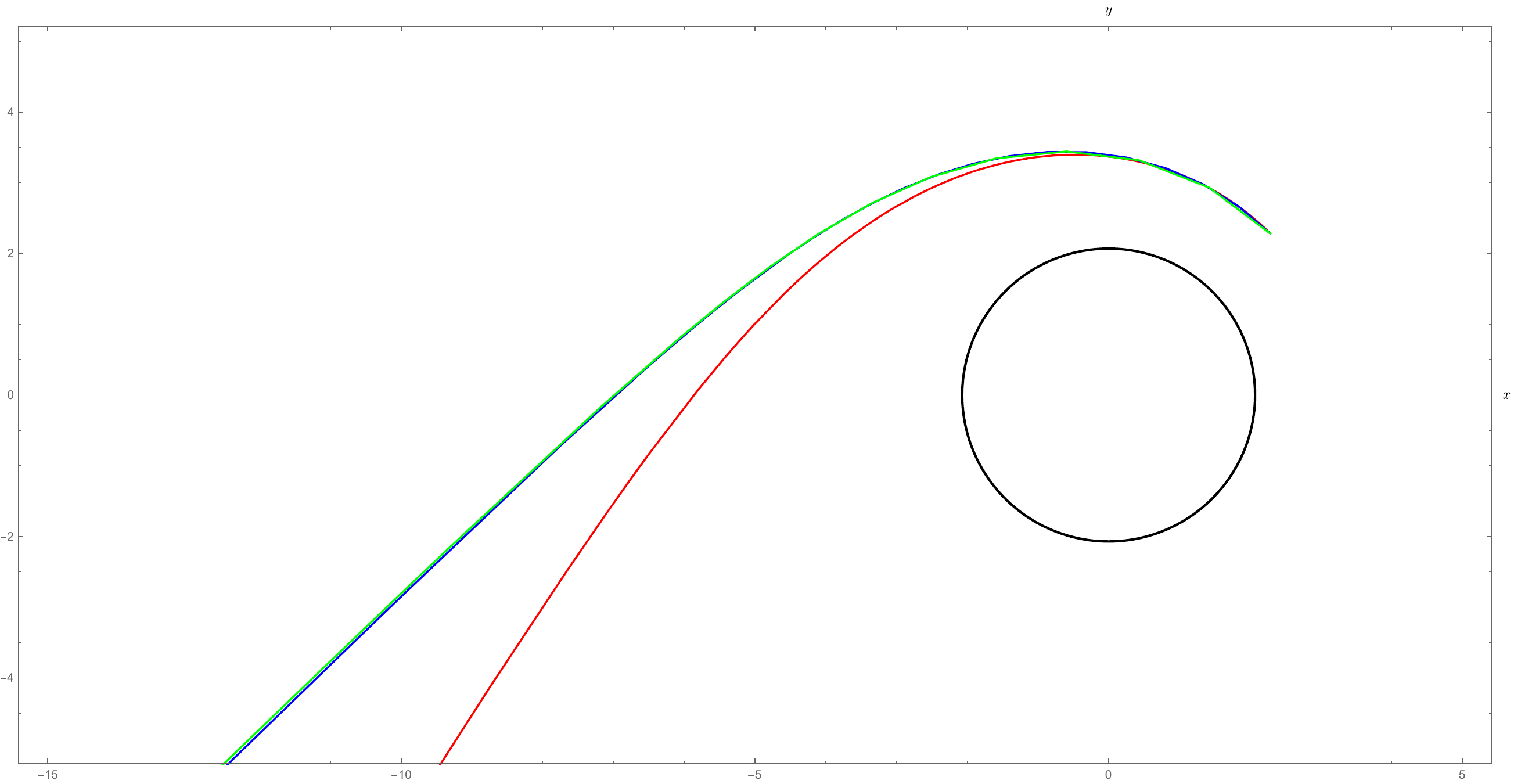}
                \subcaption{$\eta=-1$}
        \label{fig:tr2}
\end{minipage}
\caption{Trajectories of massive particles for different initial velocities.
Parameters: $M=1$, $q=0.6$}
\label{fig:traj}
\end{figure}

The corresponding timelike trajectories are illustrated in Fig.(\ref{fig:traj}) for $L=10$, $M=1$, $q=0.6$, and $\alpha=0.2$. Depending on the particle energy, three distinct classes of motion arise. For energies above the potential barrier, particles follow plunge trajectories and eventually cross the horizon. At the critical energy corresponding to the maximum of $V_{\text{eff}}$, the particle remains on an unstable circular orbit. Any infinitesimal perturbation drives the particle either toward the black hole or toward larger radial distances. For energies below the barrier, the motion becomes bounded between two turning points, leading to oscillatory radial motion around a stable circular orbit.

Figures (\ref{fig:tr1}) and (\ref{fig:tr2}) further illustrate the instability of the circular orbit located at the maximum of the effective potential. In both the Maxwell and phantom sectors, a small change in the initial velocity is sufficient to destabilize the orbit, producing either capture by the black hole or outward migration.

We now turn to circular geodesics. The conditions
\begin{equation}
  \dot{r}=0;  \qquad \ddot{r}=0,   
\end{equation}

determine the specific angular momentum and specific energy associated with circular motion in the equatorial plane. Solving these conditions yields Eqs.(\ref{43}) and (\ref{45}) for $L$ and $E$, respectively

\begin{equation}
L^{2}=\frac{r^{4}}{2\alpha }\frac{1+\frac{2\alpha M}{r^{3}}-\sqrt{1+\frac{%
8\alpha M}{r^{3}}-\eta \frac{4\alpha q^{2}}{r^{4}}}}{\frac{3M}{r}-\frac{%
2\eta q^{2}}{r^{2}}-\sqrt{1+\frac{8\alpha M}{r^{3}}-\eta \frac{4\alpha q^{2}%
}{r^{4}}}},  \label{43}
\end{equation}%
\begin{equation}
E^{2}=-\frac{r^{4}}{4\alpha ^{2}}\frac{\sqrt{1+\frac{8\alpha M}{r^{3}}-\eta 
\frac{4\alpha q^{2}}{r^{4}}}\left( 1+\frac{2\alpha }{r^{2}}-\sqrt{1+\frac{%
8\alpha M}{r^{3}}-\eta \frac{4\alpha q^{2}}{r^{4}}}\right) ^{2}}{\frac{3M}{r}%
-\eta \frac{2q^{2}}{r^{2}}-\sqrt{1+\frac{8\alpha M}{r^{3}}-\eta \frac{%
4\alpha q^{2}}{r^{4}}}}.  \label{45}
\end{equation}

To find the radius of the innermost stable circular orbit (ISCO), one must utilize the marginal stability condition:
\begin{equation}
\frac{d^{2}}{dr^{2}}V_{\text{eff}}\left( r\right) =0.
\end{equation}

This specific orbit is crucial because it separates regions of stable circular motion from unstable ones. Combining this relation with the circular orbit equations leads to

\begin{equation}
\left. r\mathcal{F}\left( r\right) ^{\prime \prime }\mathcal{F}\left(
r\right) +3\mathcal{F}\left( r\right) ^{\prime }\mathcal{F}\left( r\right)
-2r\mathcal{F}\left( r\right) ^{\prime 2}\right\vert _{r=r_{\text{ISCO}}}=0. \label{iscoeq}
\end{equation}
Since Eq.(\ref{iscoeq}) cannot be solved analytically in closed form, the ISCO parameters must be computed numerically. The corresponding values of $r_{\mathrm{ISCO}}$, $L_{\mathrm{ISCO}}$, and $E_{\mathrm{ISCO}}$ are summarized in Table\ref{tab:isco}.

The results indicate that, in the Maxwell sector, the ISCO radius decreases monotonically with increasing $\alpha$. The corresponding angular momentum and energy also decrease, implying that higher curvature corrections allow stable circular orbits to exist closer to the black hole. The phantom sector exhibits the same qualitative trend; however, the ISCO radius remains larger for all values of $\alpha$. Moreover, both $L_{\mathrm{ISCO}}$ and $E_{\mathrm{ISCO}}$ exceed their Maxwell counterparts. These features suggest that the phantom field shifts stable orbital configurations outward and reduces the overall gravitational binding of the system.

\begin{figure}[H]
\centering
\includegraphics[scale=0.2]{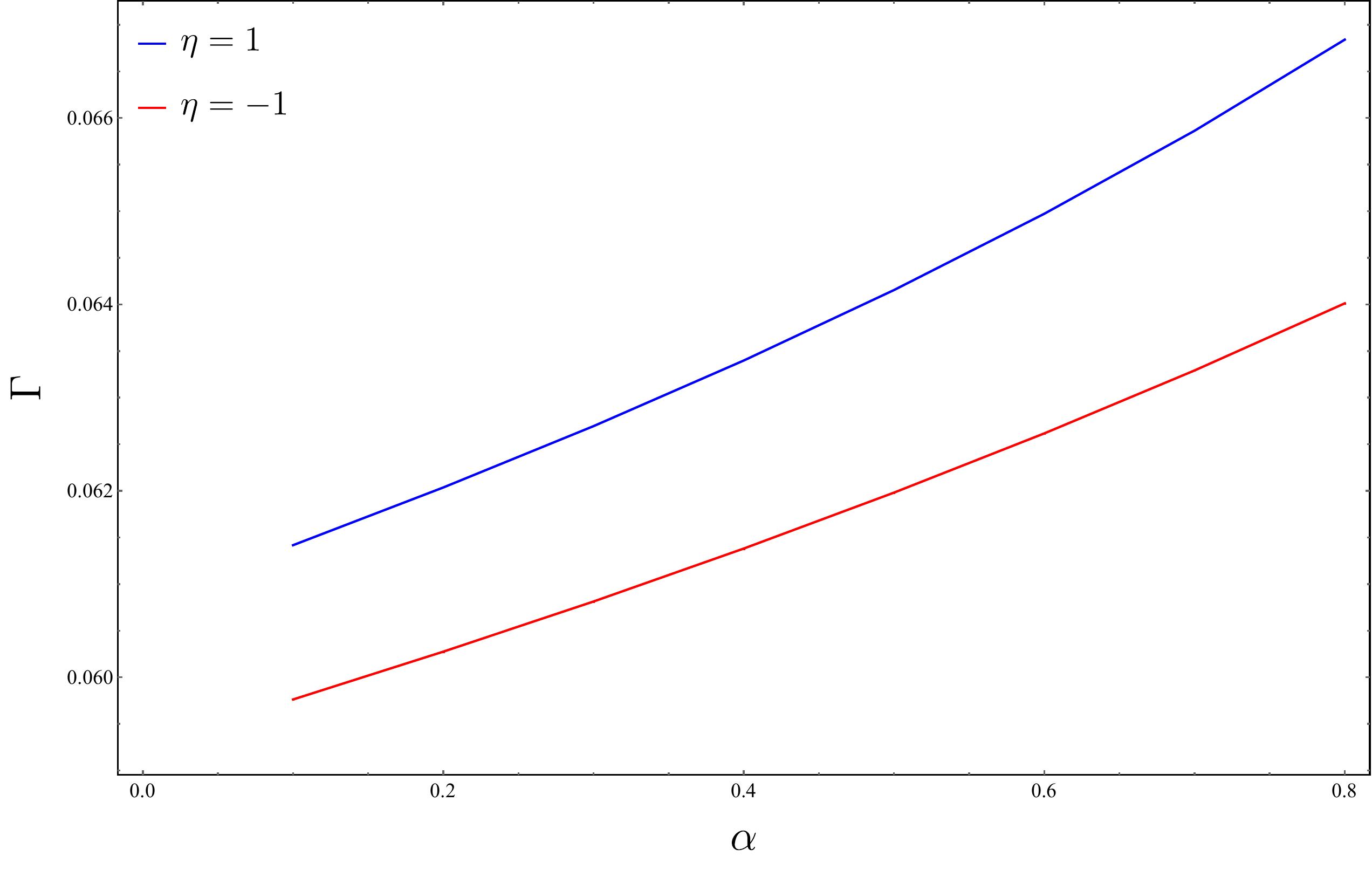}
\caption{Accretion disk energy efficiency versus the GB coupling parameter $\alpha$ for a phantom black hole in 4D EGB gravity ($M=1$, $q=0.5$).}
\label{fig:eff}
\end{figure}

\begin{table}[H]
\centering
\begin{tabular}{c|ccc|ccc}
\hline\hline
\multicolumn{1}{c|}{} & \multicolumn{3}{c|}{$\eta=1$} & \multicolumn{3}{c}{$%
\eta=-1$} \\ \cline{2-7}
$\alpha$ & $r_{\text{ISCO}}$ & $L_{\text{ISCO}}$ & $E_{\text{ISCO}}$ & $r_{%
\text{ISCO}}$ & $L_{\text{ISCO}}$ & $E_{\text{ISCO}}$ \\ \hline\hline
0.1 & 5.53491 & 3.35487 & 0.938583 & 6.30466 & 3.37083 & 0.940238 \\ 
0.2 & 5.45987 & 3.34367 & 0.937964 & 6.24769 & 3.36026 & 0.939725 \\ 
0.3 & 5.38111 & 3.33205 & 0.937306 & 6.18891 & 3.34934 & 0.939187 \\ 
0.4 & 5.29808 & 3.31997 & 0.936602 & 6.12817 & 3.33803 & 0.938619 \\ 
0.5 & 5.21013 & 3.30738 & 0.935846 & 6.06528 & 3.32630 & 0.938019 \\ 
0.6 & 5.11640 & 3.29423 & 0.935028 & 6.00003 & 3.31411 & 0.937384 \\ 
0.7 & 5.01578 & 3.28046 & 0.934138 & 5.93218 & 3.30140 & 0.936709 \\ 
0.8 & 4.90673 & 3.26599 & 0.933160 & 5.86143 & 3.28812 & 0.935989 \\ 
\hline\hline
\end{tabular}%
\caption{ISCO parameters for different values of $\alpha $ and $%
\protect\eta $, with $M=1$ and $q=0.5$.}
\label{tab:isco}
\end{table}
The implications for thin-disk accretion can be assessed within the Novikov-Thorne framework \cite{77}. The efficiency of converting rest-mass energy into radiation is given by

\begin{equation}
\Gamma =1-E_{\text{ISCO}}.
\end{equation}
Figure (\ref{fig:eff}) illustrates the influence of the GB coupling
parameter on $\Gamma $. The efficiency increases with $\alpha $ in both
branches, indicating that higher curvature corrections enhance the
conversion of rest mass energy into radiation. In contrast, for a fixed $%
\alpha $, the Maxwell sector exhibits higher efficiency than the
phantom-field sector. The presence of the phantom field therefore reduces
the accretion efficiency, a behavior that is consistent with the decrease in
gravitational binding energy obtained from the geodesic analysis and agrees
with the stronger binding of circular orbits inferred from the ISCO analysis.

\section{Scalar Perturbations and Quasinormal Mode Formalism}

\label{sec5} 
To explore the dynamic properties and stability of phantom charged black holes under 4D EGB gravity, we investigate how a massive scalar field propagates through this fixed background geometry. The evolution of such a scalar field $\phi$ is governed by the covariant Klein-Gordon equation:

\begin{equation}
\left( \frac{1}{\sqrt{-g}}\partial _{\sigma }\sqrt{-g}g^{\sigma \rho
}\partial _{\rho }-\mu ^{2}\right) \phi =0.  \label{39}
\end{equation}%
where $\mu$ denotes the rest mass of the scalar field. We can exploit the background symmetries to decompose the field wavefunction using a separation of variables ansatz:
\begin{equation}
\phi \left( r,\theta ,\varphi ,t\right) =\sum_{\ell =0}^{\infty }\sum_{\nu
=-\ell }^{\ell }e^{-i\omega t}\frac{u_{\ell }\left( r\right) }{r}\mathcal{Y}%
_{\ell }^{\nu }\left( \theta ,\varphi \right) ,
\end{equation}

Here, $\mathcal{Y}_{\ell }^{\nu }\left( \theta ,\varphi \right)$ are the standard spherical harmonics, and $\omega$ represents the characteristic QNM frequency. This decomposition allows us to project the radial dynamics into a simplified differential form:

\begin{equation}
\left[ \mathcal{F}\left( r\right) \frac{d}{dr}\left( \mathcal{F}\left(
r\right) \frac{d}{dr}\right) +\omega ^{2}-\mathcal{U}_{\text{eff}}\left(
r\right) \right] u_{\ell }\left( r\right) =0,
\end{equation}
where the background geometry dictates the form of the effective potential barrier:

\begin{equation}
\mathcal{U}_{\text{eff}}\left( r\right) =\mathcal{F}\left( r\right) \left( 
\frac{1}{r}\frac{d}{dr}\mathcal{F}\left( r\right) +\frac{\ell \left( \ell
+1\right) }{r^{2}}+\mu ^{2}\right) .  \label{potentielQNM}
\end{equation}%
By mapping the physical radial coordinate onto the tortoise coordinate via the transformation $dx = dr / \mathcal{F}(r)$, the radial wave equation can be recast into a convenient Schrödinger-like form:

\begin{equation}
\left[ \frac{d^{2}}{dx^{2}}+\omega ^{2}-\mathcal{U}_{\text{eff}}\right]
u_{\ell }\left( x\right) =0.  \label{EQ2}
\end{equation}%
Quasinormal modes are defined by imposing purely ingoing waves at the event horizon and purely outgoing waves at spatial infinity,

\begin{equation}
u\left( x\right) \simeq e^{i\sqrt{\omega ^{2}-\mu ^{2}}x}, \qquad x\rightarrow +\infty,
\end{equation}%
and
\begin{equation}
u\left( x\right) \simeq e^{-i\omega x}, \qquad x\rightarrow -\infty.
\end{equation}
These conditions guarantee that the system remains physically isolated, meaning no radiation leaks outward from the interior of the horizon, and no external waves impinge on the system from spatial infinity.

\begin{figure}[H]
\begin{minipage}[t]{0.5\textwidth}
        \centering
        \includegraphics[width=\textwidth]{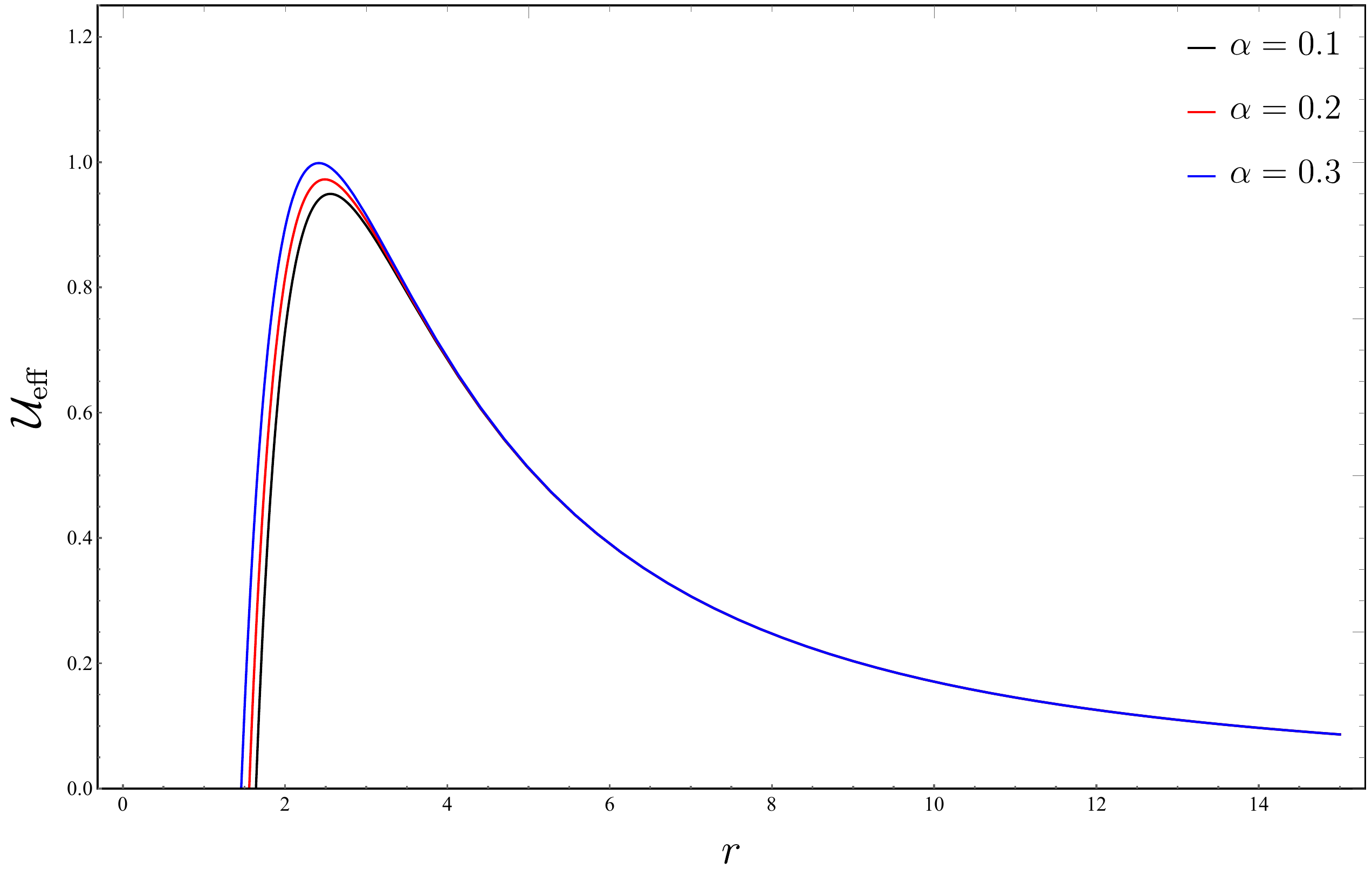}
                \subcaption{$\eta=1$}
        \label{fig:u1}
\end{minipage}%
\begin{minipage}[t]{0.5\textwidth}
        \centering
        \includegraphics[width=\textwidth]{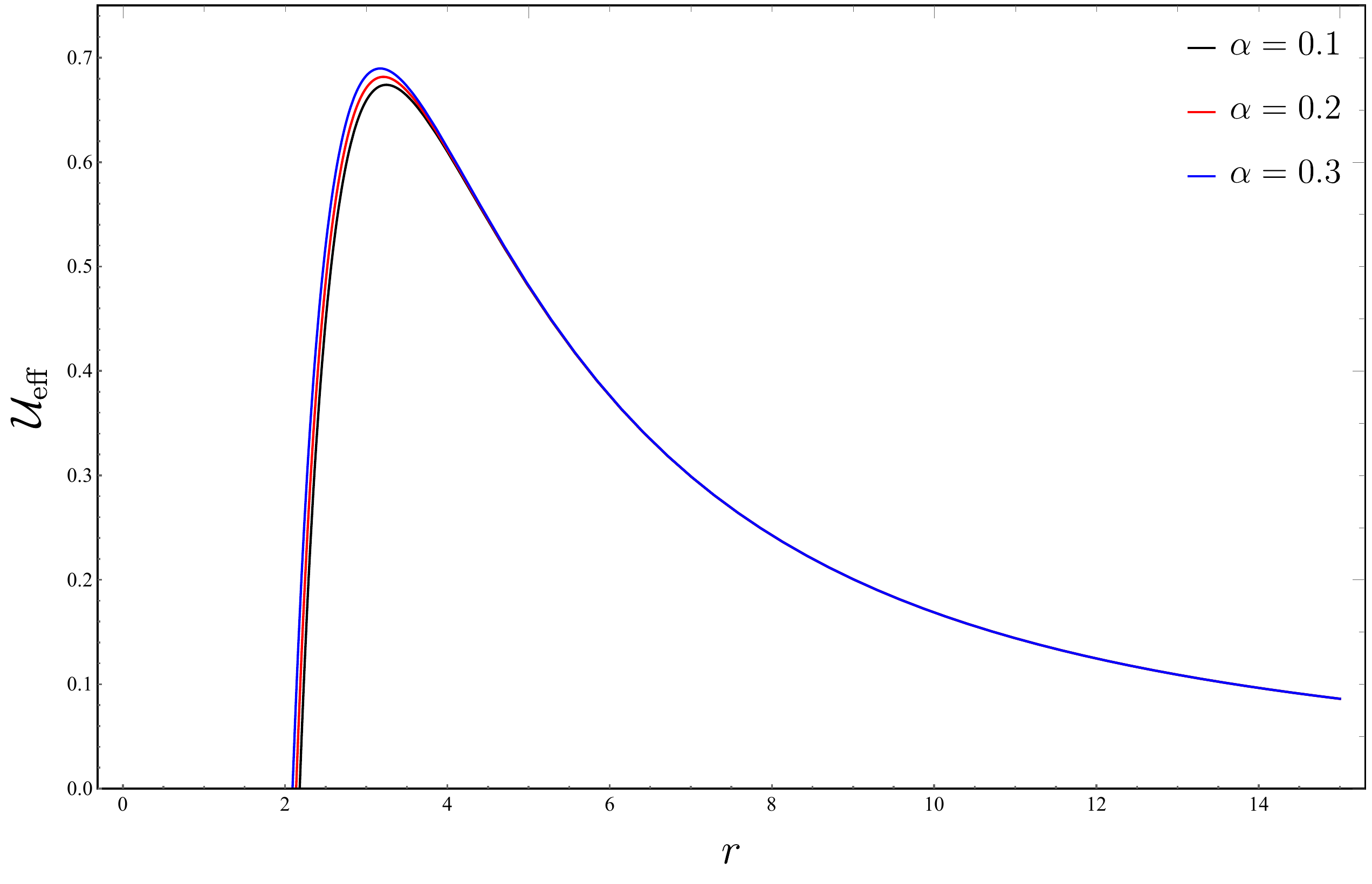}
                \subcaption{$\eta=-1$}
        \label{fig:u2}
        \end{minipage}
\caption{Radial dependence of the scalar effective potential for various values of the GB coupling $\alpha$. Parameters: $M=1$, $q=0.7$, $\ell =4$ and $\protect\mu =0.1$.}
\label{fig:ueffB}
\end{figure}

\begin{figure}[H]
\begin{minipage}[t]{0.5\textwidth}
        \centering
        \includegraphics[width=\textwidth]{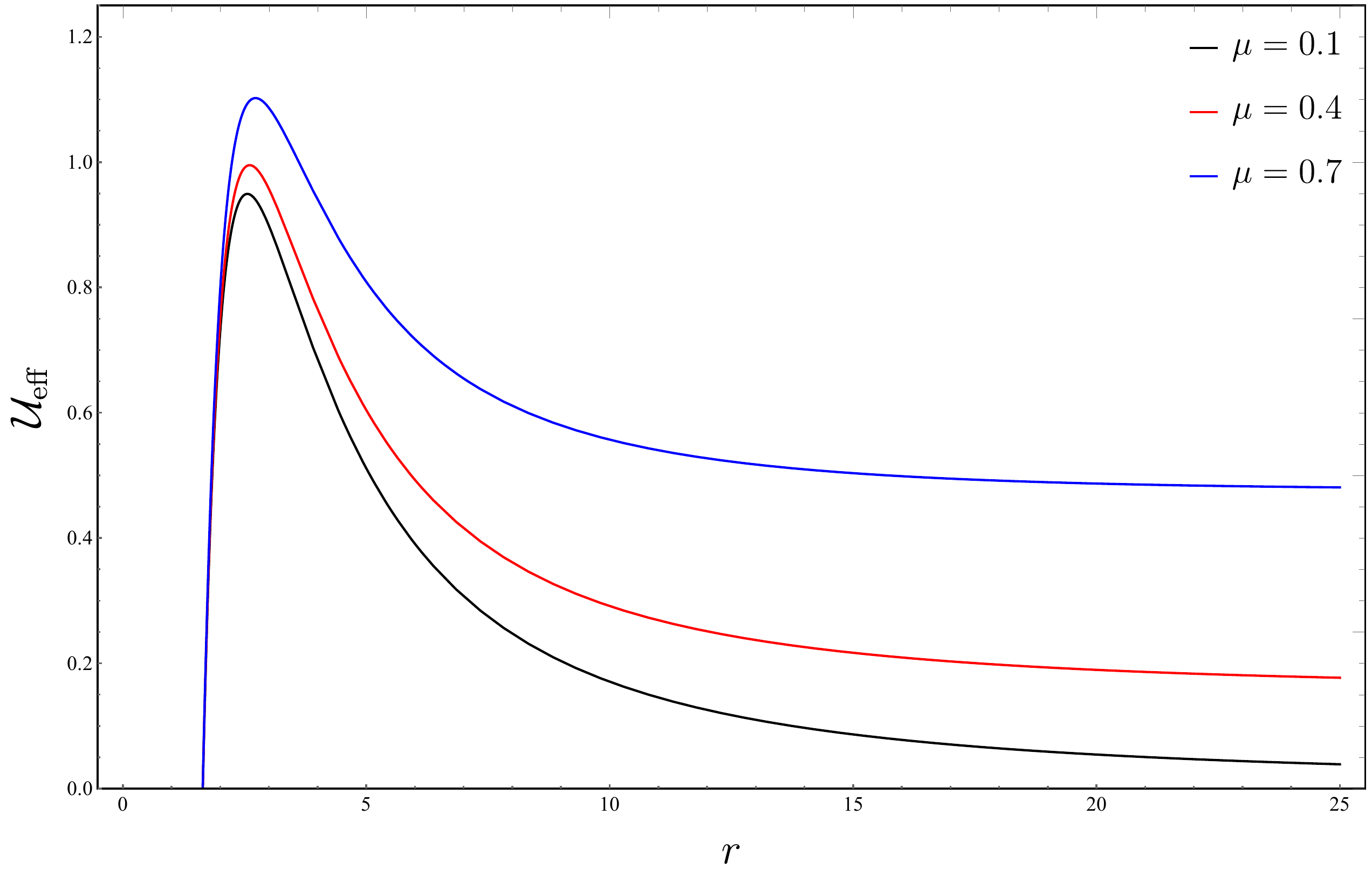}
                \subcaption{$\eta=1$}
        \label{fig:u3}
\end{minipage}%
\begin{minipage}[t]{0.5\textwidth}
        \centering
        \includegraphics[width=\textwidth]{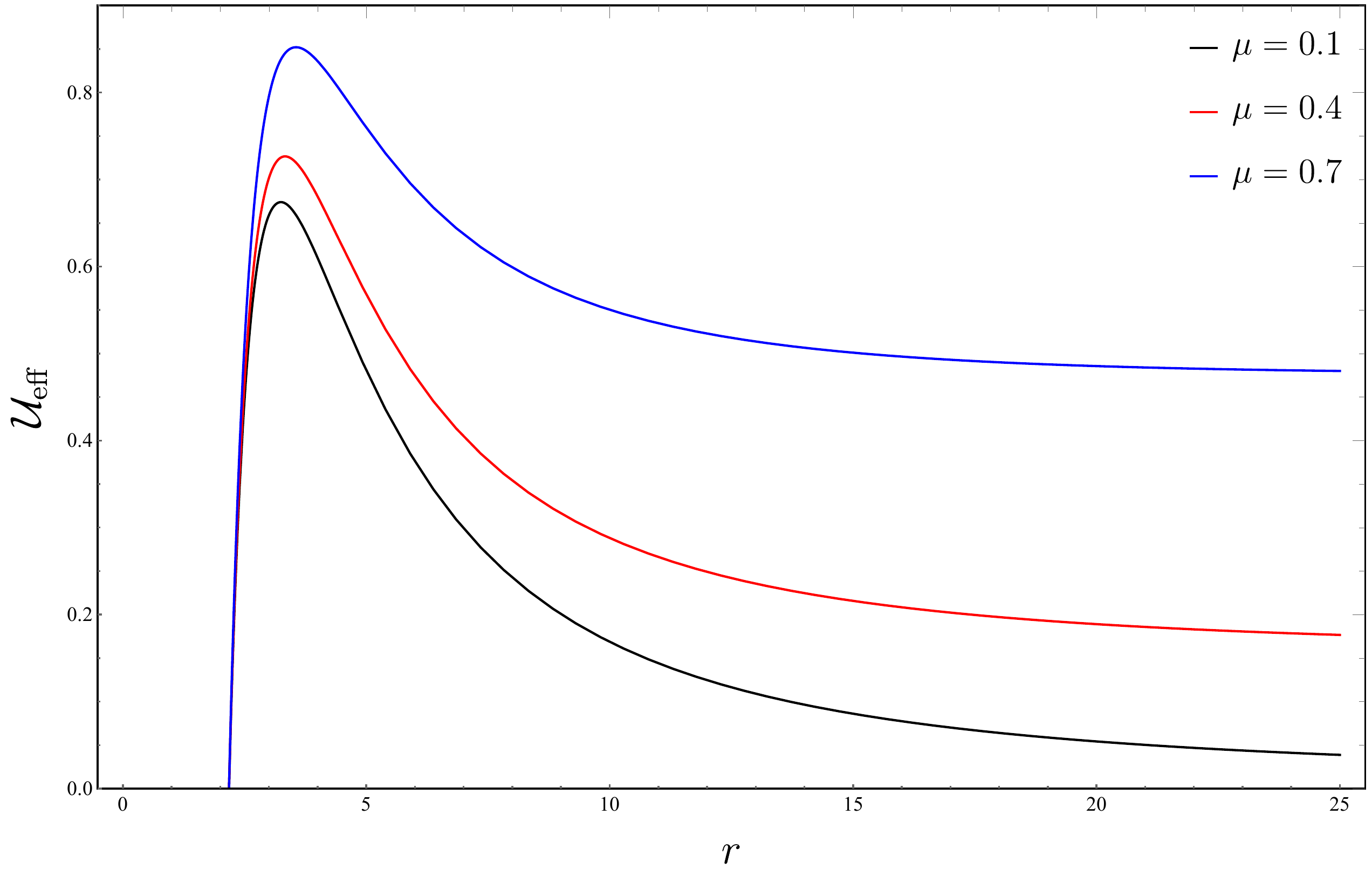}
                \subcaption{$\eta=-1$}
        \label{fig:u4}
\end{minipage}
\caption{Radial dependence of the scalar effective potential for various scalar field masses $\mu$, with $M=1$, $q=0.7$,$\ell =4$ and $\protect\alpha =0.1$.}
\label{fig:ueffB2}
\end{figure}

\begin{figure}[H]
\begin{minipage}[t]{0.5\textwidth}
        \centering
        \includegraphics[width=\textwidth]{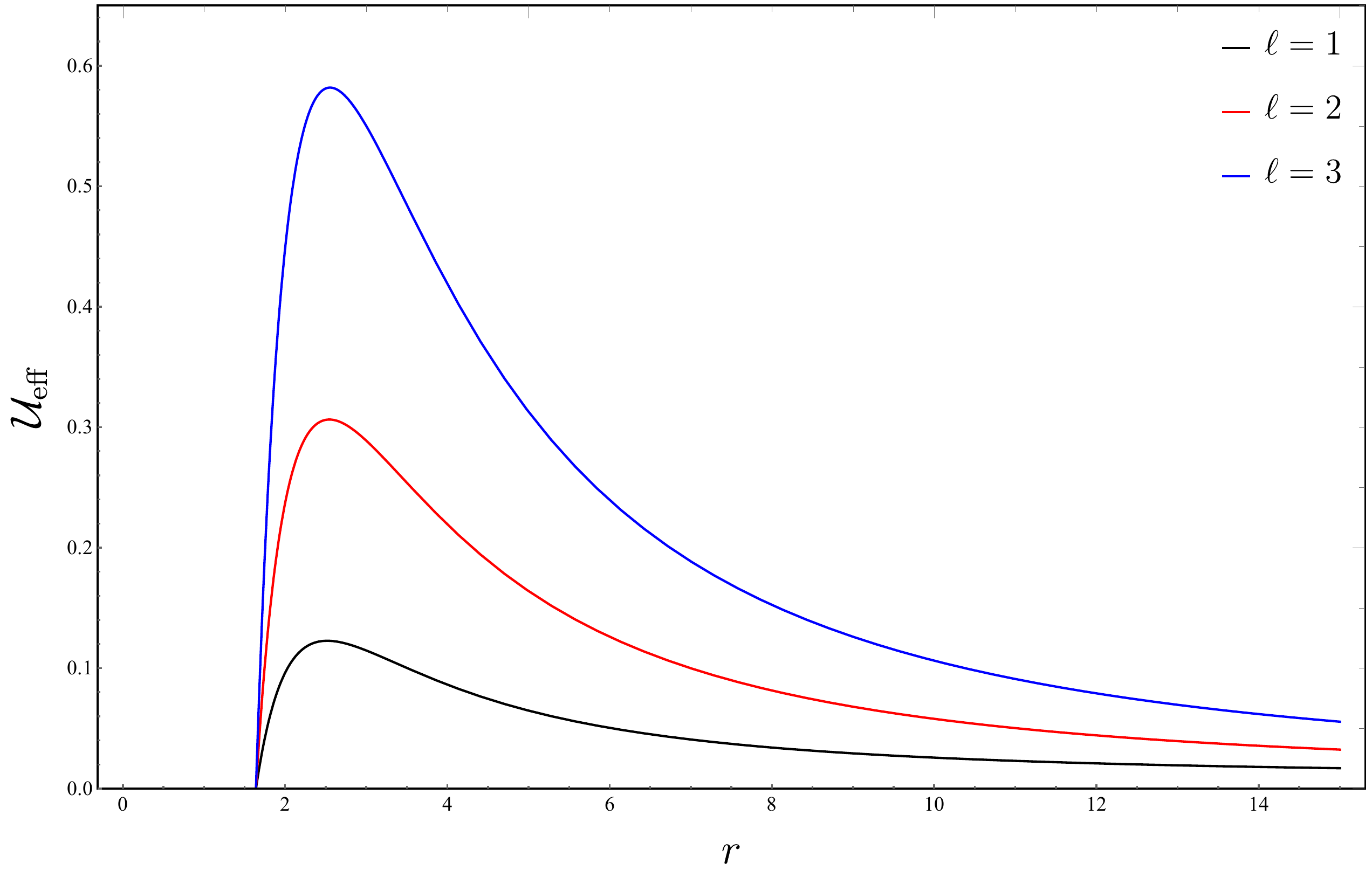}
                \subcaption{$\eta=1$}
        \label{fig:u5}
\end{minipage}%
\begin{minipage}[t]{0.5\textwidth}
        \centering
        \includegraphics[width=\textwidth]{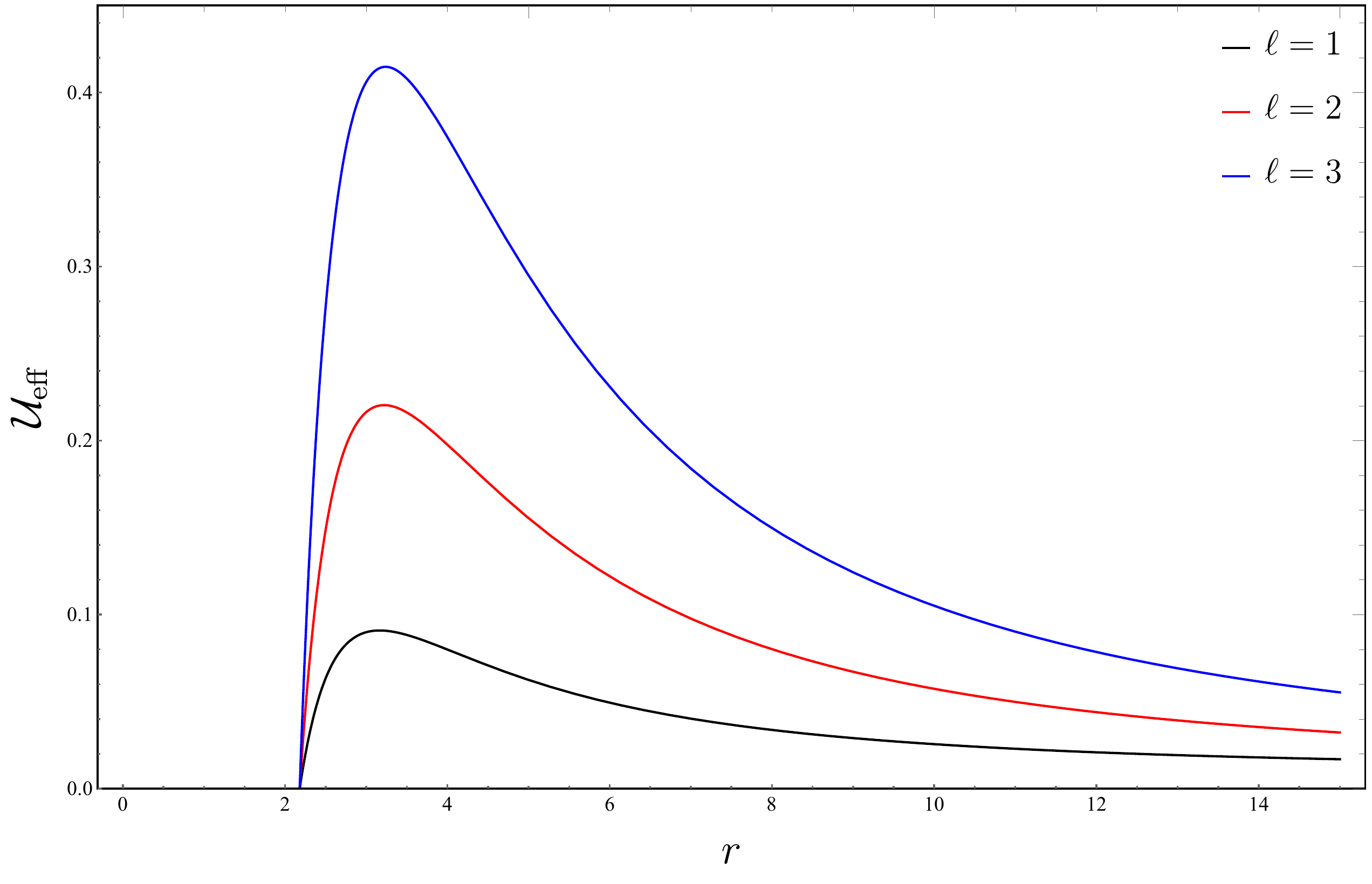}
                \subcaption{$\eta=1$}
        \label{fig:u6}
\end{minipage}
\caption{Variation of the scalar effective potential with radial distance, for different values of the multipole number $\ell $, with $M=1$, $q=0.7$, $\alpha =0.1$ and $\mu=0.1$.} 
\label{fig:ueffB3}
\end{figure}

The effective potential $\mathcal{U}_{\text{eff}}$ encodes the influence of the spacetime geometry on scalar field propagation and therefore plays a central role in determining the quasinormal spectrum. Its radial behavior for different model parameters is presented in Figs.(\ref{fig:ueffB})-(\ref{fig:ueffB3}). For all configurations examined here, $\mathcal{U}_{\text{eff}}$ remains positive outside the event horizon and exhibits a single barrier structure. No potential well capable of supporting unstable bound states is observed, providing strong evidence for the dynamical stability of the black hole against scalar perturbations.

Fig. (\ref{fig:ueffB}) demonstrates how the GB coupling strength $\alpha$ alters the profile of the effective potential. For both Maxwell and phantom configurations, larger values of $\alpha$ increase the peak of the barrier, indicating that higher curvature terms act to inhibit field transmission. Furthermore, for fixed $\alpha$, the standard Maxwell sector yields a higher barrier than its phantom counterpart. This suggests that scalar field perturbations are trapped more strongly by ordinary charged geometries.

Figure (\ref{fig:ueffB}) illustrates the dependence of the effective potential on the GB coupling parameter $\alpha$. For both the Maxwell and phantom  sectors, increasing $\alpha$ raises the height of the potential barrier. This behavior reflects the stronger influence of higher curvature corrections on scalar-field propagation. For fixed $\alpha$, the Maxwell branch exhibits a larger potential barrier than the phantom branch, implying stronger trapping of scalar perturbations in the ordinary charged geometry.

The effect of the scalar field mass is presented in Fig.(\ref{fig:ueffB2}). The potential barrier increases monotonically with $\mu$, reflecting the additional mass contribution to the effective potential. Unlike the massless case, the asymptotic value of the potential is nonvanishing,

\begin{equation*}
\lim_{r\rightarrow \infty }\mathcal{U}_{\text{eff}}=\mu ^{2}.
\end{equation*}
As in the previous case, the potential in the Maxwell sector remains systematically higher than that of the phantom sector.

Finally, Fig.(\ref{fig:ueffB3}) shows the dependence on the multipole number $\ell$. Larger values of $\ell$ enhance the centrifugal term $\ell(\ell+1)/r^2$, leading to a higher and narrower potential barrier. Consequently, scalar perturbations with larger angular momentum experience stronger confinement around the black hole.

\subsection{Wentzel-Kramers-Brillouin method}

Quasinormal frequencies can be efficiently evaluated using semi-analytical techniques based on the Wentzel–Kramers–Brillouin (WKB) approximation. This technique has been widely employed in studies of perturbations and scattering problems in black hole spacetimes, where it provides accurate estimates of resonance spectra based on the structure of the effective potential barrier \cite{Vishveshwara,Kokkotas,Hegde}. The method becomes applicable once the perturbation equation is reduced to a one-dimensional, Schrödinger-type wave equation, such as Eq.(\ref{EQ2}), where the underlying dynamics are governed by a single potential barrier. Within this formulation, the QNM problem is mapped onto a quantum-tunneling scenario characterized by matching wave solutions across the turning points of the potential. The WKB framework was first developed in \cite{Schutz}, while later works refined the approach by incorporating higher-order corrections and enhancing its numerical precision \cite{Iyer,Konoplya,KonoplyaRA,Hamil1,Hamil2,Hamil3,Hamil4}.

By constructing approximate wave solutions in distinct regions of the potential and connecting them near the turning points via suitable matching conditions, one can impose the continuity of the logarithmic derivative of the wave function to extract the QNM spectrum. Around the maximum of the effective potential, located at $x=x_{0}$, the WKB quantization condition governing the complex QNM frequencies takes the generalized form
\begin{equation}
\frac{i\left( \omega ^{2}-V_{0}\right) }{\sqrt{-2V_{0}^{\prime \prime }}}%
-\sum_{i=2}^{N}\Lambda _{i}=n+\frac{1}{2}.
\end{equation}%
In this expression, $V_{0}$ and $V_{0}^{\prime \prime }$ denote the value
and second derivative of the potential $\mathcal{U}_{\text{eff}}$ at $%
x=x_{0} $, $\Lambda _{i}$ are constants arising from higher-order WKB
corrections, and $n=0,1,2,\dots$ is the overtone number.

This formalism is highly reliable for black hole spacetimes characterized by smooth effective potentials with a single dominant peak \cite{VBolokhov,YZhao,Skvortsova,MSkvortsova,SVBolokhov}. In particular, the first-order WKB approximation reproduces the eikonal regime and becomes increasingly accurate in the large multipole limit ($\ell \rightarrow \infty 
$), where the perturbation wavelength is much shorter than the characteristic variation scale of the potential barrier. Higher-order corrections substantially improve the accuracy for the fundamental modes and low overtones satisfying ($\ell >n$) \cite{BAhmed,Oglialoro}. Nevertheless, the convergence of the WKB series is not always uniform, meaning that increasing the expansion order does not necessarily guarantee a systematic enhancement of numerical precision.

\subsection{Temporal evolution}

To verify the QNM obtained via the WKB approximation, we employ the finite-difference method \cite{Melgar}. This approach allows us to directly monitor the dynamical evolution of linear perturbations in the black hole background. After an appropriate rescaling of the temporal coordinate, the master wave equation, Eq.(\ref{39}), can be recast as
\begin{equation}
\left[ \frac{\partial ^{2}}{\partial x^{2}}-\frac{\partial ^{2}}{\partial
t^{2}}-\mathcal{U}_{\text{eff}}\right] \Phi \left( x,t\right) =0. \label{t61}
\end{equation}%
By introducing a discrete grid defined by $t=t_{0}+j\Delta t$ and $x=x_{0}+k\Delta x$,
the finite-difference discretization of Eq.(\ref{t61}) yields the integration scheme
\begin{equation}
\Phi _{k}^{j+1}=\Phi _{k}^{j-1}+\left( \frac{\Delta t}{\Delta x}\right)
^{2}\left( \Phi _{k+1}^{j}+\Phi _{k-1}^{j}\right) +\left( 2-2\left( \frac{%
\Delta t}{\Delta x}\right) ^{2}-\left( \Delta t\right) ^{2}\left( \mathcal{U}%
_{\text{eff}}\right) _{k}\right) \Phi _{k}^{j}.
\end{equation}

The initial perturbation is modeled as a Gaussian wave packet profiles at $t=0$,
\begin{equation}
\Phi \left( x,t\right) =\exp \left[ -\frac{\left( x-c_{b}\right) ^{2}}{%
2\sigma }\right] ,
\end{equation}

supplemented by the initial condition $\left. \Phi \left( x,t\right) =0\right\vert _{t<0}$, where $c_{b}$ and $\sigma $ denote the center and width of the initial wave
packet, respectively.

By iterating this finite-difference scheme forward in time for a fixed grid ratio $\frac{%
\Delta t}{\Delta x}$, we numerically evolve the time-domain profiles of the perturbation field. To guarantee the numerical stability of the time evolution, the Courant–Friedrichs–Lewy stability criterion
\begin{equation}
\frac{\Delta t}{\Delta x}<1.
\end{equation}%
Subsequently, the Prony method is employed to extract the complex quasinormal frequencies from the late-time numerical signal. In this framework, the ringdown profile is modeled as a linear combination of exponentially damped sinusoids \cite{Berti}:

\begin{equation}
\Phi \left( t\right) =\sum_{s}C_{s}e^{-i\omega _{s}t}.
\end{equation}

where $C_{s}$ are complex amplitudes and $\omega _{s}$ are the quasinormal frequencies. Fitting the numerical data to this ansatz yields the quasinormal spectrum with high accuracy.

\subsection{Numerical results}

The quasinormal spectrum was computed using the sixth-order WKB approximation and verified through finite-difference method combined with the Prony method. Since the WKB approach is most accurate for $\ell>n$, we focus primarily on this regime.

\begin{figure}[H]
\begin{minipage}[t]{0.5\textwidth}
        \centering
        \includegraphics[width=\textwidth]{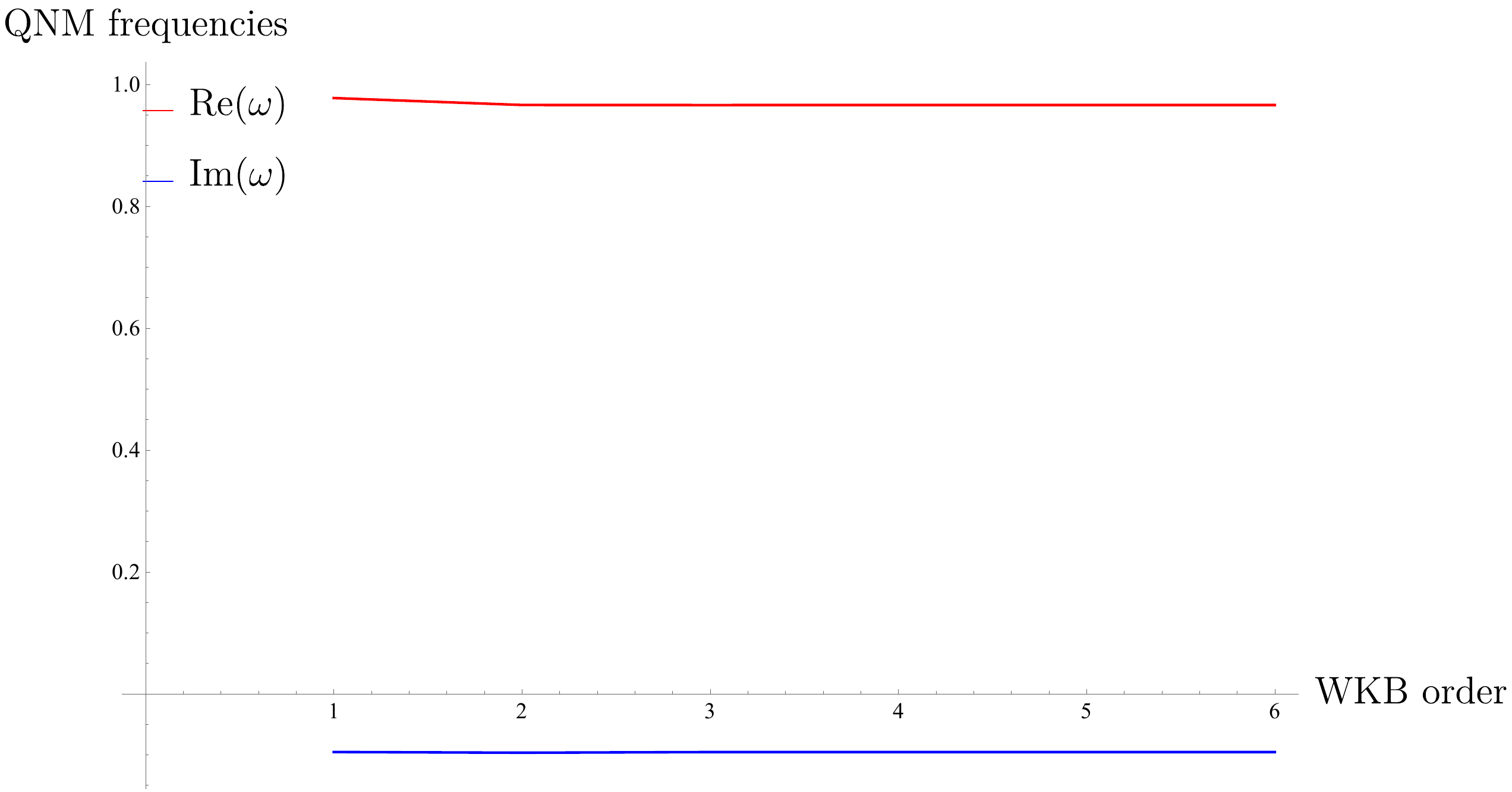}
                \subcaption{$\eta=1$}
        \label{fig:Q1}
\end{minipage}
\begin{minipage}[t]{0.5\textwidth}
        \centering
        \includegraphics[width=\textwidth]{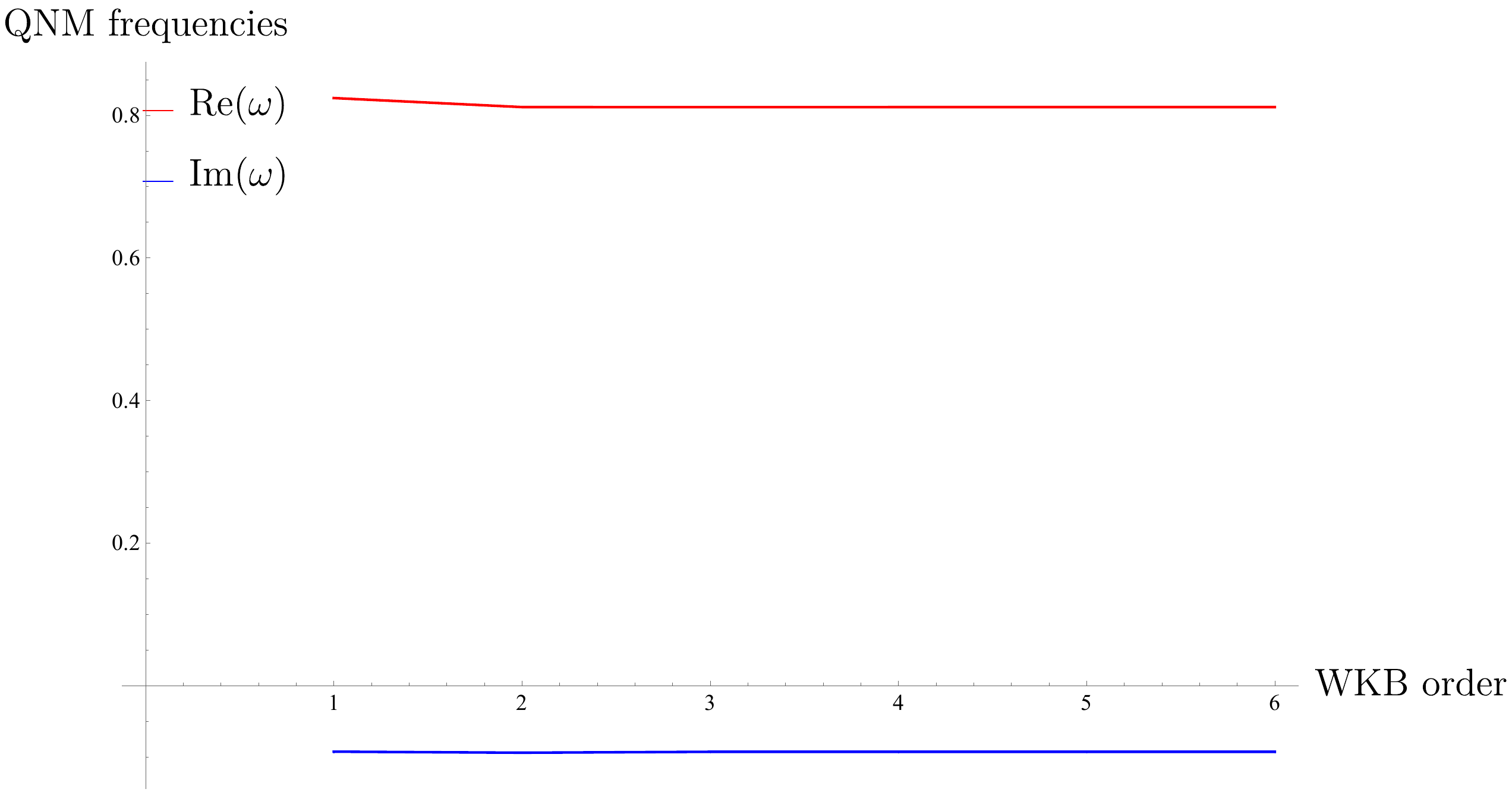}
                \subcaption{$\eta=-1$}
        \label{fig:Q2}
\end{minipage}
\caption{Dependence of the QNM frequencies on the WKB order for the black-hole spacetime described by  (\ref{20}). Parameters: $M=1$, $q=0.7$ and $\protect\alpha =0.1$, $%
\ell =4$, $n=0$.}
\label{fig:11}
\end{figure}

Figure(\ref{fig:11}) and Table \ref{tab:Comparison} demonstrate the rapid convergence of the WKB expansion. The discrepancy between third- and sixth-order calculations decreases with increasing multipole number, confirming the reliability of the method for low-lying modes.

\begin{table}[H]
\centering
\begin{tabular}{c|c|c|c|c}
\hline\hline
$\ell$ & $n$ & \multicolumn{1}{c|}{WKB6} & \multicolumn{1}{c|}{WKB3} & $|%
\text{WKB6}-\text{WKB3}|$ \\ \hline
1 & 0 & $0.326559 - 0.096763i$ & $0.324722 - 0.096503i$ & 0.001855 \\ 
2 & 0 & $0.538808 - 0.095889i$ & $0.538331 - 0.095845i$ & 0.000479 \\ 
& 1 & $0.522815 - 0.291556i$ & $0.521556 - 0.291632i$ & 0.001261 \\ 
3 & 0 & $0.752228 - 0.095688i$ & $0.752044 - 0.095677i$ & 0.000184 \\ 
& 1 & $0.740386 - 0.289088i$ & $0.739879 - 0.289119i$ & 0.000508 \\ 
& 2 & $0.718377 - 0.488204i$ & $0.718381 - 0.487249i$ & 0.000955 \\ 
4 & 0 & $0.966042 - 0.095611i$ & $0.965954 - 0.095607i$ & 0.000088 \\ 
& 1 & $0.956695 - 0.288068i$ & $0.956447 - 0.288081i$ & 0.000248 \\ 
& 2 & $0.938798 - 0.484102i$ & $0.938892 - 0.483708i$ & 0.000405 \\ 
& 3 & $0.913890 - 0.685663i$ & $0.915259 - 0.683035i$ & 0.002964 \\ 
5 & 0 & $1.180040 - 0.095573i$ & $1.179990 - 0.095571i$ & 0.000050 \\ 
& 1 & $1.172330 - 0.287550i$ & $1.172190 - 0.287556i$ & 0.000140 \\ 
& 2 & $1.157360 - 0.481961i$ & $1.157440 - 0.481773i$ & 0.000204 \\ 
& 3 & $1.135980 - 0.680238i$ & $1.137020 - 0.678911i$ & 0.001685 \\ 
& 4 & $1.109430 - 0.883516i$ & $1.112140 - 0.878940i$ & 0.005318 \\ 
\hline\hline
\end{tabular}%
\caption{QNM frequencies computed using the third- and sixth-order WKB methods for different values of $\ell $ and $n$ .
The parameters are fixed to $M=1$,$q=0.7$,$\alpha=0.1$,$\protect\eta =1$,$\mu =0$} 
\label{tab:Comparison}
\end{table}
Table\ref{tab:I} summarizes the influence of the GB coupling on the quasinormal spectrum. In both electromagnetic sectors, increasing $\alpha$ leads to larger values of $\operatorname{Re}(\omega)$ and smaller values of $|\operatorname{Im}(\omega)|$. Thus, higher curvature corrections increase the oscillation frequency while reducing the damping rate of the perturbations. Moreover, the Maxwell branch consistently yields larger frequencies than the phantom-field branch. This behavior can be attributed to the different contributions of the electromagnetic field to the effective potential. In the Maxwell case, the potential barrier becomes higher, enhancing the confinement of scalar perturbations, whereas the phantom field lowers the barrier and consequently reduces both the oscillation frequency and the damping rate.

\begin{table}[H]
\centering
\begin{tabular}{c|c|c|c|c}
\hline\hline
\multicolumn{1}{c|}{\multirow{2}{*}{$\alpha$}} &  
\multicolumn{2}{c|}{$\eta = 1$} & 
\multicolumn{2}{c}{$\eta = -1$} \\ 
\cmidrule(lr){2-3} \cmidrule(lr){4-5} & WKB6 & $\Delta$ & WKB6 & $\Delta$ \\ 
\midrule
0.1 & $0.510723 - 0.0959884i$ & $2.78 \times 10^{-5}$ & $0.468167 -
0.0941199i$ & $2.52 \times 10^{-5}$ \\ 
0.2 & $0.515781 - 0.0938395i$ & $5.45 \times 10^{-5}$ & $0.471431 -
0.0927841i$ & $4.96 \times 10^{-5}$ \\ 
0.3 & $0.521190 - 0.0914471i$ & $7.90 \times 10^{-5}$ & $0.474837 -
0.0913690i$ & $7.74 \times 10^{-5}$ \\ 
0.4 & $0.527012 - 0.0887355i$ & $9.74 \times 10^{-5}$ & $0.478405 -
0.0898572i$ & $1.06 \times 10^{-4}$ \\ 
0.5 & $0.533320 - 0.0855974i$ & $1.07 \times 10^{-4}$ & $0.482156 -
0.0882284i$ & $1.25 \times 10^{-4}$ \\ 
0.6 & $0.540197 - 0.0818718i$ & $8.31 \times 10^{-5}$ & $0.486114 -
0.0864583i$ & $1.44 \times 10^{-4}$ \\ 
0.7 & $0.547715 - 0.0773025i$ & $4.49 \times 10^{-5}$ & $0.490307 -
0.0845167i$ & $1.55 \times 10^{-4}$ \\ 
0.8 & $0.555867 - 0.0714739i$ & $6.19 \times 10^{-5}$ & $0.494766 -
0.0823652i$ & $1.26 \times 10^{-4}$ \\ 
0.9 & $0.563866 - 0.0635085i$ & $7.38 \times 10^{-4}$ & $0.499528 -
0.0799535i$ & $9.85 \times 10^{-5}$ \\ \hline\hline
\end{tabular}%
\caption{QNM dependence on GB coupling $\alpha $ for both the
ordinary Maxwell sector and the phantom sector. Parameters: $M=1$, $q=0.5$,$%
\ell =2$, $n=0$}
\label{tab:I}
\end{table}

\begin{figure}[H]
\begin{minipage}[t]{0.32\textwidth}
        \centering
        \includegraphics[width=\textwidth]{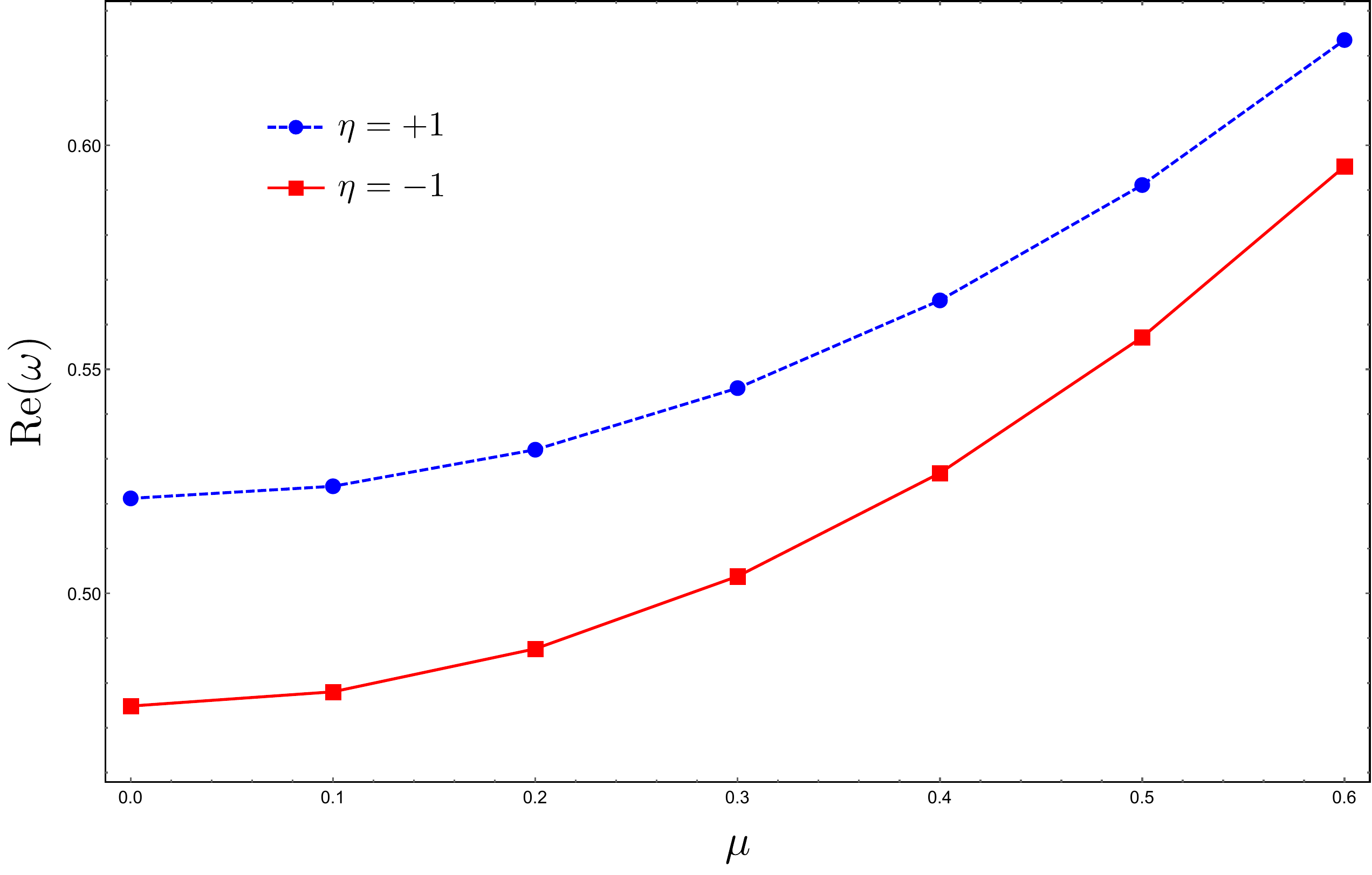}
        \label{fig:Q11}
\end{minipage}
\begin{minipage}[t]{0.32\textwidth}
        \centering
        \includegraphics[width=\textwidth]{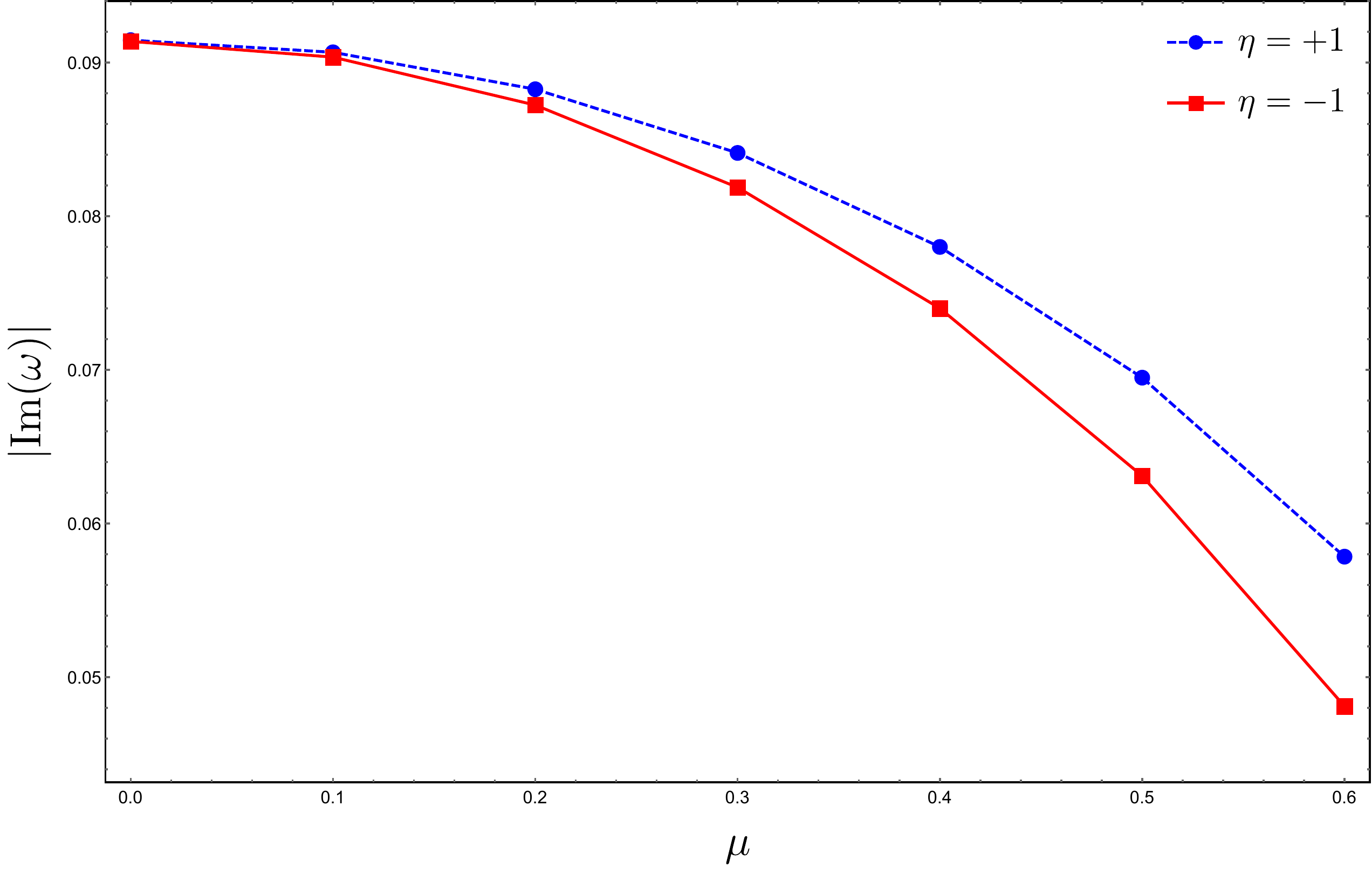}
        \label{fig:Q22}
\end{minipage}
\begin{minipage}[t]{0.32\textwidth}
        \centering
        \includegraphics[width=\textwidth]{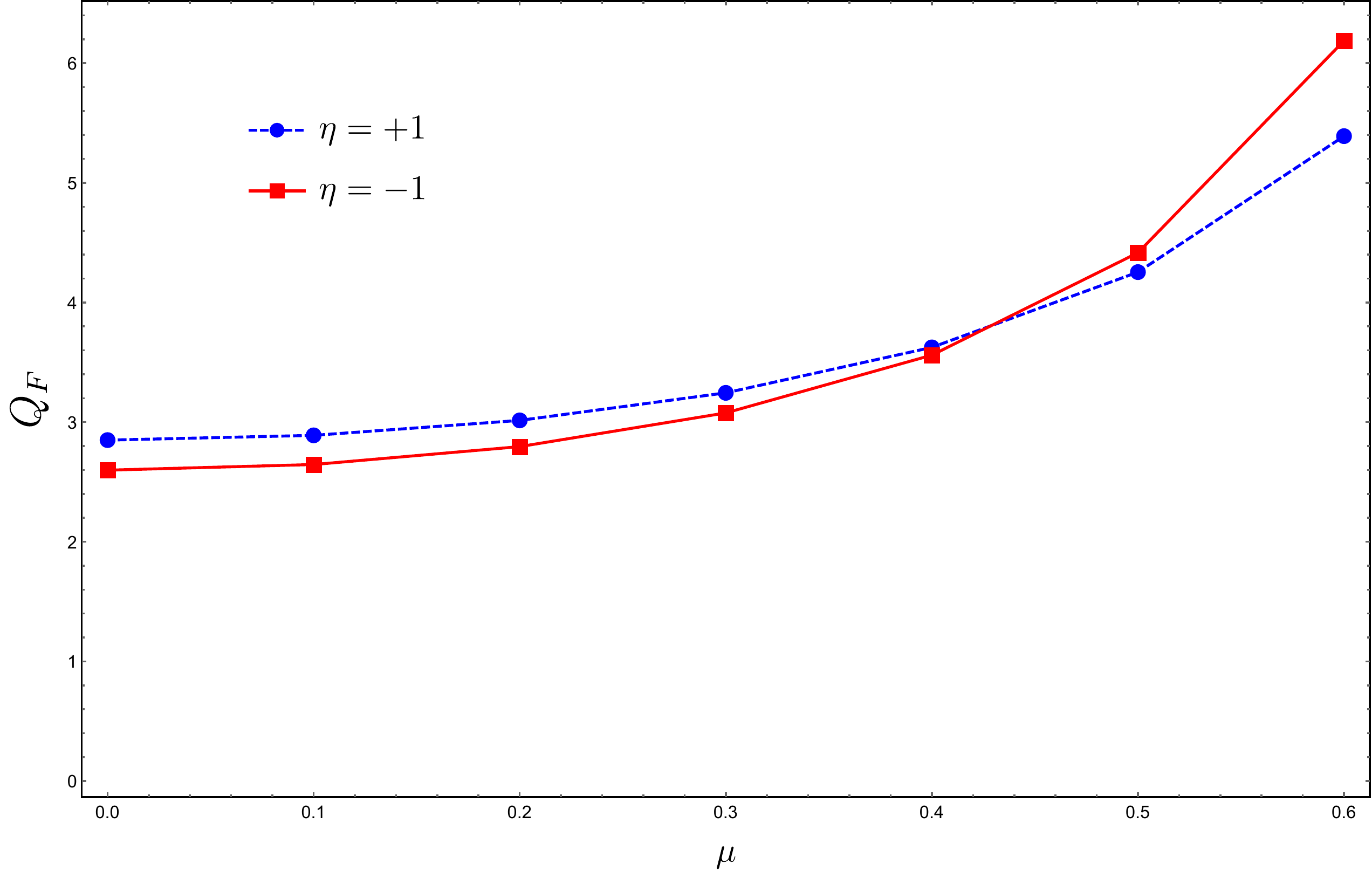}
        \label{fig:33}
\end{minipage}
\caption{QNM frequencies as functions of the scalar field mass $\mu$. Parameters:$%
M=1$, $q=0.5$, $\protect\alpha =0.3$, $\ell =2$, $n=0$}
\label{qnmmass}
\end{figure}

As shown in Fig.(\ref{fig:ueffB2}), increasing the scalar field mass $\mu$ causes the peak of the potential to grow. Beyond a threshold value, however, the peak height drops below the asymptotic value $\mu^2$, a regime in which QNMs cannot exist. With a further increase in $\mu$, the peak vanishes entirely, transforming the potential barrier into a potential step. Consequently, QNMs exist exclusively when the maximum of the effective potential, $\mathcal{U}_{\text{eff}}\left( x=x_{0}\right) $, exceeds $\mu^{2}$.
The corresponding QNM frequencies are displayed in Fig.(\ref{qnmmass}). In both the Maxwell and phantom field sectors, the oscillation frequency increases with $\mu$, while the damping rate decreases. Consequently, massive scalar perturbations become longer lived and are characterized by larger quality factors, indicating more persistent oscillatory behavior.

\begin{figure}[H]
\begin{minipage}[t]{0.32\textwidth}
        \centering
        \includegraphics[width=\textwidth]{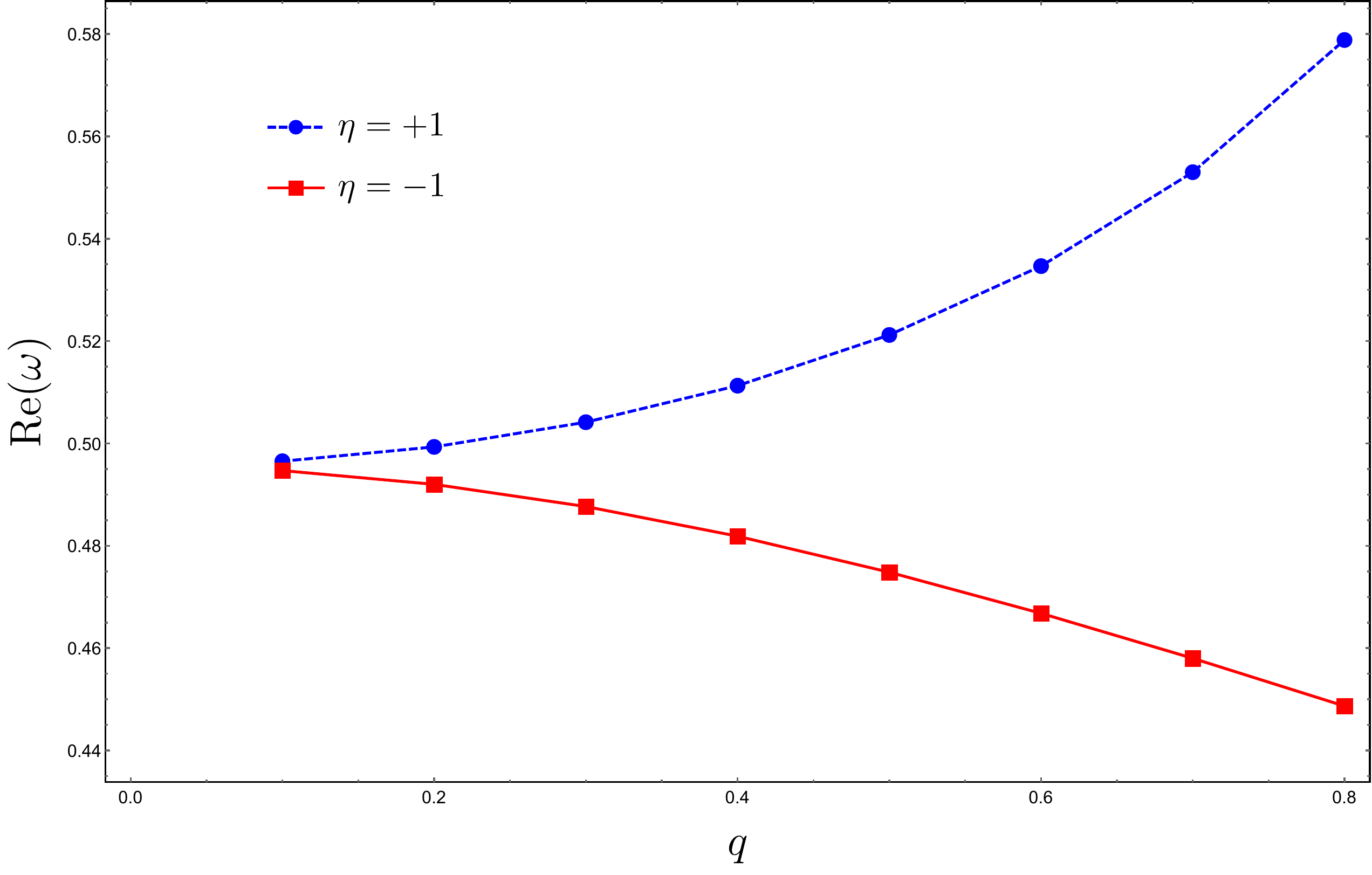}
        \label{fig:Q15}
\end{minipage}
\begin{minipage}[t]{0.32\textwidth}
        \centering
        \includegraphics[width=\textwidth]{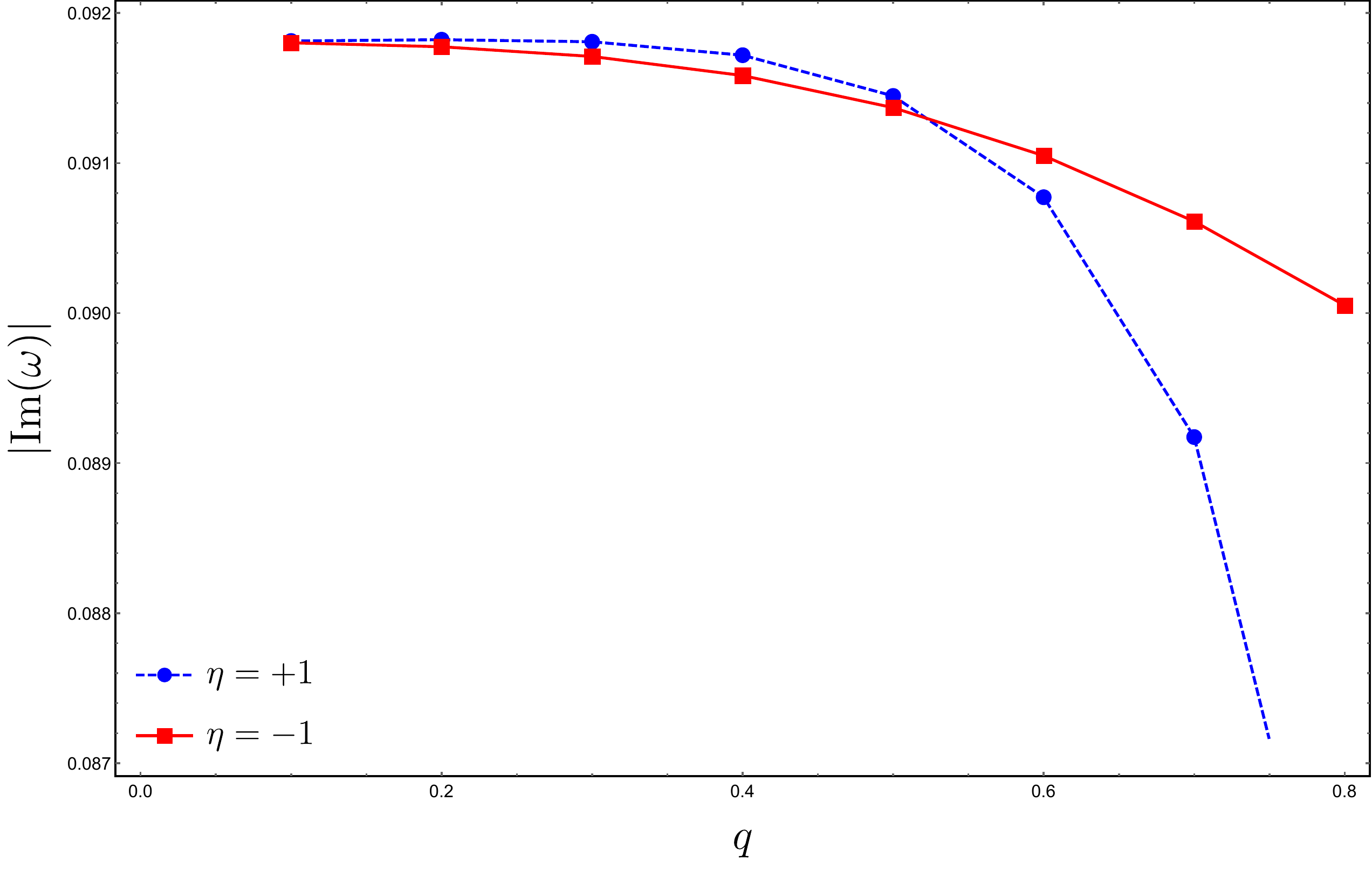}
        \label{fig:Q25}
\end{minipage}
\begin{minipage}[t]{0.32\textwidth}
        \centering
        \includegraphics[width=\textwidth]{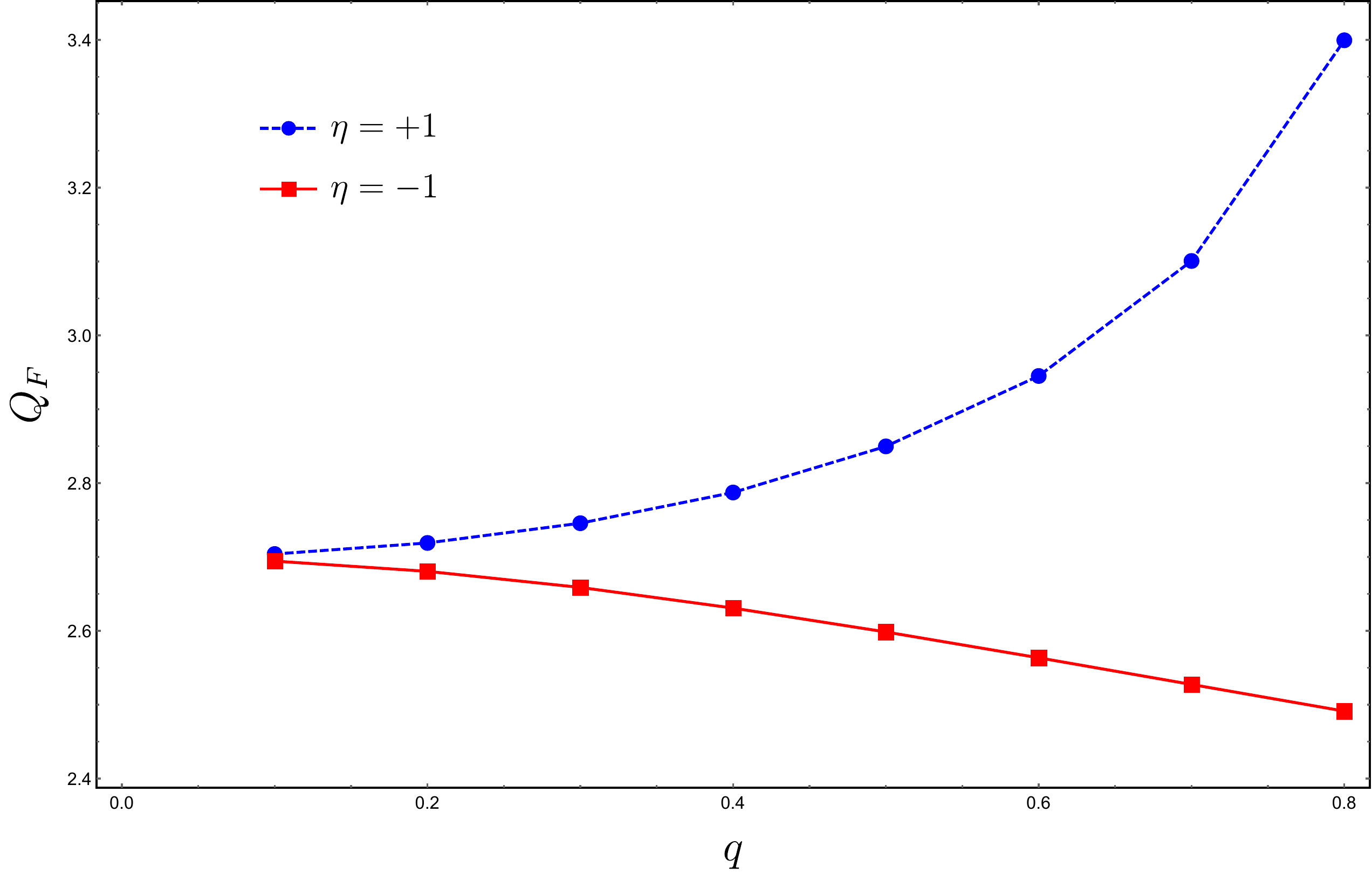}
        \label{fig:35}
\end{minipage}
\caption{QNM dependence on the electric charge $q$. Parameters: $M=1$, $%
\mu =0$, $\alpha =0.3$, $\ell =2$, $n=0$}
\label{fig:charge}
\end{figure}

Fig.\ref{fig:charge} illustrates the QNMs as a function of the electric charge $q$ for both the Maxwell and phantom sectors, exhibiting distinct behaviors in the real and imaginary parts of the frequency. In the Maxwell sector, the real part $\operatorname{Re}(\omega)$ increases with $q$, whereas it decreases in the phantom sector. This divergence stems from the opposing contributions of the electromagnetic fields to the effective potential: the Maxwell field enhances the effective restoring force, while the phantom field weakens it due to the reversed sign of its energy-momentum tensor. In both sectors, the magnitude of the imaginary part $\lvert\operatorname{Im}(\omega) \rvert$ decreases as $q$ increases, indicating slower damping and longer-lived perturbations for highly charged configurations. Consequently, the quality factor increases with $q$ in the Maxwell case, leading to more sustained oscillations, whereas it decreases in the phantom case, reflecting a suppression of the oscillatory behavior. Overall, these results demonstrate that the underlying nature of the electromagnetic sector dictates both the oscillation frequency and the damping rate, yielding qualitatively distinct dynamical responses.

\begin{figure}[H]
\begin{minipage}[t]{0.5\textwidth}
        \centering
        \includegraphics[width=\textwidth]{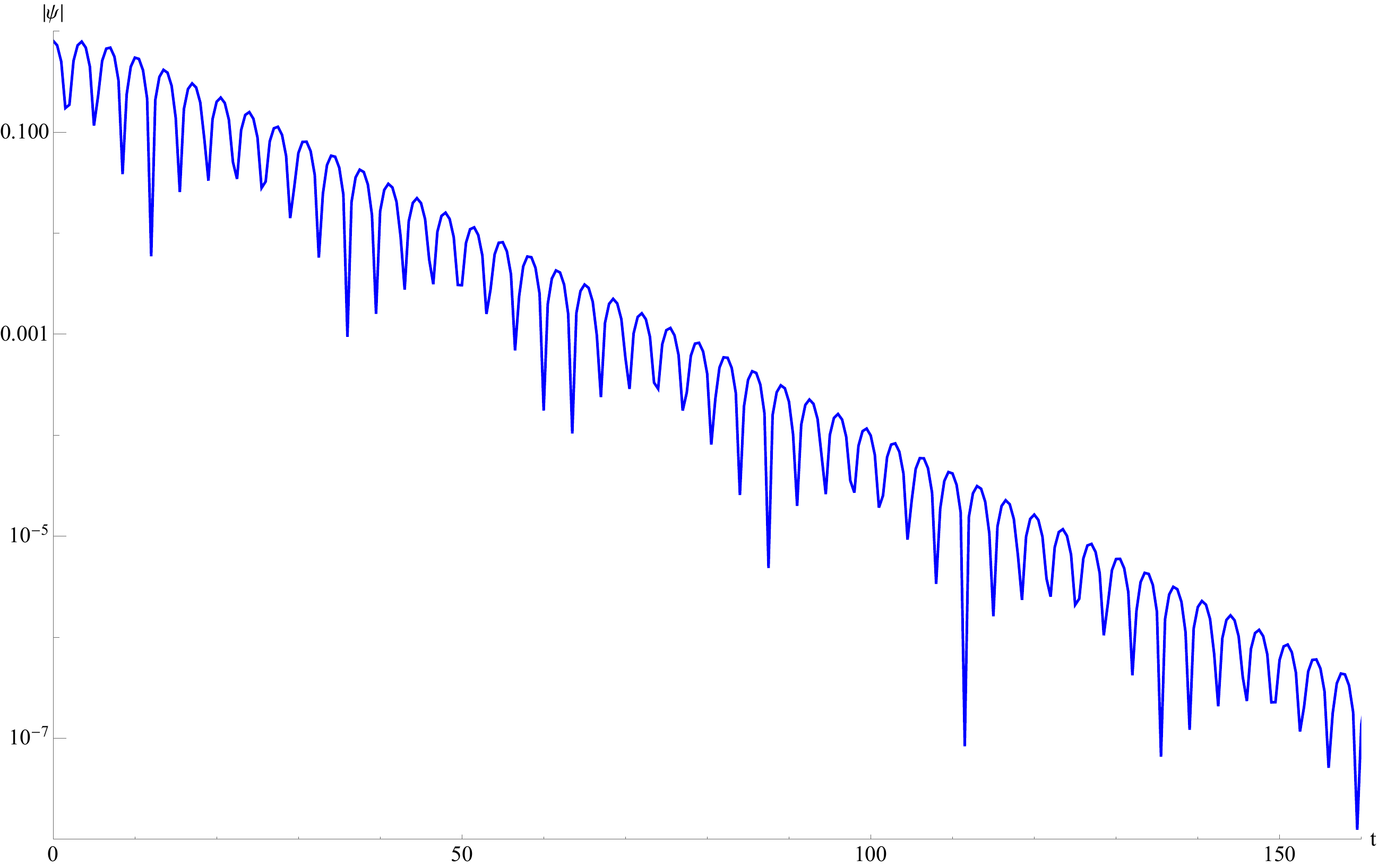}
        \subcaption{$\eta=+1$}
        \label{fig:TQ15}
\end{minipage}
\begin{minipage}[t]{0.5\textwidth}
        \centering
        \includegraphics[width=\textwidth]{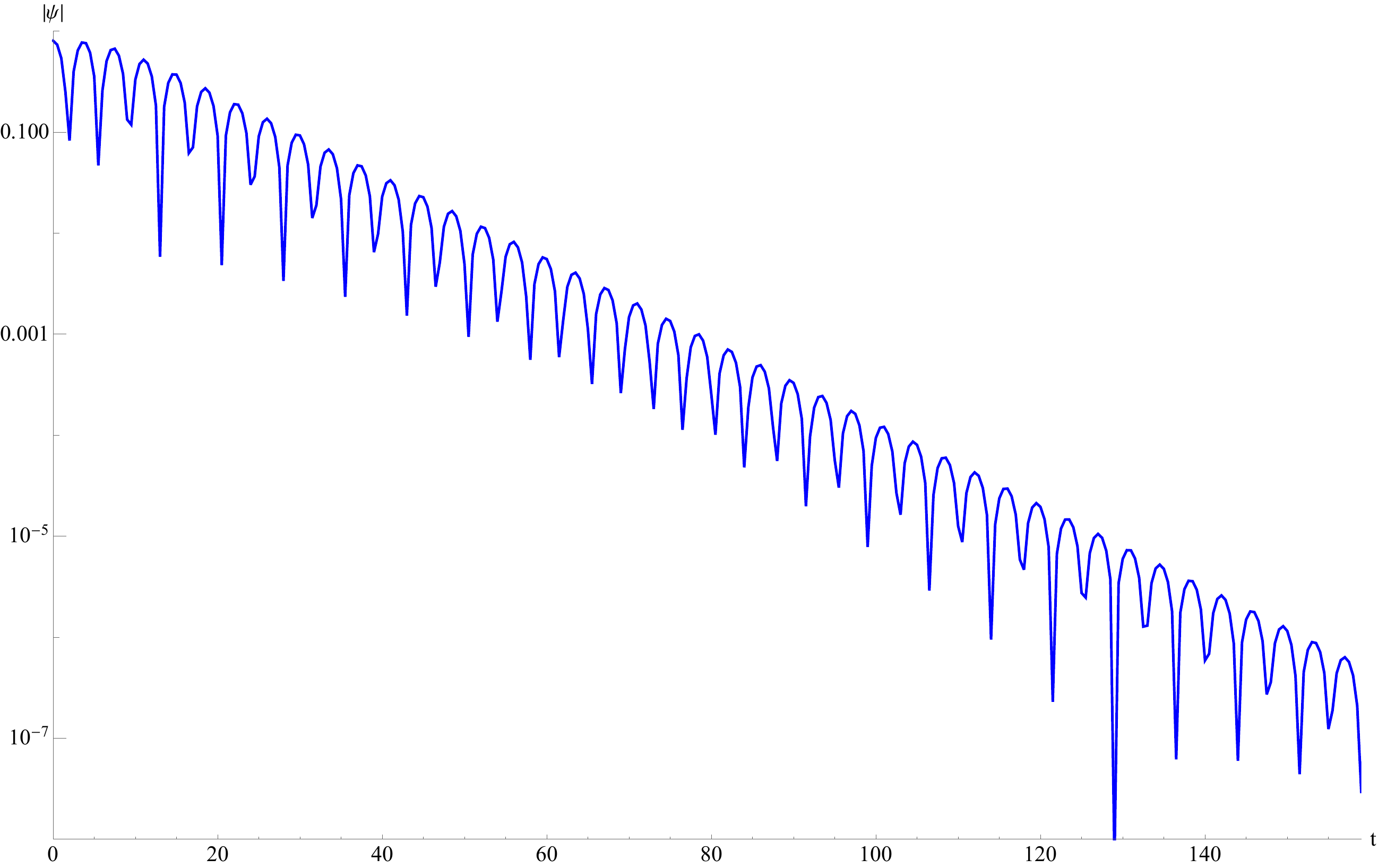}
        \subcaption{$\eta=-1$}
        \label{fig:TQ25}
\end{minipage}
\caption{The time evolutions of massless perturbations at $M=1$, $q=0.5$, $%
\protect\alpha =0.1$, $\ell =4$. Left: the WKB-6 data $\omega %
=0.91573-0.0956722i$ with the extraction of the frequency from the finite
difference method combined with the Prony fitting technique $\protect\omega %
=0.917404-0.0890426i$ leading to a relative error of about 0.74\%. Right:
the WKB-6 result $\omega =0.839625-0.0937623i$ with the extraction
of the frequency from the finite difference method combined with the Prony
fitting technique $\omega =0.841761-0.0880312i$ leading to a
relative error of about 0.72\%.}
\label{fig:fed}
\end{figure}

The time domain evolution presented in Fig.\ref{fig:fed} is in good agreement with the WKB results. The dominant quasinormal frequencies extracted via the Prony method are $\omega = 0.917404 - 0.0890426i$ for the Maxwell field and $\omega = 0.841761 - 0.0880312i$ for the phantom field. These values are highly consistent with the corresponding WKB predictions of $\omega = 0.91573 - 0.0956722i$ and $\omega = 0.839625 - 0.0937623i$, yielding relative deviations of approximately 0.74\% and 0.72\%, respectively. This close agreement between the two independent approaches confirms the accuracy of the computed quasinormal mode spectrum and validates the reliability of the numerical implementation.

Table \ref{tab:extract} presents a comparison between the QNM frequencies obtained via the sixth-order WKB approximation and those extracted using the finite difference method combined with the Prony fitting technique for $\ell =0$ and $n=0$. The relative discrepancy ranges from approximately 1.46\%--5.97\%. This deviation is attributed to the reduced accuracy of the WKB method in the $\ell =n$, regime, where its underlying eikonal-type assumptions become less reliable.

\begin{table}[H]
\centering%
\begin{tabular}{l|l|l|l|l}
\hline\hline
\multicolumn{5}{c}{$q=0.5$} \\ \hline
\rowcolor{lightgray} $\eta$ & $\alpha$ & WKB  & Finite difference method & Relative errors \\ \hline
+1 & 0.1 & 0.117644 - 0.102235i & 0.116278 - 0.0979886i & 2.86\% \\ 
 & 0.2 & 0.120193 - 0.0999828i & 0.121449 - 0.0906481i & 5.97\% \\ \hline
-1 & 0.1 & 0.107631 - 0.099495i & 0.110552 - 0.0997659i & 1.99\% \\ 
 & 0.2 & 0.109268 - 0.0990435i & 0.111243 - 0.0978883i & 1.55\% \\ 
\hline\hline
\multicolumn{5}{c}{$\alpha=0.1$} \\ \hline
\rowcolor{lightgray} $\eta$ & $q$ & WKB  & Finite difference method & Relative errors \\ \hline
+1 & 0.6 & $0.120634 - 0.102254i$ & $0.121737 - 0.099956i$ & 1.61\% \\ 
 & 0.7 & $0.125052 - 0.100963i$ & $0.124495 - 0.0986921i$ & 1.46\% \\ \hline
-1 & 0.6 & $0.105865 - 0.0987657i$ & $0.108572 - 0.0986059i$ & 1.87\% \\ 
 & 0.7 & $0.103916 - 0.0978949i$ & $0.107006 - 0.0983379i$ & 2.19\% \\ 
\hline\hline
\end{tabular}%
\caption{QNM extracted from 6th-order WKB method and the finite difference
method combined with the Prony fitting technique for $M=1$, $\ell =0$ and $n=0$}
\label{tab:extract}
\end{table}
In the eikonal regime, an expansion with respect to the small parameters $%
\left( \alpha ,q^{2}\right) $ yields the location of the maximum of the
effective potential as

\begin{equation}
r_{\max }=3m-\frac{2\eta q^{2}}{3m}-\frac{4\alpha }{9m}-\frac{4\alpha \eta
q^{2}}{81m^{3}}.
\end{equation}

Substituting this result into the first-order WKB approximation, the
quasinormal frequencies in the eikonal limit can be expressed as

\begin{eqnarray}
\omega  &=&\frac{1}{3m\sqrt{3}}\left( \ell +\frac{1}{2}+i\left( n+\frac{1}{2}%
\right) \right) +\left( \frac{\eta q^{2}}{18\sqrt{3}m^{3}}+\frac{2\alpha }{81%
\sqrt{3}m^{3}}+\frac{5\alpha \eta q^{2}}{486\sqrt{3}m^{5}}\right) \left(
\ell +\frac{1}{2}\right)   \notag \\
&&+i\left( \frac{\eta q^{2}}{54\sqrt{3}m^{3}}-\frac{4\alpha }{81\sqrt{3}m^{3}%
}-\frac{44\alpha \eta q^{2}}{729\sqrt{3}m^{5}}\right) \left( n+\frac{1}{2}%
\right) .
\end{eqnarray}

In the eikonal limit $\ell \rightarrow \infty $, quasinormal modes are governed by the properties of unstable photon orbits. Following \cite{66}, the QNM frequencies can be expressed as

\begin{equation}
\omega =\Omega _{c}\left( \ell +\frac{1}{2}\right) -i\lambda \left( n+\frac{1%
}{2}\right) .
\end{equation}%
where $\Omega _{c}$ denotes the angular frequency of the circular, and $\lambda$measures its instability. For a static spherically symmetric spacetime,
\begin{equation}
\Omega _{c}=\frac{1}{R_{\text{Sh}}}=\frac{\sqrt{f\left( r_{\text{p}}\right) }%
}{r_{\text{p}}},
\end{equation}

with $r_{\text{p}}$ representing the photon-sphere radius, and $R_{\text{Sh}}$ the corresponding critical impact parameter. The instability of the orbit is quantified by

\begin{equation}
\lambda =\sqrt{\frac{f\left( r_{\text{p}}\right) }{2r_{\text{p}}^{2}}\left[
2f\left( r_{\text{p}}\right) -r_{\text{p}}^{2}f\left( r_{\text{p}}\right)
^{\prime \prime }\right] },
\end{equation}

These relations provide a geometric interpretation of the eikonal QNM spectrum in terms of photon sphere properties. Although this correspondence is known to admit exceptions in certain gravitational theories \cite{70,71}, it remains valid for the class of black hole solutions considered here.

The QNM spectrum also contains information about wave scattering. In particular, the grey body factor  $\Upsilon _{\ell }\left(
\Omega \right) $ characterizes the transmission of radiation through the effective potential barrier. Near a resonance frequency, it can be approximated by the Breit–Wigner form \cite{83}

\begin{equation}
\Upsilon _{\ell }\left( \Omega \right) =\left[ 1+\exp \left( \frac{2\pi
\left( \Omega ^{2}-\operatorname{Re}\left( \omega _{0}\right) ^{2}\right) }{4\operatorname{Re}
\left( \omega _{0}\right) \operatorname{Im}\left( \omega _{0}\right) }\right) %
\right] ^{-1}.
\end{equation}%
Accordingly, the real part of the quasinormal frequency determines the resonance location, while the imaginary part controls its width. This relation allows the grey body factors to be estimated directly from the QNM spectrum without performing a complete scattering analysis.
\section{Gravitational wave constraints}

\label{sec6}
The gravitational wave observations of binary black hole mergers by the
LIGO-Virgo Collaboration provide a powerful probe of deviations from general
relativity and can be used to constrain modified theories of gravity
\cite{Abbott3,BAbbott4,140}. Here, we derive an order-of-magnitude bound on the parameter
combination $\beta \equiv \alpha + \eta q^2$ using the heuristic argument
proposed in \cite{140}. Our purpose is not to replace a full
numerical-relativity analysis in 4DEGB phantom gravity framework, but rather to estimate the
constraining power of current gravitational-wave observations.

For a black hole of mass $M_{i}$, the horizon radius is 
\begin{equation}
r_{+}(M_{i})=M_{i}+\sqrt{M_{i}^{2}-\left( \alpha +\eta q^{2}\right) }.
\label{eq:74}
\end{equation}

As a representative example, we consider the GW150914 event. The
gravitational-wave frequency at the peak of the signal is $f_{\text{GW}}
\simeq 150~\mathrm{Hz}$, while the measured chirp mass is $M_c \sim 30
M_{\odot}$. Assuming an equal-mass binary, these values correspond to
component masses $M_1 \simeq M_2 \simeq 35 M_{\odot}$ \cite{Abbott3,BAbbott4,140}. Since $f_{%
\text{GW}}$ is directly inferred from the observed waveform and $M_c$ is
determined during the weak-field inspiral phase, these estimates are
expected to remain approximately valid in both general relativity and 4DEGB
gravity.

Using the Newtonian relation between orbital frequency and separation, the
binary separation at the time of peak emission is estimated as 
\begin{equation}
R = \left( \frac{M_1+M_2}{\pi^{2} f_{\text{GW}}^{2}}\right)^{1/3} \sim 350~%
\mathrm{km}.
\end{equation}

To compare the orbital separation with the size of the horizons, we define 
\begin{equation}
\mathcal{R} = \frac{R}{r_{+}(M_1)+r_{+}(M_2)}.
\end{equation}

Requiring that the two horizons remain distinct at the peak of the signal, $%
\mathcal{R}>1$, yields a lower bound on negative values of $\beta$: 
\begin{equation}
\beta \gtrsim -10^{10}~\mathrm{m}^{2}.
\end{equation}

An upper bound follows from the reality condition of the horizon radius in
Eq.~(\ref{eq:74}), namely $M_{i}^{2}-\beta \geq 0$. For $M_{i}=35M_{\odot }$%
, this condition implies 
\begin{equation}
\alpha +\eta q^{2}\lesssim 10^{9}~\mathrm{m}^{2}.
\end{equation}

Combining the two constraints, we obtain 
\begin{equation}
-10^{10}~\mathrm{m}^{2}\lesssim \alpha +\eta q^{2}\lesssim 10^{9}~\mathrm{m}%
^{2}.  \label{eq:77}
\end{equation}

The bound in Eq.~(\ref{eq:77}) is obtained from a simple phenomenological
argument and should therefore be interpreted as an order-of-magnitude
estimate. Nevertheless, it indicates that current gravitational wave
observations already constrain the parameter combination $\alpha +\eta q^{2}$
to be significantly smaller than the characteristic horizon scale of
stellar-mass black holes $M^{2}\sim 10^{9}~\mathrm{m}^{2}$ for $35M_{\odot }$%
. A more robust determination would require waveform modeling and
numerical relativity simulations within 4DEGB phantom gravity framework.
\section{Conclusions}

\label{sec7}
We have studied electrically charged black holes in regularized 4D EGB gravity coupled to a phantom electromagnetic field. The resulting solutions provide a useful framework for investigating the combined influence of higher curvature effects and exotic matter sources on black-hole physics.

The presence of the GB coupling and the phantom field produces significant modifications of the spacetime geometry. In contrast to the ordinary Maxwell sector, the phantom configuration remains free of branch singularities for positive values of the physical parameters considered here. The horizon structure is also strongly affected, with the phantom charge increasing the event horizon radius, whereas the standard electromagnetic charge produces the opposite behavior.

The thermodynamic analysis reveals a clear distinction between the two electromagnetic sectors. While the ordinary charged solution exhibits an extremal configuration together with a thermodynamic phase transition associated with a divergent heat capacity, no analogous critical behavior is found in the phantom case. Instead, the heat capacity remains negative throughout the physical domain, indicating persistent thermodynamic instability.

The influence of the phantom field extends to the dynamics of test particles. Stable circular orbits are displaced toward larger radii, accompanied by a reduction in accretion efficiency relative to the standard Maxwell branch. These results suggest that phantom field weakens the effective gravitational binding in the vicinity of the black hole.

We have also investigated the behavior of the geometry under massive scalar perturbations. The effective potential remains positive outside the event horizon, supporting the linear stability of the background. The QNM spectrum shows that the GB coupling increases the oscillation frequencies and reduces the damping rates, whereas the phantom field suppresses both quantities. 

The combined effects of higher curvature corrections and phantom field therefore generate distinctive signatures in the thermodynamic, geodesic, and perturbative properties of the black hole. These features may provide additional avenues for testing departures from General Relativity in strong gravity environments.

\section*{Conflict of Interests}

The author declare no such conflict of interest.

\section*{Data Availability Statement}

No data were generated or analyzed in this study.


\begin{thebibliography}{99}
\bibitem{Austin} A. Joyce, B. Jain, J. Khoury and Mark Trodden, \href{https://doi.org/10.1016/j.physrep.2014.12.002}%
{Physics Reports \textbf{568},1 (2015)}

\bibitem{Leor} L. Barack et al.,\href{https://doi.org/10.1088/1361-6382/ab0587}%
{Class. Quantum. Grav. \textbf{36}, 143001 (2019)}

\bibitem{Abbott} B.\thinspace P. Abbott et al.,\href{https://doi.org/10.1103/PhysRevLett.116.061102}%
{Phys. Rev. Lett. \textbf{116}, 061102 (2016)}

\bibitem{BAbbott} B.\thinspace P. Abbott et al., \href{https://doi.org/10.1103/PhysRevLett.116.241103}%
{Phys. Rev. Lett. \textbf{116}, 241103(2016)}

\bibitem{BPAbbott} B.\thinspace P. Abbott et al.,\href{https://doi.org/10.1103/PhysRevLett.118.221101}%
{Phys. Rev. Lett. \textbf{118}, 221101 (2017)}

\bibitem{RAbbott} B.\thinspace P. Abbott et al.,\href{https://doi.org/10.1103/PhysRevLett.119.141101}%
{Phys. Rev. Lett. \textbf{119}, 141101 (2017)}

\bibitem{Lovelock} D. Lovelock, \href{https://doi.org/10.1063/1.1665613}{J.
Math. Phys. \textbf{12}, 498 (1971)}

\bibitem{Gustavo} G. Dotti, R. J. Gleiser,\href{https://doi.org/10.1088/0264-9381/22/1/L01}%
{Class. Quantum Grav. \textbf{22}, L1 (2025)}

\bibitem{Dott} G. Dotti, R. J. Gleiser, \href{https://doi.org/10.1103/PhysRevD.72.044018}%
{Phys. Rev. D \textbf{72}, 044018 (2005)}

\bibitem{Gleiser} R. J. Gleiser, G. Dotti, \href{https://doi.org/10.1103/PhysRevD.72.124002}%
{Phys. Rev. D \textbf{72}, 124002 (2005)}

\bibitem{Beroiz} M. Beroiz, G. Dotti, R. J. Gleiser, \href{https://doi.org/10.1103/PhysRevD.76.024012}%
{Phys. Rev. D 76, 024012 (2007)}

\bibitem{Zhidenko} R. A. Konoplya, A. Zhidenko, \href{https://doi.org/10.1103/PhysRevD.77.104004}%
{Phys. Rev. D \textbf{77}, 104004 (2008)}

\bibitem{PCandelas} P. Candelas, G. T. Horowitz, A. Strominger, E. Witten, 
\href{https://doi.org/10.1016/0550-3213(85)90602-9}{Nucl. Phys. B \textbf{258%
}, 46 (1985)}

\bibitem{Barton} B. Zwiebach, \href{https://doi.org/10.1016/0370-2693(85)91616-8}%
{Phys. Lett. B \textbf{156}, 315 (1985)}

\bibitem{Bruno} B. Zumino, \href{https://doi.org/10.1016/0370-1573(86)90076-1}%
{Physics Reports \textbf{137}, 109 (1986)}

\bibitem{Gross} D. J. Gross and J. H. Sloan, \href{https://doi.org/10.1016/0550-3213(87)90465-2}%
{Nucl. Phys. B \textbf{291}, 41 (1987)}

\bibitem{Lin} D. Glavan, C. Lin, \href{https://doi.org/10.1016/j.physletb.2020.135468}%
{Phys. Lett.B \textbf{805}, 135468 (2020)}

\bibitem{Tomozawa} Y.Tomozawa, \href{https://doi.org/10.48550/arXiv.1107.1424}%
{arXiv:1107.1424}

\bibitem{Cognola} G. Cognola, R. Myrzakulov, L. Sebastiani, S. Zerbini, 
\href{https://doi.org/10.1103/PhysRevD.88.024006}{Phys. Rev. D \textbf{88},
024006 (2013)}

\bibitem{Ohta} Rong-Gen Cai, Li-Ming Cao, N. Ohta,\href{https://doi.org/10.1007/JHEP04(2010)082}%
{ J. High Energ. Phys. \textbf{2010}, 82 (2010)}

\bibitem{Pang} H. Lu, Yi Pang, \href{https://doi.org/10.1016/j.physletb.2020.135717}%
{Phys. Lett. B \textbf{809}, 135717 (2020)}

\bibitem{Aoki} K. Aoki, M. A. Gorji and S. Mukohyama, \href{https://doi.org/10.1016/j.physletb.2020.135843}%
{Phys. Lett. B \textbf{810}, 135843 (2020)}

\bibitem{CLiu} C. Liu, T. Zhu and Q. Wu, \href{https://doi.org/10.1088/1674-1137/abc16c}%
{Chin. Phys. C \textbf{45}, 015105 (2021)}

\bibitem{Guo} M. Guo and P. C. Li, \href{https://doi.org/10.1140/epjc/s10052-020-8164-7}%
{Eur. Phys. J. C \textbf{80}, 588 (2020)}

\bibitem{Wei} SW. Wei, YX. Liu, \href{https://doi.org/10.1140/epjp/s13360-021-01398-9}%
{Eur. Phys. J. Plus \textbf{136}, 436 (2021)}

\bibitem{Zubair} M. Zubair, M. A. Raza,\href{https://doi.org/10.1016/j.dark.2023.101200}%
{Phys. Dark Universe \textbf{40}, 101200 (2023)}

\bibitem{Malafarina} D. Malafarina, B. Toshmatov and N. Dadhich, \href{https://doi.org/10.1016/j.dark.2020.100598}%
{Phys. Dark Universe \textbf{30}, 100598 (2020)}

\bibitem{Mansoori} S. A. H. Mansoori, \href{https://doi.org/10.1016/j.dark.2021.100776}%
{Phys. Dark Universe \textbf{31}, 100776 (2021)}

\bibitem{XHGe} X. H. Ge and S. J. Sin, \href{https://doi.org/10.1140/epjc/s10052-020-8288-9}%
{Eur. Phys. J. C \textbf{80}, 695 (2020)}

\bibitem{Rayimbaev} J. Rayimbaev, A. Abdujabbarov, B. Turimov and F.
Atamurotov, \href{10.1016/j.dark.2020.100715}{Physics of the Dark Universe \textbf{30}, 100715 (2020)}


\bibitem{Chakraborty} S. Chakraborty and N. Dadhich, \href{https://doi.org/10.1016/j.dark.2020.100658}%
{Phys. Dark Universe \textbf{30}, 100658 (2020)}

\bibitem{Odintsov} S.D. Odintsov, V.K. Oikonomou, \href{https://doi.org/10.1016/j.physletb.2020.135437}%
{Phys.Lett B \textbf{805}, 135437 (2020)}

\bibitem{KYang} Z. C. Lin, K. Yang, S. W. Wei, Y. Q. Wang and Y. X. Liu, 
\href{https://doi.org/10.1140/epjc/s10052-020-08612-5}{Eur.Phys. J. C 
\textbf{80}, 1033 (2020)}

\bibitem{BAhmedov} S. Shaymatov, J. Vrba, D. Malafarina, B. Ahmedov and Z.
Stuchlik, \href{https://doi.org/10.1016/j.dark.2020.100648}{Phys. Dark
Universe \textbf{30}, 100648 (2020)}

\bibitem{RKumar} S. Ul Islam, R. Kumar, S. G. Ghosh, \href{https://doi.org/10.1088/1475-7516/2020/09/030}%
{JCAP \textbf{2020},30 (2020)}

\bibitem{SGGhosh} S. G. Ghosh, D. V. Singh, R.Kumar, S. D. Maharaj, \href{https://doi.org/10.1016/j.aop.2020.168347}%
{Ann. Phys.\textbf{424}, 168347 (2021)}

\bibitem{Larranaga} J. M. Ladino1, E. Larranaga, \href{https://doi.org/10.1007/s10773-023-05440-7}%
{Int. J.Theo. Phys. \textbf{62}, 209 (2023)}

\bibitem{Hamil} B.Hamil and T. Birkandan, \href{https://doi.org/10.1016/j.nuclphysb.2025.117145}%
{Nucl. Phys. B \textbf{1020}, 117145 (2025)}
\bibitem{Ayoub} B.Hamil, \href{https://doi.org/10.1142/S0217751X2650017X}%
{Int. J. Mod. Phys. \textbf{41}, 2650017 (2026)}
%

\bibitem{Caldwell} R.R. Caldwell, \href{https://doi.org/10.1016/S0370-2693(02)02589-3}{Phys. Lett. B \textbf{545}, 23 (2002)}

\bibitem{Nojiri} S. Nojiri, S.D. Odintsov, \href{https://doi.org/10.1016/S0370-2693(03)00594-X}{Phys. Lett. B \textbf{562}, 147 (2003)}


\bibitem{Elizalde} E. Elizalde, S. Nojiri, S.D. Odintsov, \href{https://doi.org/10.1103/PhysRevD.70.043539}{Phys. Rev. D 
\textbf{70}, 043539 (2004)}

\bibitem{Melchiorri} A.Melchiorri, L. Mersini, C.J. Odman, M. Trodden, \href{https://doi.org/10.1103/PhysRevD.68.043509}{Phys.
Rev. D \textbf{68}, 043509 (2003)}

\bibitem{Clement} G. Cl\'{e}ment, J.C. Fabris, M.E. Rodrigues, \href{https://doi.org/10.1103/PhysRevD.79.064021}{Phys. Rev. D 
\textbf{79}, 064021 (2009)}

\bibitem{ronnikov} K.A. Bronnikov, R.A. Konoplya, A. Zhidenko, \href{https://doi.org/10.1103/PhysRevD.86.024028}{Phys. Rev. D
\textbf{86}, 024028 (2012)}

\bibitem{Jamil} M.Jamil, I. Hussain, M. U. Farooq, \href{https://doi.org/10.1007/s10509-011-0762-2}{Astrophys. Space Sci. 
\textbf{335}, 339 (2011)}

\bibitem{Jardim} D.F. Jardim, M.E. Rodrigues, S.J.M. Houndjo, \href{https://doi.org/10.1140/epjp/i2012-12123-x}{Eur. Phys. J.
Plus \textbf{127}, 123 (2012)}

\bibitem{Izquierdo} G. Izquierdo, D. Pavon, \href{10.1016/j.physletb.2005.12.040}{Phys. Lett. B \textbf{633}, 420
(2006)}

\bibitem{Sadjadi} H.M. Sadjadi, \href{10.1016/j.physletb.2006.12.029}{Phys. Lett. B \textbf{645}, 108 (2007)}


\bibitem{Ding} C.Ding, C. Liu, Y. Xiao, L. Jiang, R. G. Cai, \href{https://doi.org/10.1103/PhysRevD.88.104007}{Phys. Rev. D 
\textbf{88}, 104007 (2013)}
\bibitem{Maria} B. Hamil, B.C. Lütfüoğlu, \href{https://doi.org/10.1140/epjc/s10052-025-14047-7}{Eur. Phys. J. C \textbf{85}, 313 (2025)}
\bibitem{MJamil} M.Jamil, D.Momeni, K.Bamba, R.Myrzakulov, \href{https://doi.org/10.1142/S0218271812500654}{Int. J. Mod.
Phys. D \textbf{21}, 1250065 (2012)}

\bibitem{Quevedo} H. Quevedo, M.N. Quevedo, A. Sanchez, \href{https://doi.org/10.1140/epjc/s10052-016-3949-4}{Eur. Phys. J. C \textbf{76},
110 (2016)}

\bibitem{Zotos} A.S. Mohamed, E.E. Zotos, \href{10.1016/j.ascom.2024.100881}{Astron. Comput. \textbf{48}, 100862 (2024)}

\bibitem{MAkbar} M.Jamil, M. Akbar, \href{https://doi.org/10.1007/s10714-010-1024-2}{Gen. Relativ. Gravit. \textbf{43}, 1061 (2011)}

\bibitem{Eslam} B.Eslam Panah, M.E.Rodrigues, \href{https://doi.org/10.1140/epjc/s10052-023-11402-4}{Eur. Phys. J. C \textbf{83}, 237
(2023)}

\bibitem{BEslam} B.Eslam-Panah, M.E.Rodrigues, \href{https://doi.org/10.1140/epjc/s10052-024-13485-z}{Eur. Phys. J. C \textbf{84}, 1125
(2024)}


\bibitem{Lanczos} C. Lanczos, \href{https://doi.org/10.2307/1968467}{Ann. of
Math. \textbf{39}, 842 (1938).}

\bibitem{Wheeler} J.T. Wheeler, \href{https://doi.org/10.1016/0550-3213(86)90388-3}%
{Nucl. Phys.B \textbf{268}, 737 (1986)} 
\bibitem{77} I. D. Novikov and K. S. Thorne, in Black Holes (Les n Black Holes (Les
Astres Occlus) (1973) pp. 343-450

\bibitem{Vishveshwara} C.V. Vishveshwara, \href{https://doi.org/10.1038/227936a0}%
{Nature \textbf{227}, 936 (1970)}

\bibitem{Kokkotas} K.D. Kokkotas, B.G. Schmidt, \href{https://doi.org/10.12942/lrr-1999-2}%
{Living Rev. Relativ. \textbf{2}, 2 (1999). }

\bibitem{Hegde} S. S. Hegde, V. Subramanyan, B. Bradlyn, and S.
Vishveshwara, \href{https://doi.org/10.1103/PhysRevLett.123.156802}{Phys.
Rev. Lett. \textbf{123}, 156802 (2019)}

\bibitem{Schutz} R. A. Konoplya,\href{https://doi.org/10.1103/RevModPhys.83.793}%
{Rev. Mod. Phys. \textbf{83}, 793 (2011)}

\bibitem{Iyer} S. Iyer and C. M. Will, \href{https://doi.org/10.1103/PhysRevD.35.3621}%
{Phys. Rev. D \textbf{35}, 3621 (1987)}

\bibitem{Konoplya} R. A. Konoplya, \href{https://doi.org/10.1103/PhysRevD.68.024018}%
{Phys. Rev. D \textbf{68}, 024018 (2003)}

\bibitem{KonoplyaRA} R. A.Konoplya, \href{https://doi.org/10.30970/jps.08.93}%
{J. Phys. Stud. \textbf{8}, 93 (2004)}

\bibitem{Hamil1} B.Hamil, B. C. Lutfuoglu, \href{https://doi.org/10.1002/prop.202400105}%
{Fortsch. Phys \textbf{73}, 2400105 (2025) }

\bibitem{Hamil2} B.Hamil, B. C. Lutfuoglu, \href{https://doi.org/10.1016/j.aop.2024.169861}%
{Ann. Phys. \textbf{472}, 169861 (2025)}

\bibitem{Hamil3} B.Hamil, B. C. Lutfuoglu, \href{https://doi.org/10.1016/j.dark.2024.101484}%
{Dark Universe \textbf{44}, 101484 (2024)}

\bibitem{Hamil4} B.Hamil, B. C. Lutfuoglu, \href{https://doi.org/10.1007/s10714-025-03478-y}%
{Gen Relativ Gravit \textbf{57}, 140 (2025)}

\bibitem{VBolokhov} S. V. Bolokhov and V. D. Ivashchuk, \href{https://doi.org/10.1140/epjc/s10052-022-10578-5}{Eur. Phys. J. C 
\textbf{82}, 624 (2022)}

\bibitem{YZhao} Y. Zhao, X. Ren, A. Ilyas, E. N. Saridakis, and Y.-F.Cai, \href{https://doi.org/10.1088/1475-7516/2022/10/087}{
JCAP \textbf{10}, 087 (2022)}

\bibitem{Skvortsova} M. Skvortsova, \href{ https://doi.org/10.1002/prop.202400132}{Fortsch. Phys. \textbf{72}, 2400132 (2024)}

\bibitem{MSkvortsova} M. Skvortsova, \href{ https://doi.org/10.1002/prop.202400036}{Fortsch. Phys. \textbf{72}, 2400036 (2024)}




\bibitem{SVBolokhov} S.V.Bolokhov, M. Skvortsova, \href{https://doi.org/10.1140/epjc/s10052-026-15624-0}{Eur. Phys. J. C \textbf{86}, 374 (2026)}

\bibitem{BAhmed} R. A. Konoplya, D. Ovchinnikov, and B. Ahmedov, \href{https://doi.org/10.1103/PhysRevD.108.104054}{Phys. Rev.
D \textbf{108}, 104054 (2023)}

\bibitem{Oglialoro} A. M. Bonanno, R. A. Konoplya, G. Oglialoro, and A.
Spina, \href{https://doi.org/10.1088/1475-7516/2025/12/042}{JCAP \textbf{12}, 042 (2025)}

\bibitem{Melgar} B. Cuadros-Melgar, J. de Oliveira, C. E. Pellicer, \href{https://doi.org/10.1103/PhysRevD.85.024014}{
Phys.Rev. D \textbf{85}, 024014 (2012)}

\bibitem{Berti}  E. Berti, V. Cardoso, J. A. Gonzalez, and U.
Sperhake, \href{https://doi.org/10.1103/PhysRevD.75.124017}{Phys. Rev. D \textbf{75}, 124017 (2007)}
\bibitem{66} R. A. Konoplya, \href{https://doi.org/10.1016/j.physletb.2023.137674}{Phys. Lett. B \textbf{838}, 137674 (2023)}
\bibitem{70} G. Khanna and R. H. Price, \href{https://doi.org/10.1103/PhysRevD.95.081501}{Phys. Rev. D \textbf{95}, 081501 (2017)}
\bibitem{71}] R. A. Konoplya and Z. Stuchlík, \href{https://doi.org/10.1016/j.physletb.2017.06.015}{Phys. Lett. B \textbf{771}, 597 (2017)}
\bibitem{83} ] R. A. Konoplya and A. Zhidenko, \href{10.1016/j.physletb.2025.139288}{Phys. Lett. B \textbf{861}, 139288 (2025)}
\bibitem{Abbott3} B. Abbott et al. (LIGO Scientific, Virgo), \href{https://doi.org/10.1103/PhysRevLett.116.061102}{Phys. Rev. Lett. \textbf{116}, 061102 (2016)}
\bibitem{BAbbott4}B. Abbott et al. (LIGO Scientific, Virgo), \href{https://doi.org/10.1103/PhysRevLett.116.241102}{Phys. Rev. Lett. \textbf{116}, 241102 (2016)}
\bibitem{140}B. P. Abbott et al. (LIGO Scientific,Virgo), \href{ https://doi.org/10.1002/andp.201600209}{Annalen Phys. \textbf{529}, 1600209 (2017)}



\end{thebibliography}
\end{document}